\documentclass[onecolumn]{aastex63}
\usepackage{epsf}
\usepackage{graphicx}
\usepackage{amsmath}
\usepackage{amssymb}
\usepackage{gensymb}
\usepackage{chngcntr}
\usepackage{longtable}
\usepackage{latexsym}

\newcommand{\nth}{N$_2$H$^+$}
\newcommand{\ntd}{N$_2$D$^+$}
\newcommand{\hcn}{HC$_3$N}
\newcommand{\cch}{CCS-H}
\newcommand{\ccl}{CCS-L}

    {\pagebreak[4]\global\pdfpageattr\expandafter{\the\pdfpageattr/Rotate 90}}%
    {\pagebreak[4]\global\pdfpageattr\expandafter{\the\pdfpageattr/Rotate 0}}%
    
\received{2021 February 17}
\accepted{2021 June 7}
\submitjournal{ApJS}

\usepackage{natbib,graphicx,morefloats,multirow,pbox,booktabs, makecell}
\usepackage{hyperref}
\usepackage{soul}

\begin{document}

\title{Molecular Cloud Cores with High Deuterium Fractions: Nobeyama Mapping Survey}

\correspondingauthor{Ken'ichi Tatematsu}
\email{k.tatematsu@nao.ac.jp}

\author[0000-0002-8149-8546]{Ken'ichi Tatematsu}
\affil{Nobeyama Radio Observatory, National Astronomical Observatory of Japan,
National Institutes of Natural Sciences,
462-2 Nobeyama, Minamimaki, Minamisaku, Nagano 384-1305, Japan}
\affiliation{Department of Astronomical Science,
The Graduate University for Advanced Studies, SOKENDAI,
2-21-1 Osawa, Mitaka, Tokyo 181-8588, Japan}

\author[0000-0003-2011-8172]{Gwanjeong Kim}
\affil{Nobeyama Radio Observatory, National Astronomical Observatory of Japan,
National Institutes of Natural Sciences,
462-2 Nobeyama, Minamimaki, Minamisaku, Nagano 384-1305, Japan}

\author[0000-0002-5286-2564]{Tie Liu}
\affiliation{Shanghai Astronomical Observatory, Chinese Academy of Sciences, 80 Nandan Road, Shanghai 200030, P. R. China}
\affiliation{Korea Astronomy and Space Science Institute,
Daedeokdaero 776, Yuseong, Daejeon 305-348, South Korea}
\affiliation{East Asian Observatory, 660 N. A'ohoku Place, Hilo, HI 96720, USA}

\author[0000-0001-5175-1777]{Neal J. Evans II}
\affiliation{Department of Astronomy, The University of Texas at Austin, 2515 Speedway, Stop C1400, Austin, TX 78712$-$1205, USA}

\author{Hee-Weon Yi}
\affiliation{School of Space Research, Kyung Hee University, Seocheon-Dong, Giheung-Gu, Yongin-Si, Gyeonggi-Do, 446-701, Republic of Korea}

\author[0000-0003-3119-2087]{Jeong-Eun Lee}
\affiliation{School of Space Research, Kyung Hee University, Seocheon-Dong, Giheung-Gu, Yongin-Si, Gyeonggi-Do, 446-701, Republic of Korea}

\author{Yuefang Wu}
\affiliation{Department of Astronomy, Peking University, 100871, Beijing, China}

\author[0000-0001-9304-7884]{Naomi Hirano}
\affiliation{Academia Sinica Institute of Astronomy and Astrophysics, 11F of Astronomy-Mathematics Building, AS/NTU. No.1, Sec. 4, Roosevelt Rd, Taipei 10617, Taiwan, R.O.C.}

\author[0000-0003-4603-7119]{Sheng-Yuan Liu}
\affiliation{Academia Sinica Institute of Astronomy and Astrophysics, 11F of Astronomy-Mathematics Building, AS/NTU. No.1, Sec. 4, Roosevelt Rd, Taipei 10617, Taiwan, R.O.C.}

\author[0000-0002-2338-4583]{Somnath Dutta}
\affiliation{Academia Sinica Institute of Astronomy and Astrophysics, 11F of Astronomy-Mathematics Building, AS/NTU. No.1, Sec. 4, Roosevelt Rd, Taipei 10617, Taiwan, R.O.C.}

\author[0000-0002-4393-3463]{Dipen Sahu}
\affiliation{Academia Sinica Institute of Astronomy and Astrophysics, 11F of Astronomy-Mathematics Building, AS/NTU. No.1, Sec. 4, Roosevelt Rd, Taipei 10617, Taiwan, R.O.C.}

\author[0000-0002-7125-7685]{Patricio Sanhueza}
\affiliation{National Astronomical Observatory of Japan,
National Institutes of Natural Sciences,
2-21-1 Osawa, Mitaka, Tokyo 181-8588, Japan}
\affiliation{Department of Astronomical Science,
The Graduate University for Advanced Studies, SOKENDAI,
2-21-1 Osawa, Mitaka, Tokyo 181-8588, Japan}

\author[0000-0003-2412-7092]{Kee-Tae Kim}
\affiliation{Korea Astronomy and Space Science Institute,
Daedeokdaero 776, Yuseong, Daejeon 305-348, Republic of Korea}
\affiliation{University of Science and Technology, Korea (UST), 217 Gajeong-ro, Yuseong-gu, Daejeon 34113, Republic of Korea}

\author[0000-0002-5809-4834]{Mika Juvela}
\affiliation{Department of Physics, P.O. Box 64, FI-00014, University of Helsinki, Finland}

\author[0000-0002-5310-4212]{L. Viktor T\'{o}th}
\affiliation{Department of Astronomy, E\"{o}tv\"os Lor\'{a}nd University, P\'{a}zm\'{a}ny P\'{e}ter s\'{e}t\'{a}ny 1/A, H-1117 Budapest, Hungary}

\author[0000-0003-3453-4775]{Orsolya Feh\'{e}r}
\affiliation{Institut de Radio Astronomie Millim\'{e}trique, 300 rue de la Piscine, Domaine Universitaire 38406 Saint-Martin-d'Heres, France}

\author[0000-0002-3938-4393]{Jinhua He}
\affiliation{Yunnan Observatories, Chinese Academy of Sciences, 396 Yangfangwang, Guandu District, Kunming, 650216, China}
\affiliation{Chinese Academy of Sciences South America Center for Astronomy, National Astronomical Observatories, CAS, Beijing 100101, China}
\affiliation{Departamento de Astronom\'ia, Universidad de Chile, Casilla 36-D, Santiago, Chile}

\author{J. X. Ge}
\affiliation{Chinese Academy of Sciences, South America Center for Astrophysics (CASSACA), Camino El Observatorio 1515, 
Las Condes, Santiago, Chile}
\affiliation{Departamento de Astronom\'ia, Universidad de Chile, Casilla 36-D, Santiago, Chile}

\author[0000-0002-4707-8409]{Siyi Feng}
\affil{Academia Sinica Institute of Astronomy and Astrophysics, No.1, Sec.
4, Roosevelt Rd, Taipei 10617, Taiwan, Republic of China}
\affil{CAS Key Laboratory of FAST, National Astronomical Observatories,
Chinese Academy of Sciences, Beijing 100101, People's Republic of China}
\affil{National Astronomical Observatory of Japan, National Institutes of
Natural Sciences, 2-21-1 Osawa, Mitaka, Tokyo 181-8588, Japan}

\author{Minho Choi}
\affiliation{Korea Astronomy and Space Science Institute,
Daedeokdaero 776, Yuseong, Daejeon 305-348, South
Korea}

\author[0000-0002-5016-050X]{Miju Kang}
\affiliation{Korea Astronomy and Space Science Institute,
Daedeokdaero 776, Yuseong, Daejeon 305-348, South
Korea}

\author{Mark A. Thompson}
\affiliation{Centre for Astrophysics Research, Science \& Technology Research Institute, University of Hertfordshire, Hatfield, AL10 9AB, UK}

\author[0000-0001-8509-1818]{Gary A. Fuller}
\affiliation{Jodrell Bank Centre for Astrophysics, School of Physics and Astronomy, University of Manchester, Oxford Road, Manchester, M13 9PL, UK}

\author[0000-0003-3010-7661]{Di Li}
\affiliation{National Astronomical Observatories, Chinese Academy of Sciences, Beijing, 100012, China}

\author{Isabelle Ristorcelli}
\affiliation{RAP, CNRS (UMR5277), Universit\'e Paul Sabatier, 9 avenue du Colonel Roche, BP 44346, F-31028, Toulouse Cedex 4, France}

\author[0000-0002-7237-3856]{Ke Wang}
\affiliation{The Kavli Institute for Astronomy and Astrophysics, Peking University,
5 Yiheyuan Road, Haidian District, Beijing 100871, P. R. China}
\affiliation{European Southern Observatory, Karl-Schwarzschild-Str. 2 D-85748 Garching bei M\"{u}nchen, Germany}

\author{James Di Francesco}
\affiliation{Herzberg Astronomy \& Astrophysics, National Research Council of Canada, 5071 West Saanich Road, Victoria, BC V9E 2E7, Canada}
\affiliation{Department of Physics and Astronomy, University of Victoria, Victoria, BC V8W 2Y2, Canada}

\author[0000-0002-5881-3229]{David Eden}
\affiliation{Astrophysics Research Institute, Liverpool John Moores University, IC2, 
Liverpool Science Park, 146 Brownlow Hill, Liverpool L3 5RF, UK}

\author[0000-0002-9661-7958]{Satoshi Ohashi}
\affiliation{The Institute of Physical and Chemical Research (RIKEN), 2-1, Hirosawa, Wako-shi, Saitama 351-0198, Japan}

\author[0000-0003-2610-6367]{Ryo Kandori}
\affiliation{Astrobiology Center of NINS,
2-21-1 Osawa, Mitaka, Tokyo 181-8588, Japan}

\author{Charlotte Vastel}
\affiliation{IRAP, Universit\'{e} de Toulouse, CNRS, CNES, UPS, (Toulouse), France}

\author[0000-0003-1659-095X]{Tomoya Hirota}
\affiliation{National Astronomical Observatory of Japan,
National Institutes of Natural Sciences,
2-21-1 Osawa, Mitaka, Tokyo 181-8588, Japan}
\affiliation{Department of Astronomical Science,
The Graduate University for Advanced Studies, SOKENDAI,
2-21-1 Osawa, Mitaka, Tokyo 181-8588, Japan}

\author[0000-0003-4521-7492]{Takeshi Sakai}
\affiliation{Graduate School of Informatics and Engineering, The University of Electro-Communications, Chofu, Tokyo 182-8585, Japan}

\author[0000-0003-2619-9305]{Xing Lu}
\affiliation{National Astronomical Observatory of Japan,
National Institutes of Natural Sciences,
2-21-1 Osawa, Mitaka, Tokyo 181-8588, Japan}

\author{Quang Nguy$\tilde{\hat{e}}$n Lu'o'ng}
\affiliation{McMaster University, 1 James Street North, Hamilton, ON, L8P 1A2, Canada}
\affiliation{Korea Astronomy and Space Science Institute,
Daedeokdaero 776, Yuseong, Daejeon 305-348, South
Korea}
\affiliation{National Astronomical Observatory of Japan,
National Institutes of Natural Sciences,
2-21-1 Osawa, Mitaka, Tokyo 181-8588, Japan}
\affiliation{IBM, Canada}

\author{Hiroko Shinnaga}
\affiliation{Department of Physics, Kagoshima University, 1-21-35, Korimoto, Kagoshima, 890-0065, Japan}

\author{Jungha Kim}
\affiliation{Korea Astronomy and Space Science Institute,
Daedeokdaero 776, Yuseong, Daejeon 305-348, Republic of Korea}
\affiliation{National Astronomical Observatory of Japan,
National Institutes of Natural Sciences,
2-21-1 Osawa, Mitaka, Tokyo 181-8588, Japan}
\affiliation{Department of Astronomical Science,
SOKENDAI (The Graduate University for Advanced Studies),
2-21-1 Osawa, Mitaka, Tokyo 181-8588, Japan}

\author{JCMT Large Program ``SCOPE'' collaboration}

\begin{abstract}
We present the results of on-the-fly mapping observations of 44 fields containing 107 SCUBA-2 cores in the emission lines of molecules, N$_2$H$^+$, HC$_3$N, and CCS at 82$-$94 GHz using the Nobeyama 45-m telescope. This study aimed at investigating the physical properties of cores that show high deuterium fractions and might be close to the onset of star formation. We found that the distributions of the N$_2$H$^+$ and HC$_3$N line emissions are approximately similar to that of 850-{\micron} dust continuum emission, whereas the CCS line emission is often undetected or is distributed in a clumpy structure surrounding the peak position of the 850-{\micron} dust continuum emission. 
Occasionally (12\%), we observe the CCS emission which is an early-type gas tracer toward the young stellar object, probably due to local high excitation.
Evolution toward star formation does not immediately affect nonthermal velocity dispersion.
\end{abstract}

\section{INTRODUCTION}
The physical process of the evolution of a molecular cloud core toward the onset of star formation is not clear yet. The timescale toward the onset of star formation (start of the protostar formation), $4~\times~10^5$ yr, seems longer than the free-fall time, 10$^5$ yr for $N$(H$_2$) 
= 1~$\times~10^5$~cm$^{-3}$ \citep{2002ApJ...575..950O}. In addition, as molecular cloud cores are generally close to hydrostatic equilibrium, it is suggested that cores are not (highly) gravitationally unstable. 
Considering that they are initially stable, star formation involves a mechanism to change stable cores into unstable ones \citep{1998ApJ...494..587N}. 
\cite{1998ApJ...494..587N} suggested that the dissipation of turbulence can be such a mechanism. 
The dissipation of magnetic fields may work similarly. 
\citet{2007ApJ...669.1042G} suggested that the mass accretion may cause instability to the core. 
Star-forming cores are defined as cores associated with young stellar objects including protostars, 
whereas the other cores are classified as starless cores. 
Prestellar cores, which are a subset of starless cores, have steep radial density profiles 
(approximately proportional to $r^{-2}$ in the outer part), 
suggesting that self-gravity is important for core support 
\citep{1996A&A...314..625A}. 
Starless cores may evolve to prestellar cores, but not always.
Self-gravity becomes more important with increasing steepness of the radial density profile, leading to unstable cores. 
 
Over the past decades, many efforts have been made to study the initial condition of star formation. Recently, an all-sky survey was conducted 
with the Planck space telescope at (sub)millimeter wavelengths with a large beam (${\sim}5{\arcmin}$), which provided a catalog of ${\sim}13200$ Planck Galactic Cold Clumps (PGCCs) \citep{2011AA...536A..23P, 2016A&A...594A..28P}. Follow-up observations of the PGCCs, performed with the James Clerk Maxwell Telescope (JCMT) and the SCUBA-2 bolometer at 850~{\micron} at higher angular resolution (14$\farcs$1), produced catalogs of cores embedded in the PGCCs \citep{2015PKAS...30...79L, 2018ApJS..234...28L, 2018ApJS..236...51Y, 2019MNRAS.485.2895E}. PGCCs have cold dust temperatures ($10-20$ K), and SCUBA-2 cores inside them include those with high column densities (${\geq} 10^{22}$ cm$^{-2}$, \citet{2020ApJS..249...33K}).  Therefore,  SCUBA-2 cores in PGCCs are candidates for prestellar cores in widely different environments such as:
nearby dark clouds, giant molecular clouds, and high Galactic latitude clouds. 
As follow-up observations, we conducted ALMA observations of the SCUBA-2 cores in the dust continuum emission and molecular lines at 1.3 mm, and found substructures in starless cores and star-forming cores in the continuum \citep{2020ApJS..251...20D, 2021ApJ...907L..15S}. 
We also performed ammonia observations of the SCUBA cores to derive the rotation, kinetic, and excitation temperatures using the Effelsberg 100 m telescope \citep{feher2020}.

\begin{deluxetable*}{lCCCC}[ht!]
\tablecaption{Numbers of the 107 SCUBA-2 cores in the five regions \label{tbl:obs}}
\tablewidth{0pt}
\tabletypesize{\scriptsize}
\tablenum{1}
\tablehead{
\colhead{Region} & 
\colhead{No. of fields} & \colhead{No. of cores} &  
\colhead{No. of starless cores} &  
\colhead{No. of protostellar cores} 
}
\startdata
$\lambda$ Orionis & 1 & 2 & 0 & 2 \\
Orion A & 14 & 36 & 14 & 22 \\
Orion B & 5 & 13 & 5 & 8 \\
Galactic plane & 11 & 30 & 11 & 19 \\
High latitude & 13 & 26 & 4 & 22 \\
\hline
Total & 44 & 107 & 34 & 73 \\ 
\hline
Subtotal in the Orion region & 20 & 51 & 19 & 32 \\
\enddata
\end{deluxetable*}

Although it is difficult to assess the dynamical evolutionary stages of starless cores, 
the chemical evolution may provide estimations \citep{2017ApJS..228...12T}. For example, when starless cores evolve toward star formation, the deuterium fraction of molecules formed in the gas phase (e.g., DNC/HNC, {\ntd}/{\nth}) increases 
and attains the maximum at the onset of star formation \citep{2005ApJ...619..379C, 2006ApJ...646..258H, 2009A&A...496..731E, 2019ApJ...883..202F}. After stellar birth, the deuterium fraction decreases \citep{2011A&A...529L...7F, 
2012ApJ...747..140S, 2015A&A...579A..80G}. 
Moreover, the early-type molecules (e.g., CCS) are abundant in starless cores, whereas the late-type molecules (e.g., NH$_3$, {\nth}) are abundant in star-forming cores \citep{1992ApJ...392..551S, 1992ApJ...394..539H, 1998ApJ...506..743B, 2014PASJ...66..119O, 2016PASJ...68....3O}. 
The ortho-to-para ratio in H$_2$D$^+$ and D$_2$H$^+$  \citep{2013A&A...551A..38P,2014Natur.516..219B}, 
the CO depletion \citep{2005ApJ...619..379C,2013A&A...551A..38P,2020A&A...643A..76H,2020ApJ...901..145F}, and the $^{14}$N/$^{15}$N ratio in the molecule \citep{2020A&A...644A..29R,2020A&A...643A..76H}
are also used as chemical evolution tracers.

Using the deuterium fraction and early-type/late-type molecules,
\citet{2017ApJS..228...12T} proposed the chemical evolution factor (CEF) to evaluate the evolutionary stage of starless cores,
herein called CEF1.0. 
\cite{2020ApJ...891...36G} developed a detailed chemical model for one of the PGCCs that was studied by \cite{2017ApJS..228...12T}.
\cite{2020ApJS..249...33K} revised the CEF definition (CEF2.0), adding starless cores at distances of $<$1 kpc, 
based on a single-pointing survey of 207 SCUBA-2 cores embedded in the PGCCs with the Nobeyama 45-m telescope.  
The second version of the CEF (CEF2.0) is empirically derived to represent the evolutionary stage of starless cores using the logarithmic deuterium fraction from N$_2$D$^+$ and DNC
as an increasing function.
The CEF is defined so that the timing of the onset of star formation corresponds to CEF $\sim$ 0.
Here, we explain the difference between CEF1.0 and CEF2.0.
CEF1.0 included only nearby dark cloud cores (mostly from Taurus, but also from the Aquila, Serpens and Ophiuchus regions), but 
CEF2.0 includes also cores in giant molecular clouds (GMCs) in Orion.
CEF2.0 only includes the deuterium fraction, which seems a better chemical evolution tracer of the starless core,
whereas CEF1.0 also included $N$(N$_2$H$^+$)/$N$(CCS) and $N$(NH$_3$)/$N$(CCS) as well as the deuterium fraction.
Indeed, the chemical evolution sequence of well-known starless cores in Taurus, from L1521B to L1498, and then to L1544
suggested by \cite{2005ApJ...632..982S} though the SCUBA-2 observations, is consistently described as an increasing function in
CEF2.0.
Furthermore, CEF2.0 better describes the typical starless molecular cloud core in the Gould Belt by adding Orion GMC cores.
It is possible that CEF2.0 may not be relevant for environments outside the Gould Belt, and we need to develop environment-dependent CEFs in the future. 
Also, note that CEF2.0 is defined using single-dish observations, and the deuterium fractions of structures smaller than $\lesssim$0.01 pc 
observed with interferometers such as ALMA may be much different from that observed with single-dish telescopes \citep{2015ApJ...803...70S}.
Although we may need further revision, we adopt CEF2.0 as the best description currently available for the Gould Belt.
According to \cite{2020ApJS..249...33K}, 
CEF2.0 of the starless SCUBA-2 cores in Orion ranges from $-$61 to $-$7.
This CEF2.0 range of 
$N$({\ntd})/$N$({\nth}) runs from 0.05 to 0.4,
while that of
$N$(DNC)/$N$(HN$^{13}$C) runs from 2 to 8.
If we adopt a $^{12}$C/$^{13}$C abundance ratio of 43 obtained toward Orion A
\citep{2002ApJ...578..211S},
the corresponding $N$(DNC)/$N$(HNC) runs as high as $0.04-0.2$.
We take these ranges as typical for starless Orion cores.
The Orion  SCUBA-2 sources 
have relatively high deuterium fractions, and
are probably at the middle to late stages of the starless core phase.

We report the results of the on-the-fly (OTF) mapping observations of 44 fields including 107 SCUBA-2 cores \citep{2020ApJS..249...33K} with the Nobeyama 45-m telescope.
We selected 65 intense {\ntd} cores, 21 high-column-density cores, and their 21 neighboring cores out of 207 SCUBA-2 cores in five regions ($\lambda$ Orionis, Orion A, B, Galactic plane, and high latitude) observed in our previous single-pointing survey \citep{2020ApJS..249...33K}.  
The number of cores in each region is summarized in Table \ref{tbl:obs}.   
The employed distances were specified by \cite{2020ApJS..249...33K}.
Accurate distances to parent clouds, as available in the literature, were adopted. 
Otherwise,
if no accurate distance was available,
the distance of the cloud from the parallax-based distance estimator
of the Bar and Spiral Structure Legacy Survey was adopted \citep{2016ApJ...823...77R}, 
based on the systemic velocity of the line emission and
the sky position of the core.
The adopted distances are listed in Table \ref{tbl:field} in the Appendix.
In this study, we focus on the physical properties of Orion cores that are located at similar distances of 350$-$450~pc
\citep{2017ApJ...834..142K, 2019MNRAS.487.2977G} to draw a reliable comparison between cores by avoiding the serious beam dilution effects 
as indicated by \citet{2020ApJS..249...33K}. 
For the other cores, we only present the maps without analysis.

This paper is organized as follows. We describe our observations and data analysis in Section \ref{sec:obs},
present the results of the observations in Section \ref{sec:res},
and discuss the physical properties in Section \ref{sec:dis}. The study is summarized in Section \ref{sec:sum}.

\section{OBSERVATIONS\label{sec:obs}}

We conducted mapping observations of 44 fields of ${3\arcmin~\times~3\arcmin}$ or larger areas covering 107 SCUBA-2 cores in the
{\nth} $J = 1\rightarrow0$, {\hcn} $J = 9\rightarrow8$, CCS $J_N = 8_7\rightarrow7_6$, and CCS $J_N = 7_6\rightarrow6_5$ lines using the 45-m radio telescope
of the Nobeyama Radio Observatory (NRO) \footnote{Nobeyama Radio Observatory
is a branch of the National Astronomical Observatory of Japan,
National Institutes of Natural Sciences.} (LP177001; P.I. =  K. Tatematsu). 
The rest frequencies of these four lines, critical densities \citep{2015PASP..127..299S} 
and other relevant information are summarized in Table \ref{tbl:line}.
The critical densities of the CCS lines were calculated from the Einstein A coefficient and 
collisional cross section listed in \cite{1997ApJ...477..241W}.
Herein, we abbreviate CCS $J_N = 8_7\rightarrow7_6$ and CCS $J_N = 7_6\rightarrow6_5$ to {\cch} and {\ccl}, respectively.
Observations were performed in the OTF mapping mode \citep{2008PASJ...60..445S} from 2017 December to 2019 May. 
For the receiver frontend, the FOur-beam REceiver System on the 45-m Telescope \citep[FOREST; ][]{2016SPIE.9914E..1ZM} was used for simultaneous observations of the four lines. The half-power beam width (HPBW) and main-beam efficiency $\eta_{mb}$ at 86 GHz were 
19$\arcsec~\pm~1\arcsec$ and 50{\%}~$\pm$~4{\%}, respectively. For the receiver backend, the Spectral Analysis Machine for the 45-m telescope \citep[SAM45; ][]{2012PASJ...64...29K} was employed with a channel separation of 30.52~kHz, which corresponds to $\sim$0.1~km~s$^{-1}$ at 82~GHz.
The dump time, scan duration, and row spacing are 0.1 sec, 20 sec, and 5$\arcsec$, respectively.
The unit map size is either 3{\arcmin}~$\times~$3{\arcmin} or 4{\arcmin}~$\times~$4{\arcmin}.
When we needed to cover larger fields, we mosaicked unit maps.
Then, the scan speed was 11$\farcs$5/sec or 14$\farcs$5/sec for $3{\arcmin}~\times~3{\arcmin}$
and $4{\arcmin}~\times~4{\arcmin}$ unit maps, respectively.
In limited cases, we adopted special rectangular unit maps to cover cores efficiently.
We observed unit maps in the R.A. and  decl. directions to minimize striping effects.
The position switching mode was employed.
The typical rms noise level per channel was 0.09 K, and it took approximately 4 hrs to complete a 3{\arcmin}~$\times~$3{\arcmin} unit map.
The typical system temperature was 200 K. 
The telescope pointing calibration was performed at 1.0-1.5 hrs intervals toward SiO maser sources, which resulted in a pointing accuracy of $\lesssim$5$\arcsec$.
The patterns of the telescope main beam and error beam are given in Figure 5 of \cite{2016SPIE.9914E..1ZM}.
The maximum height of the error beam is $\lesssim$10\% of the main beam, and the neighboring error beam is located $30\arcsec-40\arcsec$ apart from the main-beam.
The FWHM size of the neighboring error beam is not very different from that of the main beam.

A total of 44 fields were mapped.
Linear baselines were subtracted from the spectral data, and the data were stacked into 5$\arcsec$ spacing pixels with the Bessel-Gauss function on the NOSTAR program \citep{2008PASJ...60..445S}. The line intensity was expressed in terms of the antenna temperature $T^{\ast}_{\rm A}$ corrected for atmospheric extinction using standard chopper wheel calibration.

We also included the SCUBA-2 data \citep{2018ApJS..236...51Y} and our previous single-pointing data toward the SCUBA-2 position \citep{2020ApJS..249...33K} 
in the analysis.

\floattable
\begin{deluxetable}{lllcccc}
\tablecaption{Observed Lines \label{tbl:line}}
\tablecolumns{7}
\tablenum{2}
\tablewidth{0pt}
\tablehead{
\colhead{Line} &
\colhead{Frequency} &
\colhead{Frequency Reference}&
\colhead{Upper Energy Level $E_u$}&
\colhead{Critical Density at 10 K}&
\colhead{HPBW} &
\colhead{Velocity Channel Width} \\
\colhead{} &
\colhead{GHz} &
\colhead{}&
\colhead{K}&
\colhead{cm$^{-3}$}&
\colhead{$\arcsec$} &
\colhead{km s$^{-1}$} 
}
\startdata
CCS $J_N$ = 7$_6\rightarrow6_5$  &81.505208   &\citet{1986ApJS...60..819C}    &15.3 &   3.1$\times10^5$   & 20 & 0.11  \\
CCS $J_N$ = 8$_7\rightarrow7_6$  &93.870107   &\citet{1990ApJ...361..318Y}     &19.9 &   4.2$\times10^5$  & 18 & 0.097 \\
HC$_3$N $J$ = 9$\rightarrow$8     &81.881462   &\citet{1998JQSRT..60..883P}  &19.7  &  1.1$\times10^5$   & 20 & 0.11 \\
N$_2$H$^+$ $J$ = 1$\rightarrow$0 &93.1737767 &\citet{1995ApJ...455L..77C}     &4.5   &   6.1$\times10^4$ & 18 & 0.098 \\
\enddata
\end{deluxetable}

\section{RESULTS \label{sec:res}}
\subsection{Molecular line distribution \label{sec:mol}}

\begin{figure}
\figurenum{1}
\epsscale{1}
\includegraphics[bb=0 0 600 600, width=10cm]{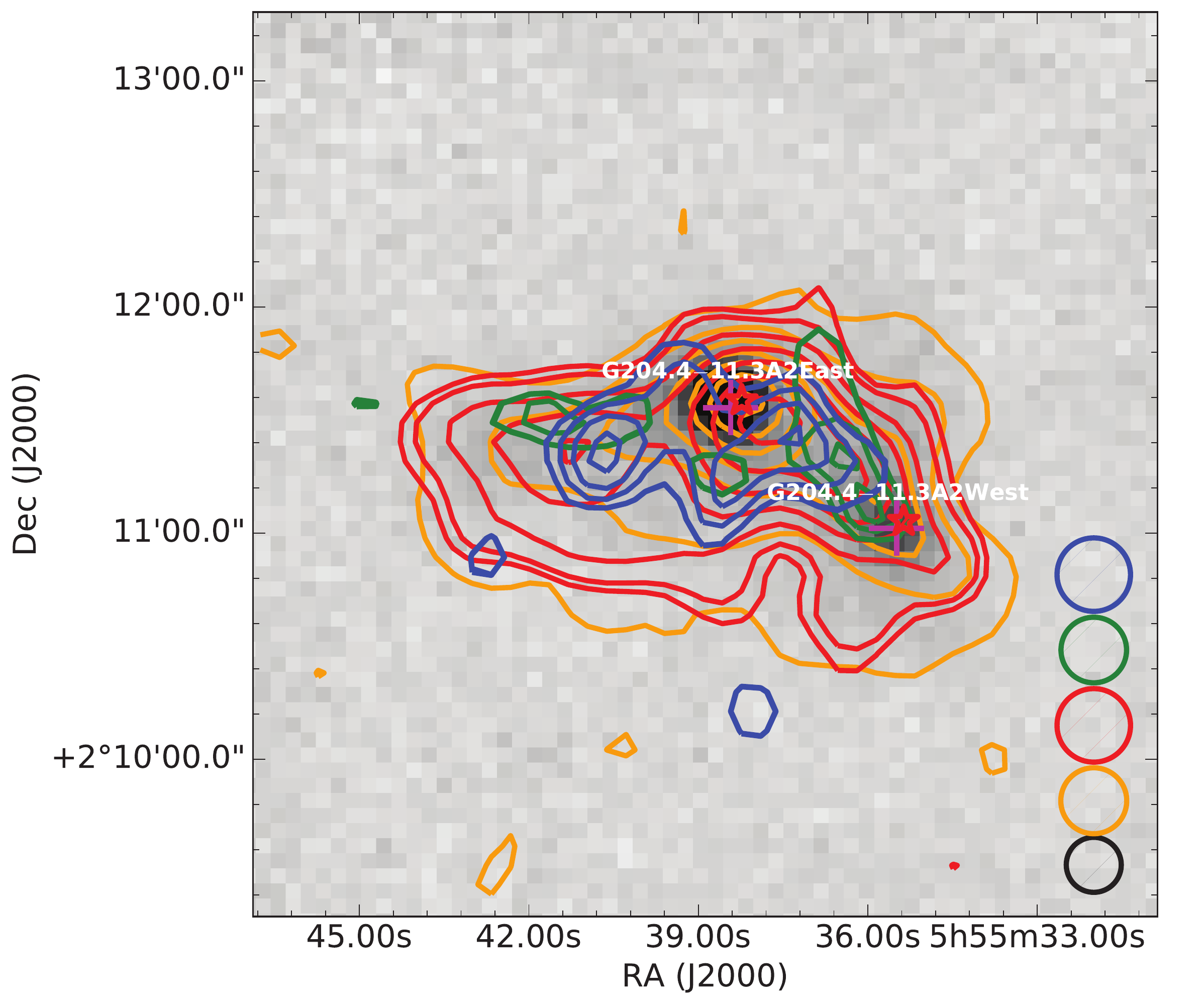}
\vspace{1cm}\caption{
NRO FOREST maps of
{\nth} (orange), {\hcn} (red), CCS-H (green), and CCS-L (blue) integrated intensities overlaid on JCMT SCUBA-2 850-{\micron} continuum emission, 
in gray scale, 
for field G204.4. The contours are drawn at 5\%, 20\%, 35\%, 50\%, 65\%, 80\%, and 95\% of the peak value above 3~$\sigma$, and also at 3~$\sigma$. 
The respective peak and $\sigma$ values of integrated intensity are listed in Table \ref{tbl:map}. The circles represent the beam sizes with corresponding colors.  
The black circle, however, corresponds to the SCUBA-2 beam size. 
The cross and star symbols represent a SCUBA-2 core and a protostar, respectively. \label{fig:850_4cont_G204}}
\end{figure}

\begin{figure}
\figurenum{2}
\epsscale{1.0}
\includegraphics[bb=0 0 600 600, width=10cm]{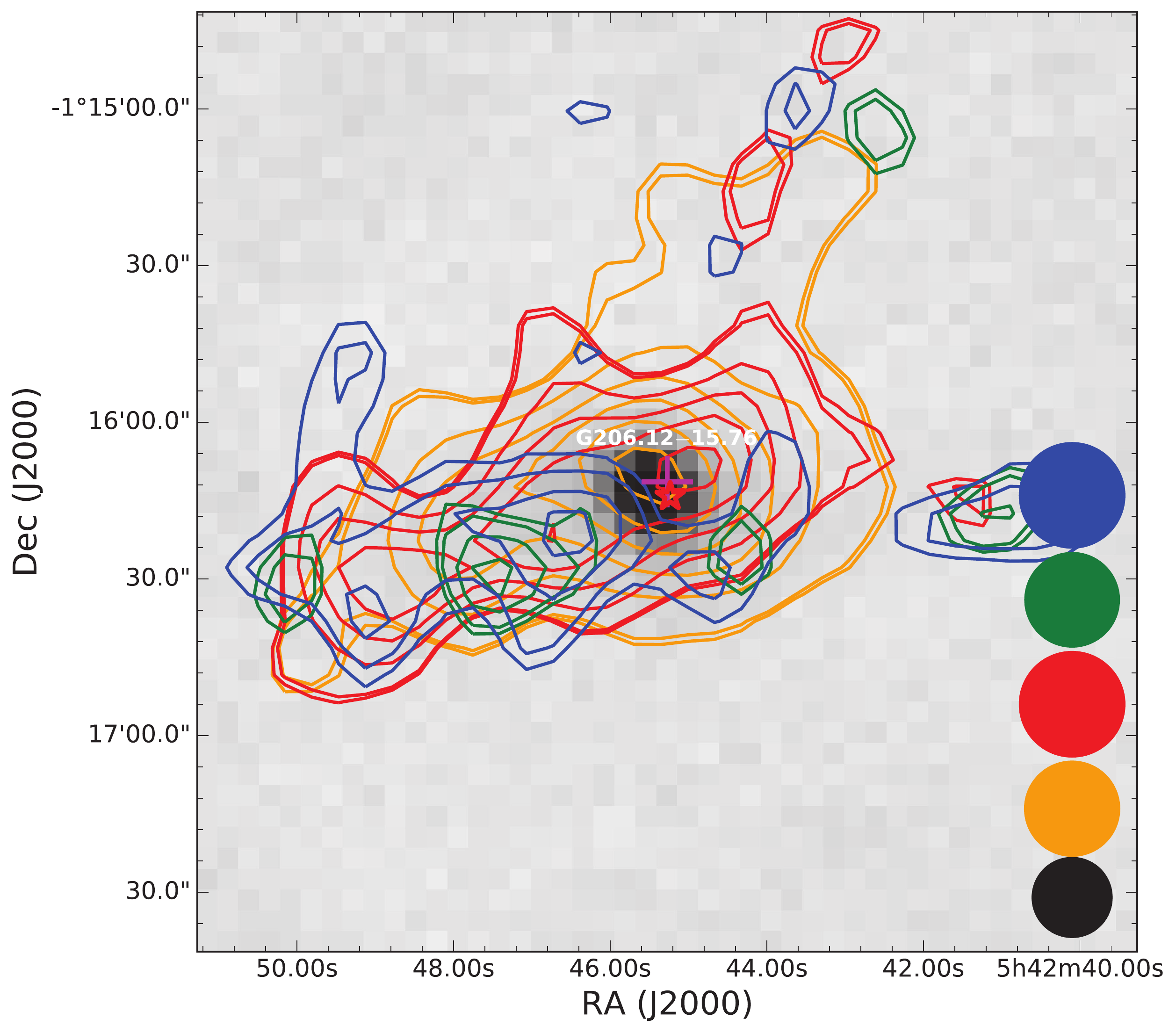}
\\
\vspace{3cm}
\caption{
Same as Figure \ref{fig:850_4cont_G204} but for field G206.12. 
\label{fig:850_4cont_G206_12}}
\end{figure}

\begin{figure}
\figurenum{3}
\epsscale{0.8}
\includegraphics[bb=0 0 600 600, width=15cm]{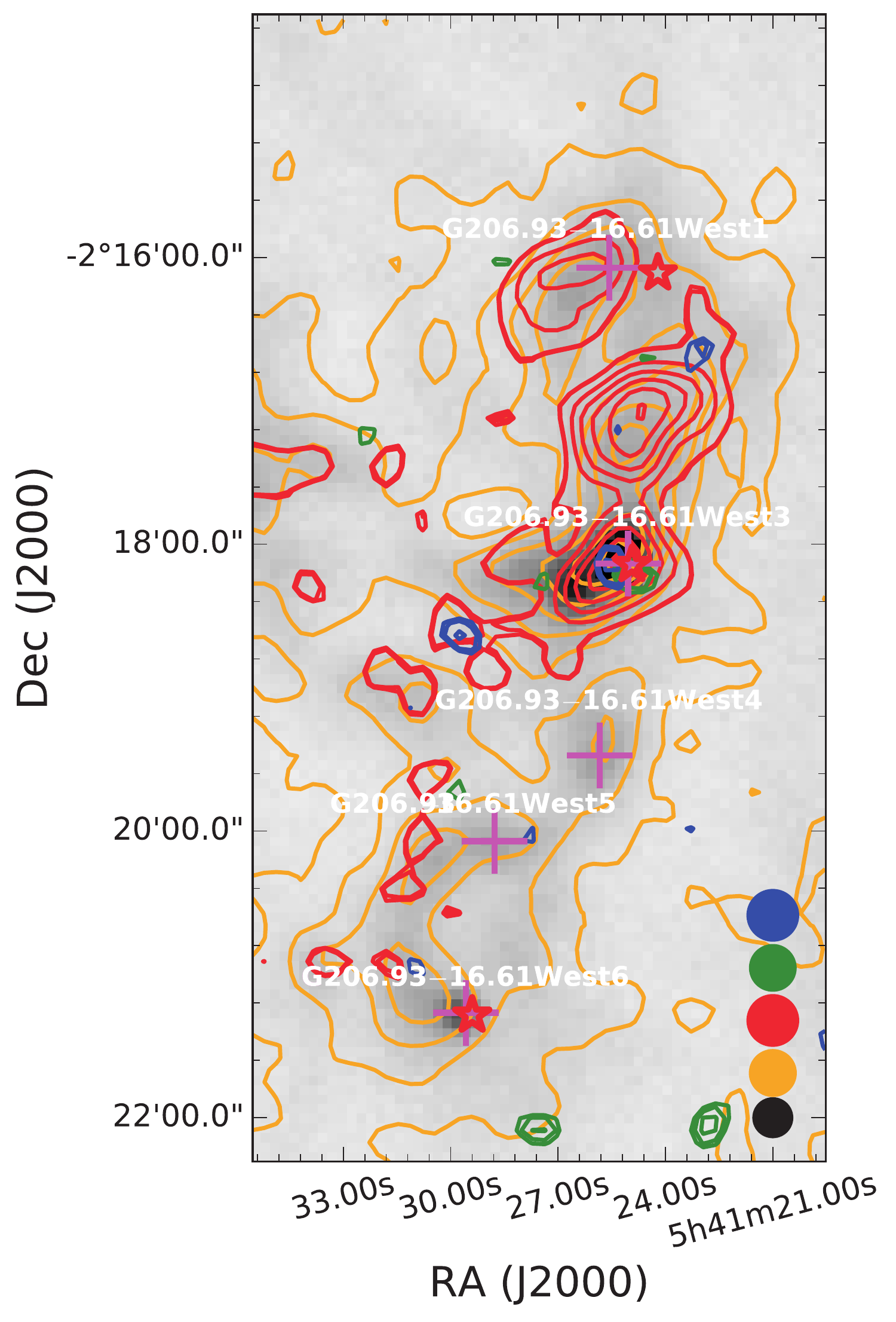}
\caption{Same as Figure \ref{fig:850_4cont_G204} but for field G206.93. \label{fig:850_4cont_G206_93}}
\end{figure}

\begin{figure}
\figurenum{4}
\epsscale{0.9}
\includegraphics[bb=0 0 600 600, width=10cm]{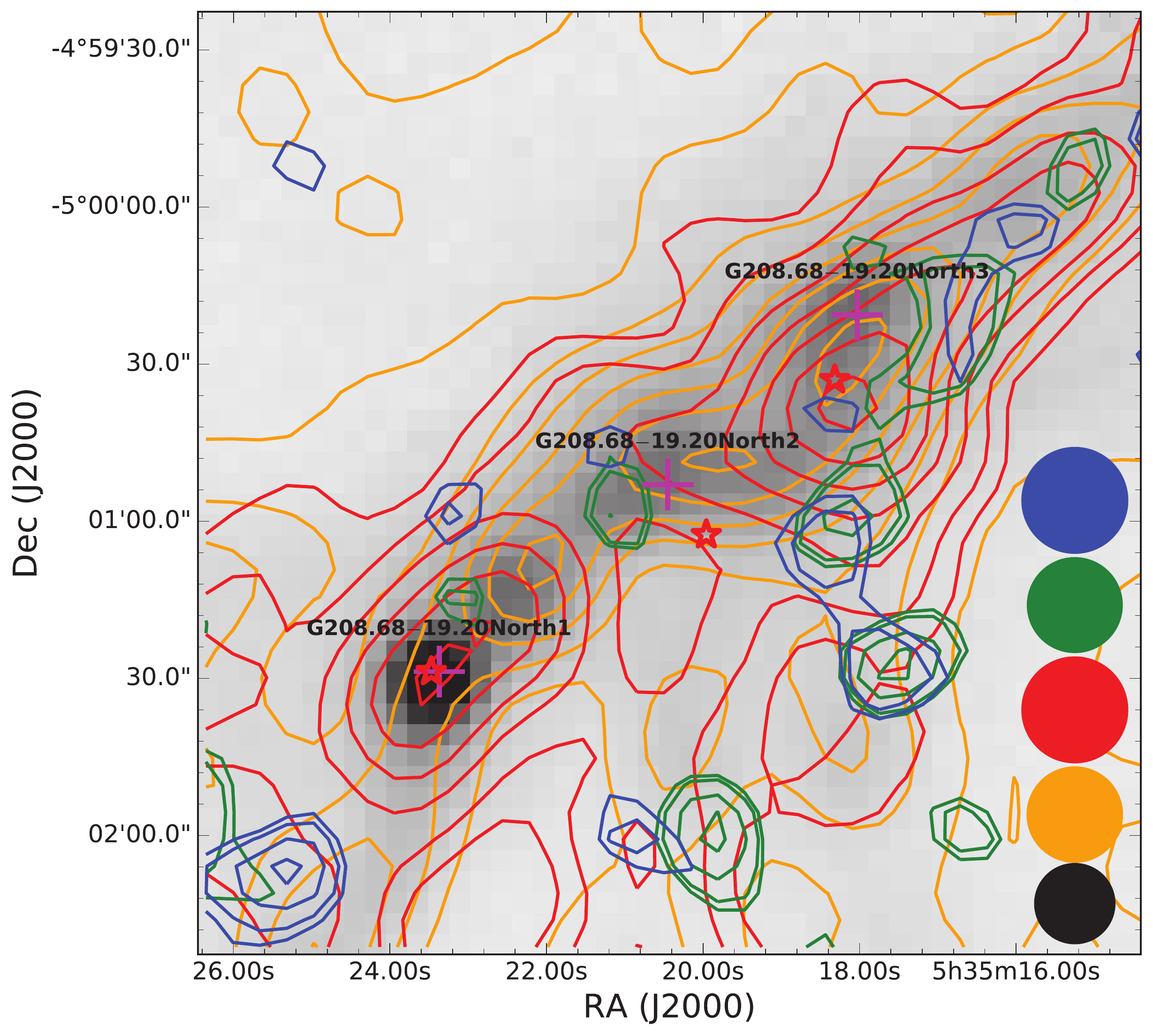}
\caption{Same as Figure \ref{fig:850_4cont_G204} but for field G208.68. \label{fig:850_4cont_G208_68}}
\end{figure}

\begin{figure}
\figurenum{5}
\epsscale{1.4}
\centering
\includegraphics[bb=0 0 600 600, width=14cm]{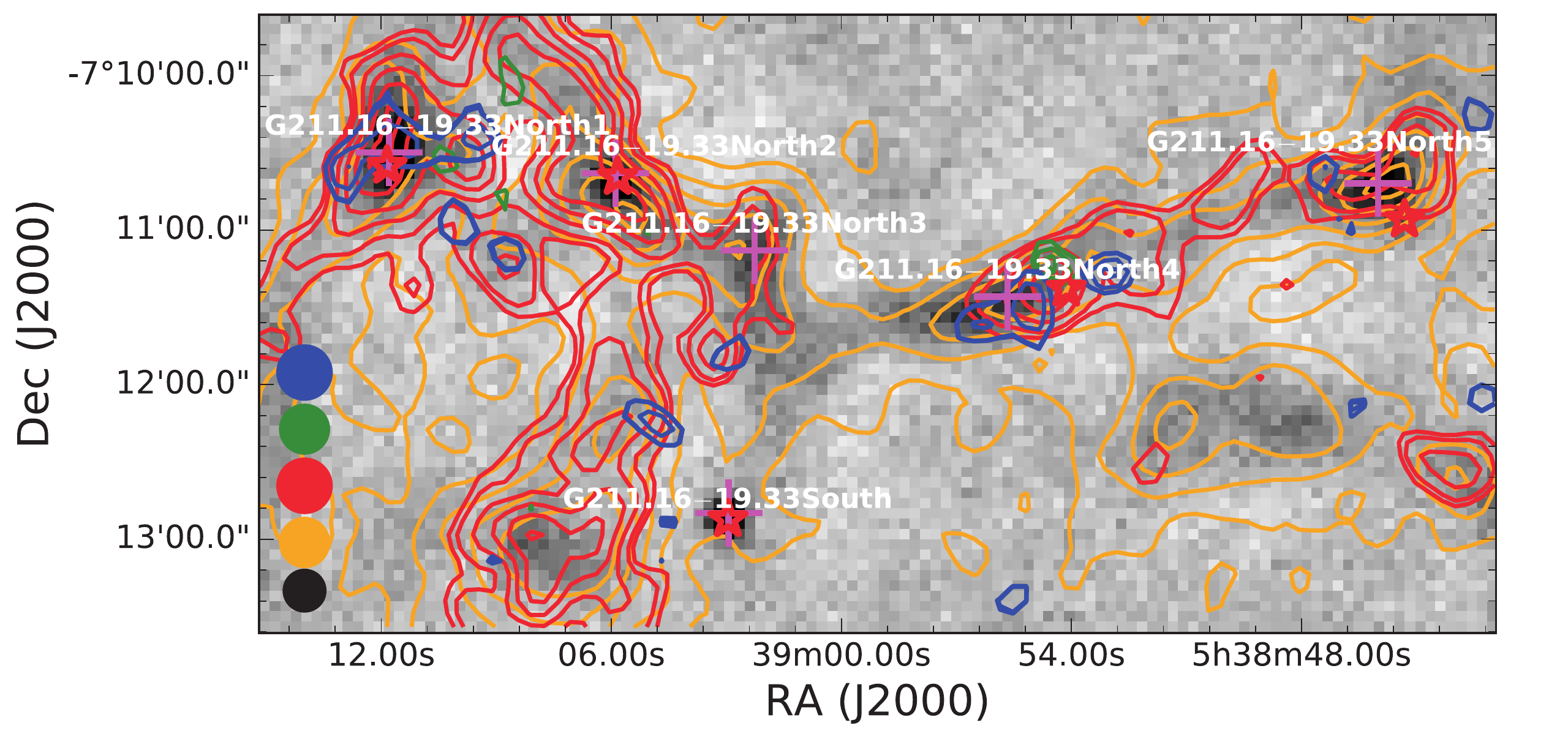}
\caption{Same as Figure \ref{fig:850_4cont_G204} but for field G211.16. \label{fig:850_4cont_G211_16}}
\end{figure}

\begin{figure}
\figurenum{6}
\epsscale{1}
\includegraphics[bb=0 0 600 600, width=10cm]{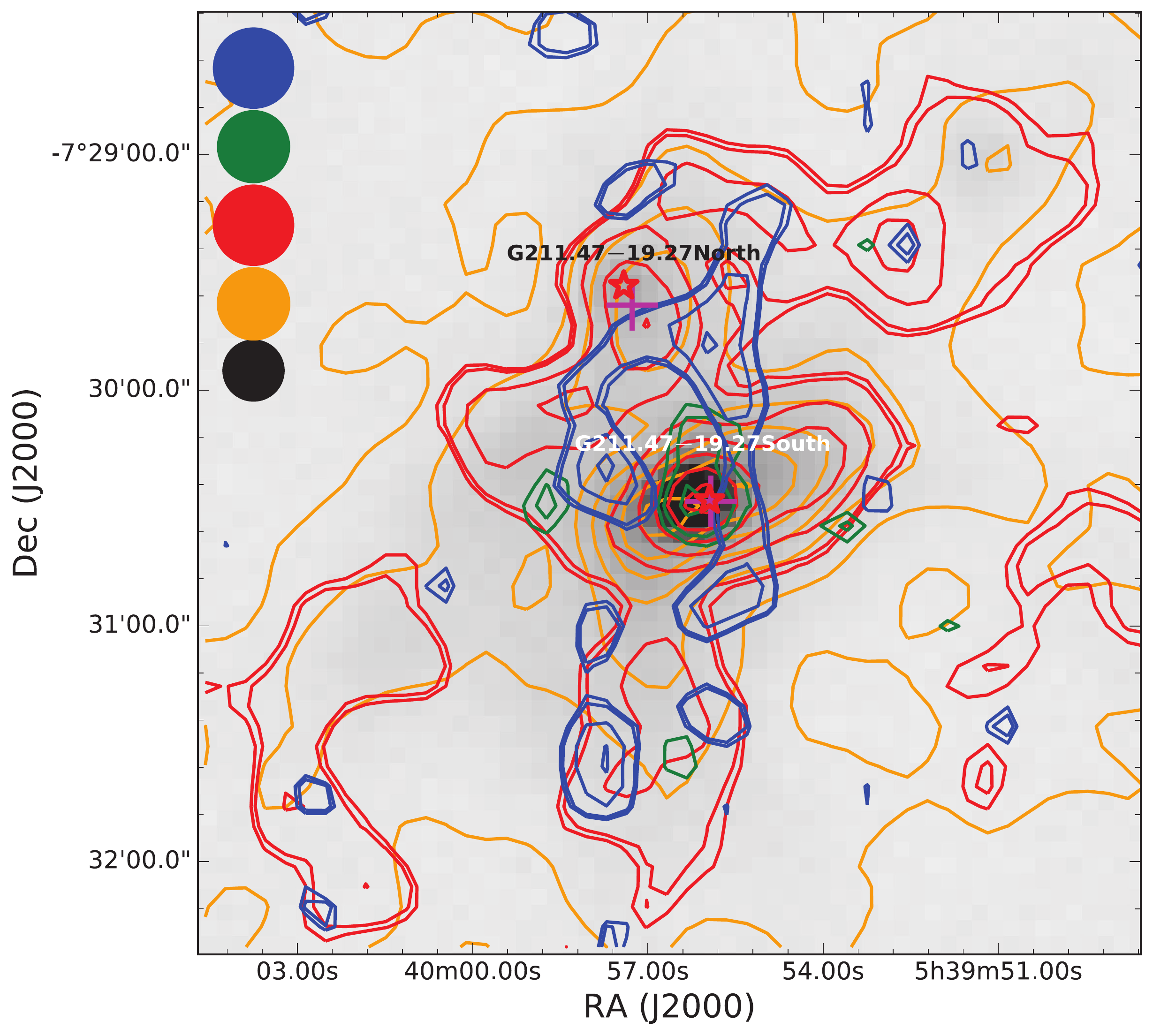}
\caption{Same as Figure \ref{fig:850_4cont_G204} but for field G211.47. \label{fig:850_4cont_G211_47}}
\end{figure}

\begin{figure}
\epsscale{0.8}
\figurenum{7}
\includegraphics[bb=0 0 600 600, width=15cm]{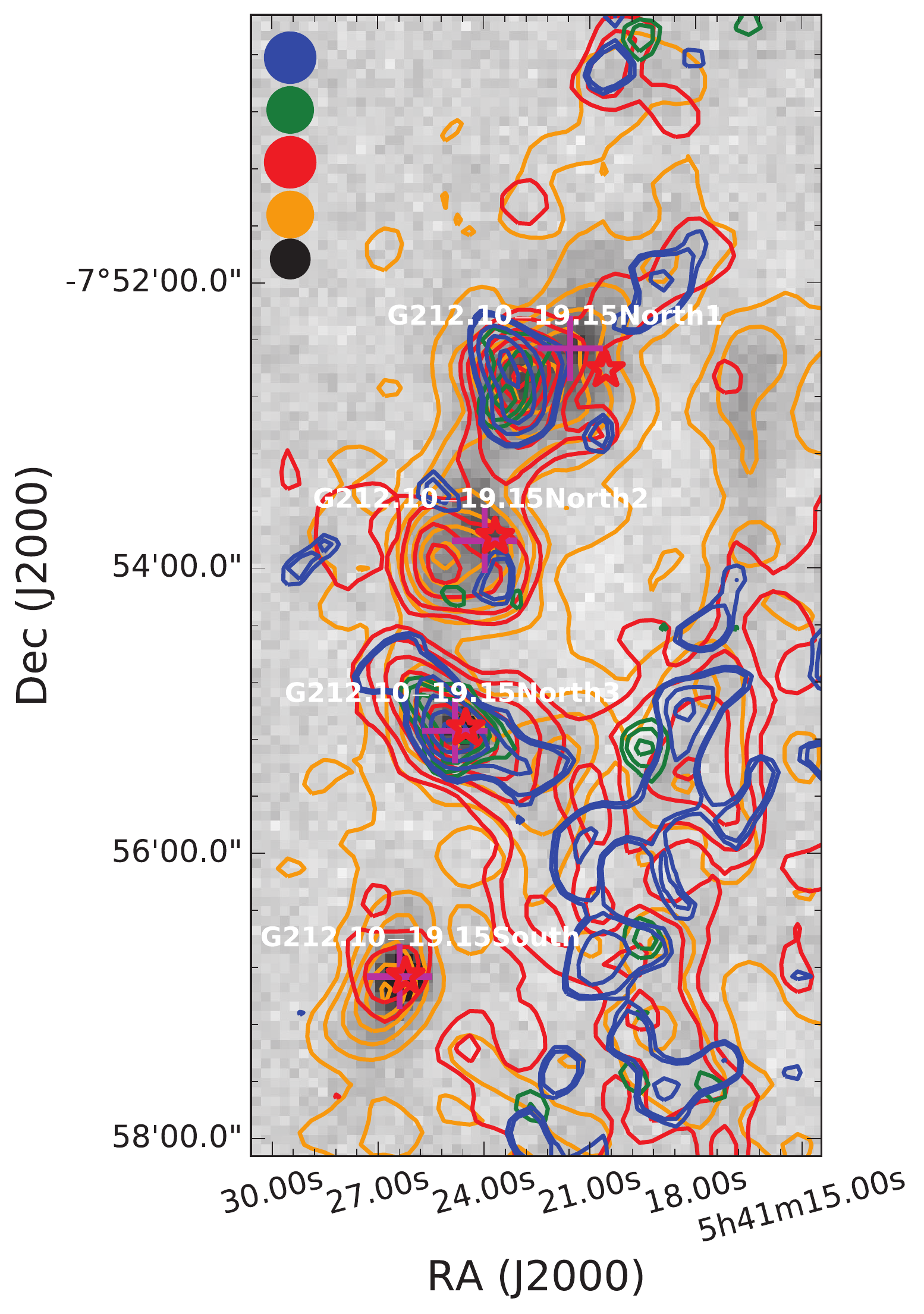}
\caption{Same as Figure \ref{fig:850_4cont_G204} but for field G212. \label{fig:850_4cont_G212_10}}
\end{figure}

Because the number of maps (44) is large, 
most maps are presented in the online 
Figure Set 19,
whereas seven representative examples are shown here as Figures \ref{fig:850_4cont_G204} to \ref{fig:850_4cont_G212_10}. 
Table \ref{tbl:maplist} indicates the relationship between the map field and Figure number.
These Figures show that, for most of the SCUBA-2 cores, the {\nth} emission distribution is similar to the 850-{\micron} dust continuum distribution. 
The former is slightly larger than the latter in distribution extent.
Difference in the beam sizes for the respective observations may explain part of this difference, but probably not all.
The {\nth} emission intensity becomes `saturated' due to high optical depths for very high densities of  $\ga  2\times10^6$ cm$^{-3}$, for which the dust continuum 
will be still sensitive.  
As a result, the dust continuum traces high-density core centers better, resulting in smaller overall sizes.
The {\hcn} emission shows a distribution similar to the 850-{\micron} dust continuum distribution in most of the cores ($\sim$2/3), 
again with slightly larger extent,
whereas in the remaining cores ($\sim$1/3), the emission is not detected or surrounds the central region of the 
850-{\micron} dust continuum emission. 

{\nth} can be destroyed by CO evaporated from the dust in warm environments ($T_{dust} \gtrsim$ 25 K) 
\citep{2004A&A...413..993J,2004ApJ...617..360L,2014PASJ...66...16T}.  
CCS is basically an early-type molecule, but may have a second abundance enhancement due to the CO depletion before protostar formation
\citep{2002ApJ...569..792L, 2003ApJ...583..789L}.
Occasionally, the CCS distribution surrounds the YSO, which reflects different stages of the chemical evolution, or 
has a intensity peak toward the YSO position due to locally high excitation.
These variations will limit the effectiveness of the {\nth}/CCS as an chemical evolution tracer in general.

In general, agreement between the SCUBA-2 cores and {\nth} distribution is good.
In some cases, the peak positions of the {\nth} emission and the 850-{\micron} continuum do not coincide with each other.
The star-forming core 
G208.68$-$19.20North1 (Figure \ref{fig:850_4cont_G208_68}; $N$(DNC)/$N$(HN$^{13}$C) = 2.6$\pm$1.8, $T_{bol}$ = 38$\pm$13 K, $T_{kin}$ = 19.7 K) 
shows such a case.
It seems that {\nth} here has been destroyed due to CO evaporated by local heating due to the YSO.  
This variation limits the effectiveness of {\nth} as a late-type gas tracer in general.
Typically, the spatial variation of  {\nth} abundance including chemical evolution will result in
the {\nth} distribution different from that of the dust continuum.
For example, the southern SCUBA-2 core G211.16$-$19.33South was
detected in the {\nth} emission, but it does not have a prominent peak there.
Instead,
a prominent {\nth} emission feature is located 
1$\arcmin$ east of the SCUBA-2 position.
This isolated {\nth} feature is an exceptional example in all the fields in this study, however.
It is likely that this isolated feature represents a starless core with less concentration.
G211.16$-$19.33North1, North2 ($T_{bol}$ = 70$\pm$20 K), North4, and North5  ($T_{bol}$ = 112$\pm$16 K),
which were detected in both the continuum and {\nth}, are star-forming cores.
Note that the starless core G211.16$-$19.33North3 detected in the continuum and {\nth}
shows infall motions in the ALMA ACA observations \citep{2020ApJ...895..119T}.

The CCS emission shows a clumpy distribution, which is frequently very different from the 850-{\micron} dust continuum distribution,
or is simply not detected in most of the cores. 
The spatial distribution of the line emission is variable from core to core. 
For the star-forming core G204.4$-$11.3A2East (Fig. \ref{fig:850_4cont_G204}; $N$({\ntd})/$N$({\nth}) = 0.27$\pm$0.07), the {\nth} and {\hcn} lines show a distribution similar to the 850-{\micron} distribution, 
whereas the CCS-H and -L lines show clumpy distributions, 
which are very different from the 850-{\micron} distribution. 
Furthermore, the CCS-L emission appears to surround the dust continuum core,
which can be explained in terms of the chemical evolution; the gas in the vicinity of the YSO is more evolved.
All four lines are detected in the star-forming core G206.93$-$16.61West3 (Fig. \ref{fig:850_4cont_G206_93})
and show distributions similar to the 850-{\micron} distribution. 
It is possible that local warmer temperatures around the YSO provide high excitation 
for both the CCS-H and -L emission lines to be detected.
The upper energy levels of the CCS-H and CCS-L are 19.9 K and 15.3 K, respectively.
Toward the starless cores G206.93$-$16.61West4 and West5 (Figure \ref{fig:850_4cont_G206_93}), 
we did not detect the CCS emission but only the {\nth} emission.
Their CEF2.0 values are $-27\pm14$ and $-34\pm3$, respectively, and they are thought to be evolved starless cores.
Furthermore, it is possible that the CCS depletion occurs in cold starless cores.
The star-forming core G212.10$-$19.15South (Figure \ref{fig:850_4cont_G212_10}; $N$({\ntd})/$N$({\nth}) = 0.39$\pm$0.06) 
with a YSO with low bolometric temperature ($T_{bol}$ = 43$\pm$12 K) does not show any CCS emission.
Another YSO with a low bolometric temperature ($T_{bol}$ = 49$\pm$ 21 K) 
associated with G211.47$-$19.27South (Figure \ref{fig:850_4cont_G211_47}) does show CCS-H emission, 
whereas CCS-L emission surrounds it as if it were avoiding the YSO.
The YSO associated with G211.16$-$19.27North1 (Figure \ref{fig:850_4cont_G211_16}) accompanies the CCS-L emission, but does not show intense CCS-H emission.
We detected both CCS-H and -L emission toward the YSO 
in G206.93$-$16.61West3 
(Figure \ref{fig:850_4cont_G206_93}; $N$({\ntd})/$N$({\nth}) = 0.04$\pm$0.02, $N$(DNC)/$N$(HN$^{13}$C) = 1.7$\pm$1.2).
The star-forming core G212.10$-$19.15North2 (Figure \ref{fig:850_4cont_G212_10}) accompanying a 
YSO having  $T_{bol} = 114\pm10$ K is associated with CCS-L emission, and also 
with very weak CCS-H emission.
12\% of the Orion starless cores show local CCS peak emission toward YSOs in either CCS-H or -L.
The CCL-H emission has a higher upper energy level (Table \ref{tbl:line}), 
but we do not see a very clear tendency that it is more concentrated on the YSO position than the CCS-L emission.

The line detection statistics, line profiles, and column density ratios toward the SCUBA-2 positions are fully presented in our single-pointing paper \citep{2020ApJS..249...33K}.
We expect that the telescope error beam will not seriously affect the observed molecular distribution. 
Because the beam efficiency is high, the line emission distribution is very clumpy with sizes of $\sim1\arcmin$,
and the apparent area filling factor of the emission is not very large.

\startlongtable
\begin{deluxetable*}{lc} 
\tablecaption{Field list \label{tbl:maplist}}
\tablewidth{0pt}
\tabletypesize{\scriptsize}
\tablenum{3}
\tablehead{
\colhead{Field} &
\colhead{Figure Number} 
}
\startdata
G192 & 19.1\\
G203 & 19.2\\
G204 & 1\\
G206.12 & 2\\
G206.21 & 19.3\\
G206.93 &  3\\
G207 & 19.4\\
G208.68 &  4\\
G208.89 &  19.5\\
G209.05 &  19.6\\
G209.29North &  19.7\\
G209.29South & 19.8\\
G209.77 & 19.9\\
G209.94North & 19.10\\
G209.94South & 19.11\\
G210 & 19.12\\
G211.16 & 5\\
G211.47 & 6\\
G211.72 &  19.13\\
G212 & 7\\
G159 & 19.14\\
G171 &  19.15\\
G172 & 19.16\\
G173 &  19.17\\
G178 &  19.18\\
G006 & 19.19\\
G001 & 19.20\\
G17 & 19.21\\
G14 & 19.22\\
G16.96 & 19.23\\
G16.36 &  19.24\\
G24 & 19.25\\
G33 & 19.26\\
G35 & 19.27\\
G34 & 19.28\\
G57 & 19.29\\
G69 & 19.30\\
G74 & 19.31\\
G82 &  19.32\\
G91 &  19.33\\
G92 &  19.34\\
G105 &  19.35\\
G93 &  19.36\\
G107 &  19.37\\
\enddata
\end{deluxetable*}

\subsection{Line emission distribution in representative cores \label{sec:representative}}

We investigated two extreme cases in detail: one case G212 where all the four lines have the same peaks,
and another case G206.12 where the distributions of the continuum, {\nth}, and HC$_3$N emission 
are different
from that of the CCS emission.

First, we considered field G212 (Figure \ref{fig:850_4cont_G212_10}). 
We detect intense emission in CCS-H and -L
toward the YSO associated with G212.10$-$19.15North3 
($N$({\ntd})/$N$({\nth}) = 0.06$\pm$0.03 and $N$(DNC)/$N$(HN$^{13}$C) = 2.8$\pm$2.0).
CCS is usually regarded as an early-type gas tracer,
but the CCS emission peak coincides with the YSO on this core.
In the SCUBA-2 peak G212.10$-$19.15North1 ($N$({\ntd})/$N$({\nth}) = 0.35$\pm$0.07), 
the four molecular lines are all distributed similarly, but their peaks are displaced by
$\sim$ 40$\arcsec$ or 0.08 pc from the SCUBA-2 peak.
It is possible that the YSO is destroying the core.
Another possibility is that the YSO has moved from its birth site due to proper motion \citep{2014PASJ...66...16T},
although we do not know the accurate age of the YSO or the vector of its proper motion with respect to the sky plane.

\begin{figure*}
\epsscale{0.7}
\figurenum{8}
\includegraphics[bb=0 0 600 600, width=12cm]{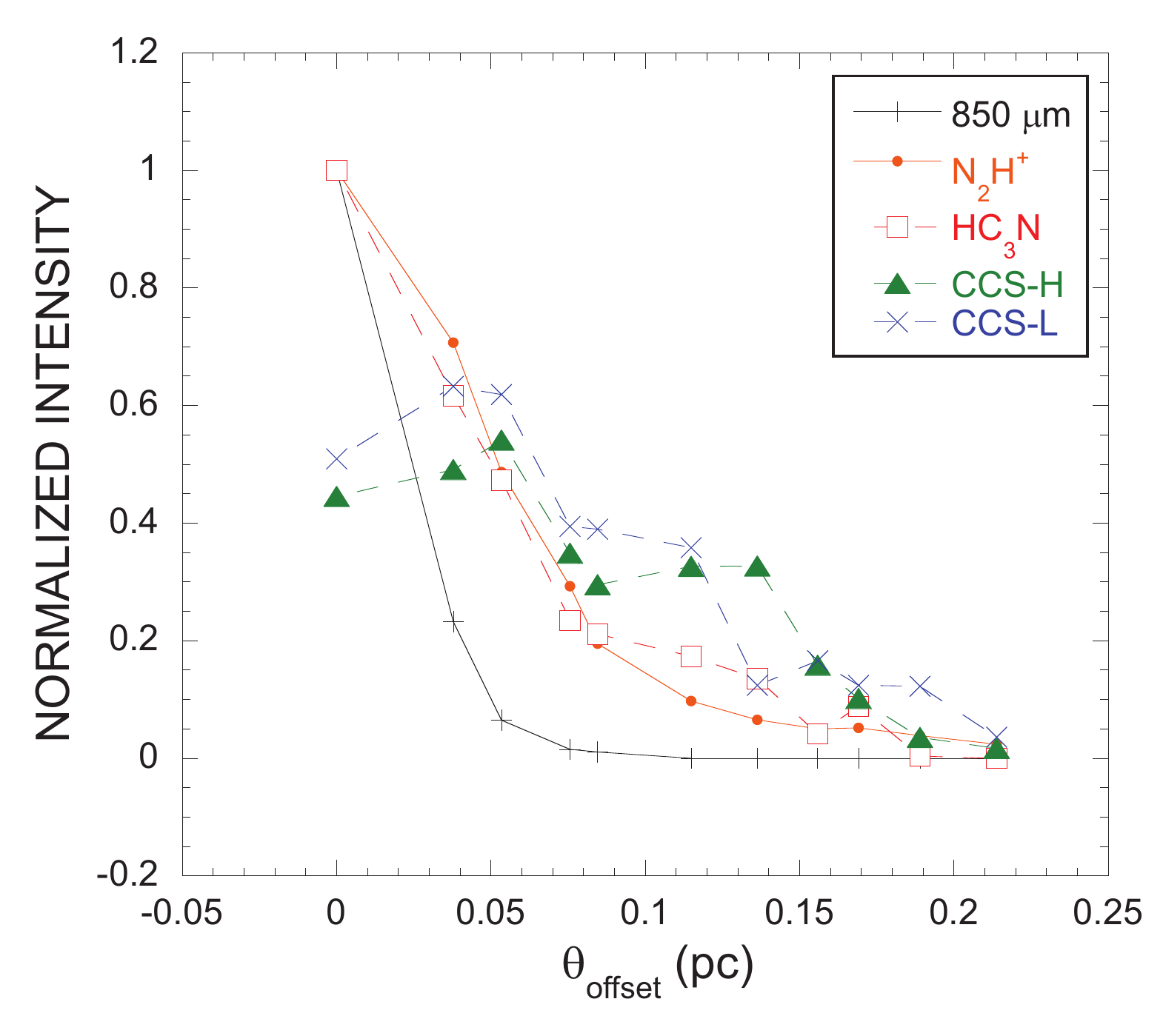}
\caption{Radial distribution of the continuum and line intensity normalized to the maximum value,
toward the star-forming core G206.12$-$15.76 (Figure \ref{fig:850_4cont_G206_12}).  \label{fig:G206_12_PixelTableAnn}}
\end{figure*}

Next, we investigated
the radial intensity distribution field G206.12 (Figure \ref{fig:850_4cont_G206_12}), which contains only one SCUBA-2 core
G206.12$-$15.76 ($N$(DNC)/$N$(HN$^{13}$C) $<$ 4.8) 
but was detected with all four lines.
This core contains a YSO ($T_{bol}$ = 35$\pm$9 K) near the SCUBA-2 position (2$\farcs$9 SSW).
We binned the continuum and integrated-intensity line maps to 20$\arcsec$ pixels, which is close to 
the NRO telescope beam size.
Figure \ref{fig:G206_12_PixelTableAnn} compares the intensity normalized for the maximum pixel value
in the SCUBA-2 850-$\mu$m continuum and the line against offset from the map center.
The values at the same offset from the map center (SCUBA-2 position) were averaged.
For 850-{\micron} continuum, {\nth} and HC$_3$N have their maxima at the core center, which is almost identical to the YSO position.
The 850-{\micron} continuum, however, decreases more sharply than those of {\nth} and HC$_3$N.
Differences in the telescope beam radius (7$\arcsec$ and $9\arcsec-10\arcsec$ for the 850-$\mu$m continuum and the lines, respectively)
may affect this difference, but also 
the molecular lines will be `saturated' with high optical depths for very high densities as explained in \S\ref{sec:res}.
CCS-H and CCS-L seem to have a depression toward the core center.
It is known that CCS also exists in evolved molecular gas as
a secondary late-stage peak due to CO depletion.
The observed emission could represent such gas.
It seems that the gas near the core center is more chemically evolved so that the CCS abundance is reduced.

\subsection{Physical properties of the SCUBA-2 core \label{sec:core}}

We investigated the physical properties of the SCUBA-2 cores using the {\nth} data
toward the SCUBA-2 position \citep{2020ApJS..249...33K}.
We adopted the hyperfine spectral fitting result for the {\nth} spectrum.
We neglected cores with two {\nth} velocity components, 
as it was difficult to identify the component that corresponded to the
SCUBA-2 emission.
The nonthermal and total velocity dispersions,  $\sigma_{\rm nt}$ and $\sigma_{\rm tot}$, are defined as \cite{1992ApJ...384..523F}:

\begin{equation}
    \sigma_{\rm nt} = \sqrt{\frac{{\Delta v_{\rm obs}}^2}{\rm 8~ln~2} - \frac{k_{\rm B}T_{\rm kin}}{\mu_{\rm obs} m_{\rm H}}}
\end{equation}

\noindent
and

\begin{equation}
    \sigma_{\rm tot} = \sqrt{\frac{{\Delta v_{\rm obs}}^2}{\rm 8~ln~2} + \frac{k_{\rm B}T_{\rm kin}}{m_{\rm H}}(\frac{1}{\mu} - \frac{1}{\mu_{\rm obs}})} ,
\end{equation}

\bigskip

\noindent
respectively, where $\Delta {v_{\rm obs}}$ is  the FWHM linewidth, 
$k_{\rm B}$ is the Boltzmann constant, and $T_{\rm kin}$ is the kinetic temperature. 
We assume that the kinetic temperature is equal to the dust temperature. 
$\mu_{\rm obs}$ is the molecular weight of the observed molecule in units of the hydrogen mass $m_{\rm H}$. 
The mean molecular weight $\mu$ per particle in units of $m_{\rm H}$ is set to 2.33.

In this study, we adopted the HWHM radius $R$~(dust) and mass $M$~(dust)  of the core from the SCUBA-2 results
of \cite{2018ApJS..236...51Y} to avoid uncertainties of the {\nth} abundance.
\cite{2018ApJS..236...51Y} expressed the core size $R$ in terms of the FWHM diameter, but we express $R$ in terms of the HWHM radius here.

The virial mass $M_{\rm vir}$ of the uniform-density sphere is derived using the following formula \citep{1992ApJ...399..551M}: 

\begin{equation}
M_{\rm vir} = \frac{5 R \sigma_{\rm tot}^2}{G} ,
\end{equation}

\noindent
where $G$ is the gravitational constant.  
The virial parameter $\alpha_{\rm vir}$ is estimated by dividing the virial mass by the SCUBA-2 core mass,

\begin{equation}            
\alpha_{vir} = \frac{M_{vir}}{M {\rm(dust)}}.
\end{equation}

\begin{figure}
\epsscale{0.7}
\figurenum{9}
\includegraphics[bb=0 0 600 600, width=12cm]{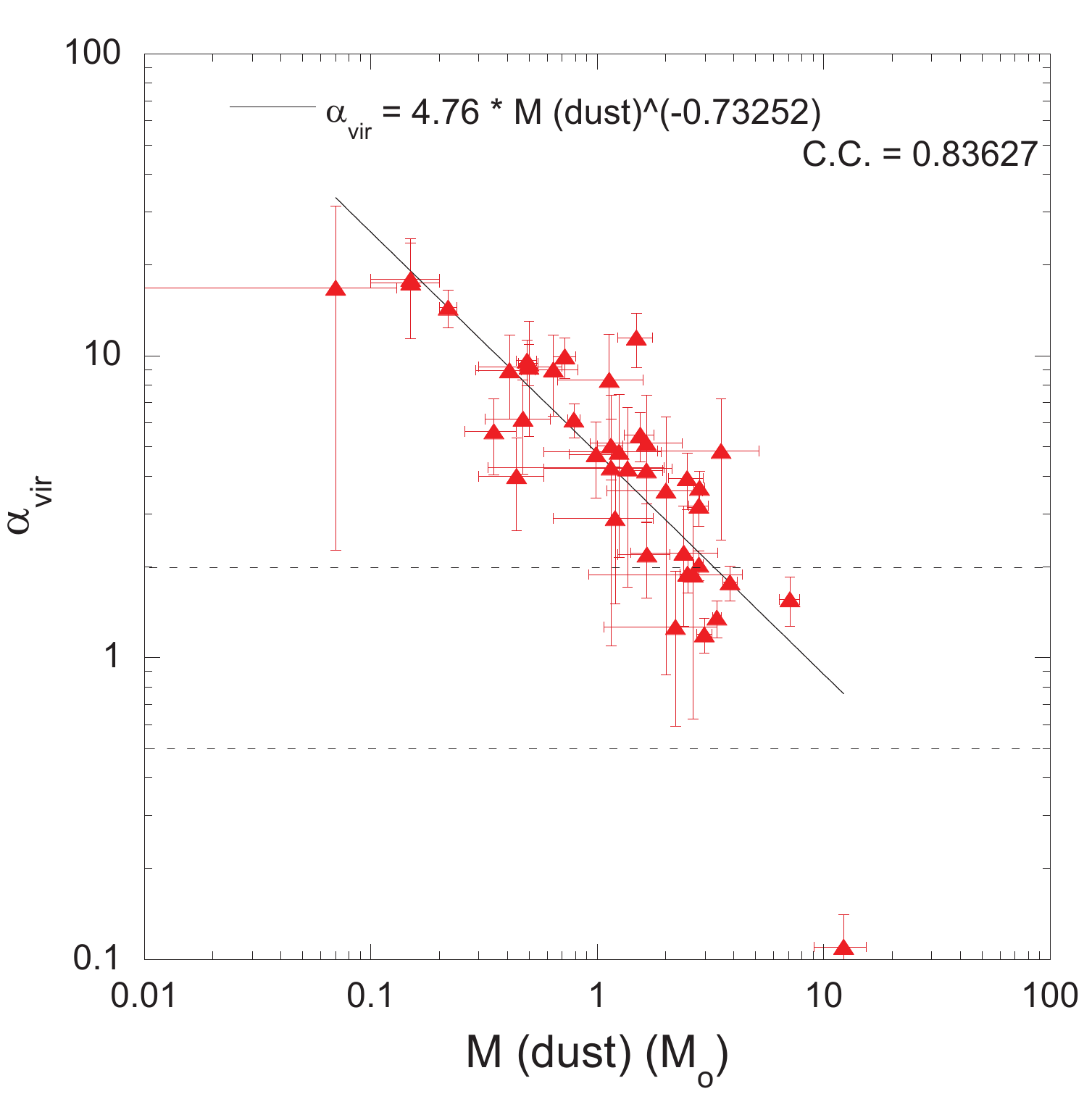}
\caption{Virial parameter $\alpha_{vir}$ vs. $M$~(dust). The straight line was computed using a least-squares program. 
C.C represents the correlation coefficient.
The horizontal dotted lines represents $\alpha_{vir}$ = 0.5 and 2.  \label{fig:alpha-M}}
\end{figure}

Table \ref{tbl:vir} lists the velocity dispersions, radii, masses, virial masses, 
and virial parameters for the SCUBA-2 cores. 
The virial parameter is as large as $\sim$5. 
Figure \ref{fig:alpha-M} plots the virial parameter as a function of SCUBA-2 core mass.
We adopted the uncertainty in the core radius to be 8\% from the uncertainty in the distance adopted by \cite{2018ApJS..236...51Y},
and derived the uncertainty in the virial parameter assuming the propagation of random errors.
From this sample, we obtained the least-squares fit to be $\alpha_{vir} \propto M$~(dust)$^{-0.7}$,
which can be well described in terms of a pressure-confined core with a form 
$\alpha_{vir} \propto M^{-2/3}$ \citep{1992ApJ...395..140B}.
For the core located near $\alpha_{vir} \sim$ unity, self-gravity is likely to  be
dominant if we assume that it is in hydrostatic equilibrium.
The core with an $\alpha_{vir}$ value that is considerably larger than unity probably needs an appropriate external pressure to bind it if 
it is not a transient object.
Similar results showing large virial parameters are obtained by \cite{2017ApJ...846..144K} and \cite{2019ApJ...874..147K} 
in Orion and in other Gould Belt clouds, respectively.
Note that core G211.47$-$19.27South has a very small virial parameter, $\alpha_{vir} = 0.1$, and was cataloged
by  \cite{2018ApJS..236...51Y} with an exceptionally small size, which is
one order of magnitude smaller than the other cores.
Its deconvolved size is smaller than the telescope beam.
We define a subcategory of  $M$~(dust) $> 2 M_{\sun}$, which seems closer to virial equilibrium.
We estimate $\alpha_{vir}$  to be
5.9$\pm$4.6 and 2.5$\pm$1.2 for all the cores and for the subcategory $M$~(dust) $> 2 M_{\sun}$, respectively.
It should be noted that the virial parameter for the structure embedded in the larger structure
can be affected by tidal forces \citep{2020ApJ...898...52M}  In addition, the virial parameter
can be affected by internal motions
(Ayushi Singh, Christopher Matzner, and the Green Bank Ammonia Collaboration, private communication).

\begin{longrotatetable}
\begin{deluxetable*}{lLLLLLLLcl} 
\tablecaption{Physical properties of {\nth} core for SCUBA-2 cores in the Orion region \label{tbl:vir}}
\tablewidth{0pt}
\tabletypesize{\scriptsize}
\tablenum{4}
\tablehead{
\colhead{SCUBA-2 core} &  
\colhead{$\Delta v {\rm(N_2H^+, SCUBA-2)}$} & \colhead{$\sigma_{\rm nt}$} & \colhead{$\sigma_{\rm tot}$} & \colhead{$R {\rm (dust)}$} &  
\colhead{$M {\rm(dust)}$} & \colhead{$M_{\rm vir}$} & \colhead{$\alpha_{\rm vir}$} & $T_{kin}$ & Filament \\
\colhead{} & \colhead{km s$^{-1}$} & \colhead{km s$^{-1}$} &\colhead{km s$^{-1}$} & \colhead{pc} &  
\colhead{$M_{\sun}$} & \colhead{$M_{\sun}$} & \colhead{} & \colhead{K} & \colhead{} 
}
\decimalcolnumbers
\startdata
G192.32$-$11.88North		&	0.66 	$\pm$	0.05 	&	0.27 	$\pm$	0.02 	&	0.37 	$\pm$	0.03 	&	0.030 			&	0.49 	$\pm$	0.05 	&	4.7 	$\pm$	0.8 	&	9.6 	$\pm$	1.7 	&	17.3			&	N	\\
G192.32$-$11.88South		&	0.54 	$\pm$	0.02 	&	0.22 	$\pm$	0.01 	&	0.33 	$\pm$	0.01 	&	0.025 			&	0.22 	$\pm$	0.02 	&	3.2 	$\pm$	0.6 	&	14.4 	$\pm$	2.1 	&	17.3			&	N	\\
G203.21$-$11.20East1		&	0.76 	$\pm$	0.04 	&	0.32 	$\pm$	0.02 	&	0.38 	$\pm$	0.02 	&	0.060 			&	2.50 	$\pm$	0.43 	&	9.8 	$\pm$	1.7 	&	3.9 	$\pm$	0.8 	&	11.2			&	Y	\\
G203.21$-$11.20East2		&	0.44 	$\pm$	0.03 	&	0.18 	$\pm$	0.01 	&	0.27 	$\pm$	0.02 	&	0.060 			&	2.65 	$\pm$	1.73 	&	5.0 	$\pm$	0.9 	&	1.9 	$\pm$	1.3 	&	11.2			&	Y	\\
G203.21$-$11.20West1		&	0.50 	$\pm$	0.01 	&	0.20 	$\pm$	0.01 	&	0.29 	$\pm$	0.01 	&	0.060 			&	2.81 	$\pm$	0.13 	&	5.7 	$\pm$	1.0 	&	2.0 	$\pm$	0.2 	&	11.2			&	Y	\\
G203.21$-$11.20West2		&	0.50 	$\pm$	0.01 	&	0.20 	$\pm$	0.01 	&	0.29 	$\pm$	0.01 	&	0.050 			&	2.51 	$\pm$	0.19 	&	4.8 	$\pm$	0.8 	&	1.9 	$\pm$	0.3 	&	11.2			&	Y	\\
G204.4$-$11.3A2East		&	0.47 	$\pm$	0.01 	&	0.19 	$\pm$	0.01 	&	0.28 	$\pm$	0.01 	&	0.040 			&	2.97 	$\pm$	0.23 	&	3.5 	$\pm$	0.6 	&	1.2 	$\pm$	0.2 	&	11.1			&	N	\\
G204.4$-$11.3A2West		&	0.83 	$\pm$	0.06 	&	0.35 	$\pm$	0.03 	&	0.40 	$\pm$	0.03 	&	0.025 			&	0.99 	$\pm$	0.24 	&	4.7 	$\pm$	0.8 	&	4.7 	$\pm$	1.3 	&	11.1			&	N	\\
G206.93$-$16.61West1		&	0.84 	$\pm$	0.03 	&	0.35 	$\pm$	0.01 	&	0.43 	$\pm$	0.02 	&	0.040 			&	1.55 	$\pm$	0.23 	&	8.5 	$\pm$	1.5 	&	5.5 	$\pm$	1.0 	&	16.8			&	Y	\\
G206.93$-$16.61West3		&	0.65 	$\pm$	0.04 	&	0.27 	$\pm$	0.02 	&	0.36 	$\pm$	0.02 	&	0.030 			&	3.38 	$\pm$	0.15 	&	4.6 	$\pm$	0.8 	&	1.4 	$\pm$	0.2 	&	16.8			&	Y	\\
G206.93$-$16.61West4		&	0.64 	$\pm$	0.04 	&	0.26 	$\pm$	0.02 	&	0.36 	$\pm$	0.02 	&	0.040 			&	1.25 	$\pm$	0.67 	&	6.0 	$\pm$	1.0 	&	4.8 	$\pm$	2.7 	&	16.8			&	Y	\\
G206.93$-$16.61West5		&	0.81 	$\pm$	0.07 	&	0.34 	$\pm$	0.03 	&	0.42 	$\pm$	0.04 	&	0.055 			&	7.10 	$\pm$	0.72 	&	11.1 	$\pm$	1.9 	&	1.6 	$\pm$	0.3 	&	16.8			&	Y	\\
G206.93$-$16.61West6		&	0.62 	$\pm$	0.07 	&	0.25 	$\pm$	0.03 	&	0.35 	$\pm$	0.04 	&	0.040 			&	1.15 	$\pm$	0.15 	&	5.8 	$\pm$	1.0 	&	5.0 	$\pm$	1.2 	&	16.8			&	Y	\\
G207.36$-$19.82North1		&	1.13 	$\pm$	0.04 	&	0.48 	$\pm$	0.02 	&	0.52 	$\pm$	0.02 	&	0.030 			&	1.13 	$\pm$	0.46 	&	9.4 	$\pm$	1.6 	&	8.3 	$\pm$	3.5 	&	11.9			&	Y	\\
G207.36$-$19.82North2		&	0.45 	$\pm$	0.02 	&	0.18 	$\pm$	0.01 	&	0.27 	$\pm$	0.01 	&	0.020 			&	0.44 	$\pm$	0.14 	&	1.8 	$\pm$	0.3 	&	4.0 	$\pm$	1.4 	&	11.9			&	Y	\\
G207.36$-$19.82North3		&	0.69 	$\pm$	0.04 	&	0.29 	$\pm$	0.02 	&	0.35 	$\pm$	0.02 	&	0.020 			&	0.47 	$\pm$	0.15 	&	2.9 	$\pm$	0.5 	&	6.2 	$\pm$	2.1 	&	11.9			&	Y	\\
G207.36$-$19.82North4		&	0.80 	$\pm$	0.06 	&	0.33 	$\pm$	0.03 	&	0.39 	$\pm$	0.03 	&	0.015 			&	0.15 	$\pm$	0.05 	&	2.7 	$\pm$	0.5 	&	18.0 	$\pm$	6.5 	&	11.9			&	Y	\\
G207.36$-$19.82South		&	0.38 	$\pm$	0.16 	&	0.15 	$\pm$	0.06 	&	0.26 	$\pm$	0.11 	&	0.095 			&	2.01 	$\pm$	0.91 	&	7.2 	$\pm$	1.2 	&	3.6 	$\pm$	2.7 	&	11.9			&	N	\\
G208.68$-$19.20North2		&	0.42 	$\pm$	0.01 	&	0.16 	$\pm$	0.01 	&	0.31 	$\pm$	0.01 	&	0.025 			&	2.22 	$\pm$	1.15 	&	2.8 	$\pm$	0.5 	&	1.3 	$\pm$	0.7 	&	19.7			&	Y	\\
G208.89$-$20.04East		&	0.38 	$\pm$	0.01 	&	0.14 	$\pm$	0.01 	&	0.30 	$\pm$	0.01 	&	0.065 			&	3.86 	$\pm$	0.29 	&	6.9 	$\pm$	1.2 	&	1.8 	$\pm$	0.2 	&	19.7			&	N	\\
G209.05$-$19.73North		&	0.38 	$\pm$	0.02 	&	0.15 	$\pm$	0.01 	&	0.28 	$\pm$	0.01 	&	0.115 			&	2.83 	$\pm$	0.15 	&	10.3 	$\pm$	1.8 	&	3.6 	$\pm$	0.5 	&	15.6			&	Y	\\
G209.05$-$19.73South		&	0.43 	$\pm$	0.03 	&	0.17 	$\pm$	0.01 	&	0.29 	$\pm$	0.02 	&	0.070 			&	1.65 	$\pm$	0.29 	&	6.9 	$\pm$	1.2 	&	4.2 	$\pm$	0.9 	&	15.6			&	Y	\\
G209.29$-$19.65South1		&	1.42 	$\pm$	0.04 	&	0.60 	$\pm$	0.02 	&	0.65 	$\pm$	0.02 	&	0.035 			&	1.49 	$\pm$	0.26 	&	17.1 	$\pm$	3.0 	&	11.5 	$\pm$	2.4 	&	17.3			&	Y	\\
G209.29$-$19.65South3		&	0.66 	$\pm$	0.03 	&	0.27 	$\pm$	0.01 	&	0.37 	$\pm$	0.02 	&	0.030 			&	0.50 	$\pm$	0.05 	&	4.7 	$\pm$	0.8 	&	9.4 	$\pm$	1.5 	&	17.3			&	Y	\\
G209.77$-$19.40East1		&	0.35 	$\pm$	0.01 	&	0.13 	$\pm$	0.01 	&	0.26 	$\pm$	0.01 	&	0.025 			&	0.35 	$\pm$	0.09 	&	2.0 	$\pm$	0.3 	&	5.6 	$\pm$	1.6 	&	13.9			&	Y	\\
G209.77$-$19.40East2		&	0.47 	$\pm$	0.01 	&	0.19 	$\pm$	0.01 	&	0.29 	$\pm$	0.01 	&	0.035 			&	1.20 	$\pm$	0.56 	&	3.5 	$\pm$	0.6 	&	2.9 	$\pm$	1.4 	&	13.9			&	Y	\\
G209.94$-$19.52North		&	0.60 	$\pm$	0.01 	&	0.25 	$\pm$	0.01 	&	0.34 	$\pm$	0.01 	&	0.065 			&	2.81 	$\pm$	0.28 	&	8.9 	$\pm$	1.5 	&	3.2 	$\pm$	0.5 	&	16.3			&	N	\\
G210.82$-$19.47North1		&	0.44 	$\pm$	0.01 	&	0.17 	$\pm$	0.01 	&	0.30 	$\pm$	0.01 	&	0.070 			&	0.72 	$\pm$	0.08 	&	7.2 	$\pm$	1.2 	&	10.0 	$\pm$	1.5 	&	16.2			&	Y	\\
G210.82$-$19.47North2		&	0.35 	$\pm$	0.01 	&	0.13 	$\pm$	0.01 	&	0.27 	$\pm$	0.01 	&	0.030 			&	0.15 	$\pm$	0.05 	&	2.6 	$\pm$	0.5 	&	17.5 	$\pm$	6.1 	&	16.2			&	Y	\\
G211.16$-$19.33North1		&	0.44 	$\pm$	0.02 	&	0.18 	$\pm$	0.01 	&	0.28 	$\pm$	0.01 	&	0.055 			&	0.79 	$\pm$	0.05 	&	4.9 	$\pm$	0.8 	&	6.1 	$\pm$	0.8 	&	12.5			&	Y	\\
G211.16$-$19.33North2		&	0.46 	$\pm$	0.01 	&	0.19 	$\pm$	0.01 	&	0.28 	$\pm$	0.01 	&	0.050 			&	0.50 	$\pm$	0.20 	&	4.6 	$\pm$	0.8 	&	9.2 	$\pm$	3.8 	&	12.5			&	Y	\\
G211.16$-$19.33North3		&	0.35 	$\pm$	0.01 	&	0.14 	$\pm$	0.01 	&	0.25 	$\pm$	0.01 	&	0.050 			&	0.41 	$\pm$	0.12 	&	3.7 	$\pm$	0.6 	&	8.9 	$\pm$	2.8 	&	12.5			&	Y	\\
G211.16$-$19.33North5		&	0.48 	$\pm$	0.01 	&	0.19 	$\pm$	0.01 	&	0.29 	$\pm$	0.01 	&	0.060 			&	0.64 	$\pm$	0.18 	&	5.8 	$\pm$	1.0 	&	9.0 	$\pm$	2.7 	&	12.5			&	Y	\\
G211.47$-$19.27North		&	0.52 	$\pm$	0.01 	&	0.21 	$\pm$	0.01 	&	0.30 	$\pm$	0.01 	&	0.035 			&	1.66 	$\pm$	0.43 	&	3.6 	$\pm$	0.6 	&	2.2 	$\pm$	0.6 	&	12.4			&	N	\\
G211.47$-$19.27South		&	1.03 	$\pm$	0.03 	&	0.43 	$\pm$	0.01 	&	0.48 	$\pm$	0.01 	&	0.005 			&	12.25 	$\pm$	3.18 	&	1.3 	$\pm$	0.2 	&	0.1 	$\pm$	0.03 	&	12.4			&	N	\\
G211.72$-$19.25North		&	0.38 	$\pm$	0.02 	&	0.15 	$\pm$	0.01 	&	0.26 	$\pm$	0.01 	&	0.015 			&	0.07 	$\pm$	0.06 	&	1.2 	$\pm$	0.2 	&	16.8 	$\pm$	14.5 	&	12.6			&	Y	\\
G211.72$-$19.25South1		&	0.40 	$\pm$	0.05 	&	0.16 	$\pm$	0.02 	&	0.26 	$\pm$	0.03 	&	0.060 			&	1.15 	$\pm$	0.82 	&	4.9 	$\pm$	0.8 	&	4.3 	$\pm$	3.2 	&	12.6			&	Y	\\
G212.10$-$19.15North1		&	0.84 	$\pm$	0.04 	&	0.35 	$\pm$	0.02 	&	0.40 	$\pm$	0.02 	&	0.090 			&	3.52 	$\pm$	1.68 	&	17.0 	$\pm$	2.9 	&	4.8 	$\pm$	2.4 	&	10.8			&	Y	\\
G212.10$-$19.15North2		&	0.69 	$\pm$	0.02 	&	0.29 	$\pm$	0.01 	&	0.35 	$\pm$	0.01 	&	0.060 			&	1.65 	$\pm$	0.72 	&	8.5 	$\pm$	1.5 	&	5.1 	$\pm$	2.3 	&	10.8			&	Y	\\
G212.10$-$19.15North3		&	0.55 	$\pm$	0.05 	&	0.23 	$\pm$	0.02 	&	0.30 	$\pm$	0.03 	&	0.055 			&	1.36 	$\pm$	0.78 	&	5.8 	$\pm$	1.0 	&	4.2 	$\pm$	2.5 	&	10.8			&	Y	\\
G212.10$-$19.15South		&	0.38 	$\pm$	0.01 	&	0.15 	$\pm$	0.01 	&	0.25 	$\pm$	0.01 	&	0.075 			&	2.41 	$\pm$	1.00 	&	5.4 	$\pm$	0.9 	&	2.2 	$\pm$	1.0 	&	10.8			&	Y	\\
\hline																																				
Mean (All)		&	0.59 	$\pm$	0.23 	&	0.24 	$\pm$	0.10 	&	0.33 	$\pm$	0.08 	&	0.047 	$\pm$	0.023 	&	1.90 	$\pm$	2.14 	&	5.9 	$\pm$	3.6 	&	5.9 	$\pm$	4.6 	&	13.9 	$\pm$	2.7 	&		\\
Median (All)		&	0.50 			&	0.20 			&	0.30 			&	0.040 			&	1.36 			&	4.9 			&	4.7 			&	12.5			&		\\
Mean (Starless)		&	0.64 	$\pm$	0.11 	&	0.26 	$\pm$	0.11 	&	0.35 	$\pm$	0.09 	&	0.048 	$\pm$	0.025 	&	1.81 	$\pm$	1.53 	&	6.8 	$\pm$	4.1 	&	5.8 	$\pm$	4.5 	&				&		\\
Median (Starless)		&	0.63 			&	0.26 			&	0.35 			&	0.040 			&	1.37 			&	5.4 			&	4.5 			&				&		\\
Mean (Star-forming)		&	0.51 	$\pm$	0.16 	&	0.21 	$\pm$	0.07 	&	0.31 	$\pm$	0.05 	&	0.045 	$\pm$	0.021 	&	2.03 	$\pm$	2.84 	&	4.6 	$\pm$	2.0 	&	6.0 	$\pm$	4.8 	&				&		\\
Median (Star-forming)		&	0.48 			&	0.19 			&	0.30 			&	0.050 			&	1.36 			&	4.8 			&	5.1 			&				&		\\
\enddata
\tablecomments{Column 1: SCUBA-2 core name, Column 2: {\nth} FWHM linewidth toward the SCUBA-2 position \citep{2020ApJS..249...33K}, Column 3: nonthermal velocity dispersion,
Column 4: total velocity dispersion, Column 5: HWHM dust continuum radius \citep{2018ApJS..236...51Y}, Column 6: mass from dust continuum \citep{2018ApJS..236...51Y}, Column 7: virial mass, Column 8: virial parameter., Column 9: kinetic temperature, Column 10: association with the filament}
\end{deluxetable*}
\end{longrotatetable}

\subsection{Line emission distribution of the Orion core \label{sec:line}}

We statistically compared the emission distributions of the four lines on the basis of their integrated-intensity maps toward Orion cores.
We selected the emission feature that was
approximately coincident with the SCUBA-2 850-{\micron} dust continuum emission through visual inspection
and
fitted a two-dimensional Gaussian to it.
We identified the most prominent feature that was the nearest to the SCUBA-2 position.
Although this description may sound somewhat imprecise, the identification is practically straightforward.

The individual core data are listed in Tables \ref{tbl:core4line1} and \ref{tbl:core4line2} in the Appendix.
Table \ref{tbl:core4summary} summarizes statistics for the offset $\theta_{\rm offset}$ of the Gaussian-fit center from the SCUBA-2 position and 
the beam deconvolved radius $R$ of the fit. 
Note that the offset of CCS-L is twice as large as those of {\nth} and CCS-H. 
Therefore, CCS-L emission is in general less correlated with the SCUBA-2 emission.
We did not observe large differences in radius between the lines.

\begin{deluxetable*}{lCCCCCCCC} 
\tablecaption{Line emission distribution summary in the Orion region \label{tbl:core4summary}}
\tablewidth{0pt}
\tabletypesize{\scriptsize}
\tablenum{5}
\tablehead{
\colhead{SCUBA-2 core} &  
\multicolumn{2}{c}{{\nth}} & \multicolumn{2}{c}{HC$_3$N} &
\multicolumn{2}{c}{CCS-H} & \multicolumn{2}{c}{CCS-L} \\
\cline{2-3}
\cline{4-5}
\cline{6-7}
\cline{8-9}
\colhead{} & 
\colhead{$\theta_{\rm offset}$} & \colhead{$R$} & 
\colhead{$\theta_{\rm offset}$} & \colhead{$R$} & 
\colhead{$\theta_{\rm offset}$} & \colhead{$R$} & 
\colhead{$\theta_{\rm offset}$} & \colhead{$R$}  
\\
\colhead{} & 
\colhead{pc} & \colhead{pc} & 
\colhead{pc} & \colhead{pc} & 
\colhead{pc} & \colhead{pc} & 
\colhead{pc} & \colhead{pc}  
}
\decimalcolnumbers
\startdata																									
mean	&	0.027 		&	0.073 		&	0.031 		&	0.065 		&	0.021 		&	0.062 		&	0.045 		&	0.081 		\\																	stdev	&	0.026 		&	0.027 		&	0.036 		&	0.026 		&	0.022 		&	0.027 		&	0.041 		&	0.040 		\\	
\enddata
\tablecomments{Column 1: Calculation, 
Column 2: offset of the {\nth} emission from the SCUBA-2 position in pc, 
Column 3: deconvolved radius of the {\nth} emission in pc, 
Column 4: offset of the HC$_3$N emission from the SCUBA-2 position in pc,
Column 5: deconvolved radius of HC$_3$N emission in pc,
Column 6: offset of the CCS-H emission from the SCUBA-2 position in pc, 
Column 7: deconvolved radius of the CCS-H emission in pc, 
Column 8: offset of the CCS-L emission from the SCUBA-2 position in pc, and
Column 9: deconvolved radius of CCS-L emission in pc.
}
\end{deluxetable*}

\begin{figure*}
\epsscale{0.7}
\figurenum{10}
\includegraphics[bb=0 0 600 600, width=14cm]{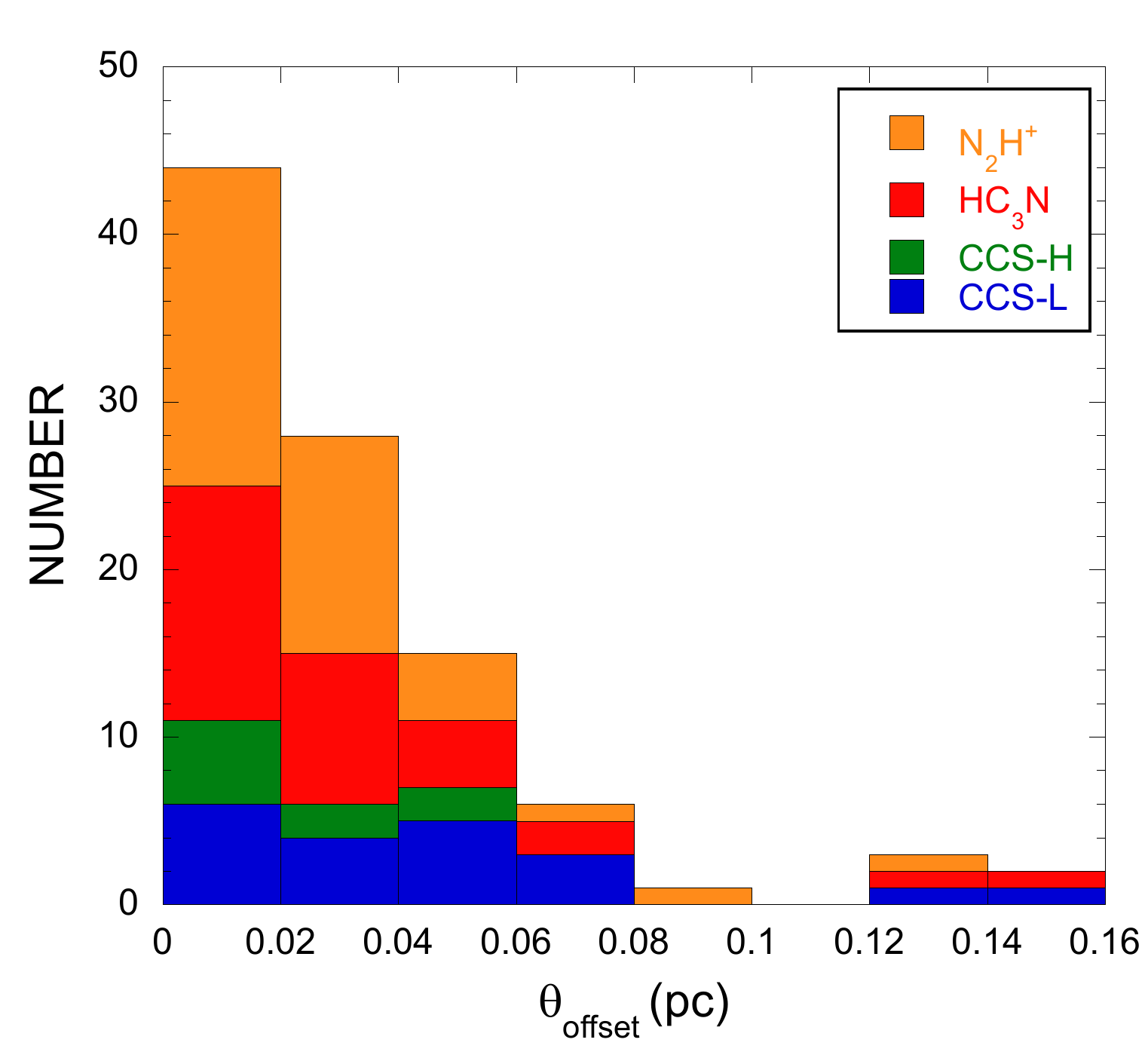}
\caption{Histogram of the line emission offset from the SCUBA-2 position.
Statistics are listed in Table \ref{tbl:core4summary}.
\label{fig:offset-scuba2}}
\end{figure*}

\begin{figure*}
\epsscale{0.7}
\figurenum{11}
\includegraphics[bb=0 0 600 600, width=14cm]{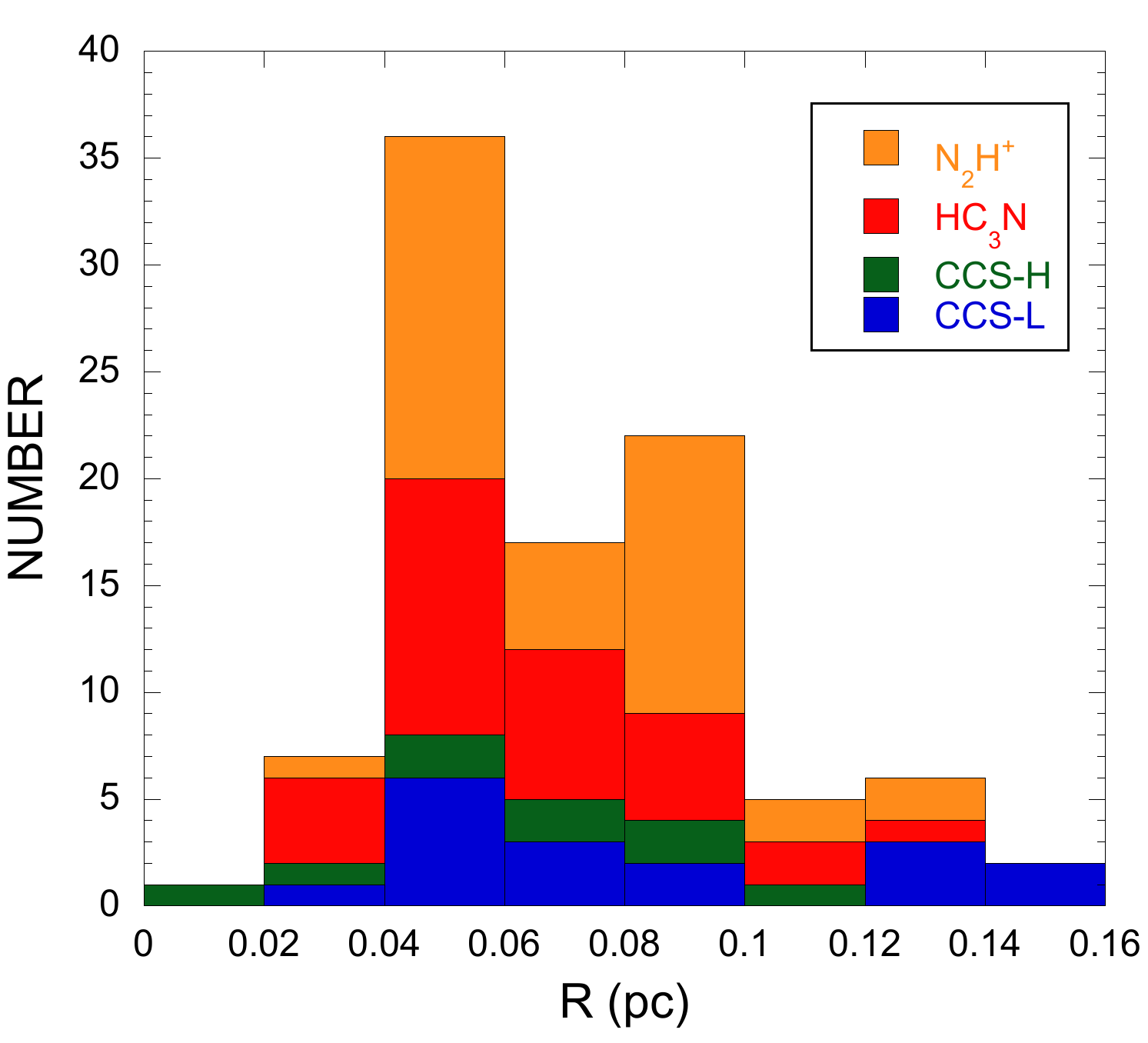}
\caption{Histogram of the deconvolved radius $R$ for each line.  
Statistics are listed in Table \ref{tbl:core4summary}.
\label{fig:R-line}}
\end{figure*}

We also illustrate the offset and emission radius by histogram.
Figure \ref{fig:offset-scuba2} provides a comparison of the offset of the line emission
fitted ellipse center from the SCUBA-2 position for the Orion cores.
The offsets in {\nth} and HC$_3$N are similarly distributed, having sharp peaks at $\theta_{\rm offset}$ = 0$-$0.02 pc,
whereas those in CCS-H and CCS-L appear to be less peaked.
The NRO telescope beam radius is 0.018 pc.
Figure \ref{fig:R-line} shows a comparison of the deconvolved radius values of the line emission
feature for the Orion cores, in which no significant differences among the lines are observed.

\subsection{Comparison between the starless and star-forming cores in the Orion region \label{sec:size}}

\begin{figure*}
\epsscale{0.7}
\figurenum{12}
\includegraphics[bb=0 0 600 600, width=14cm]{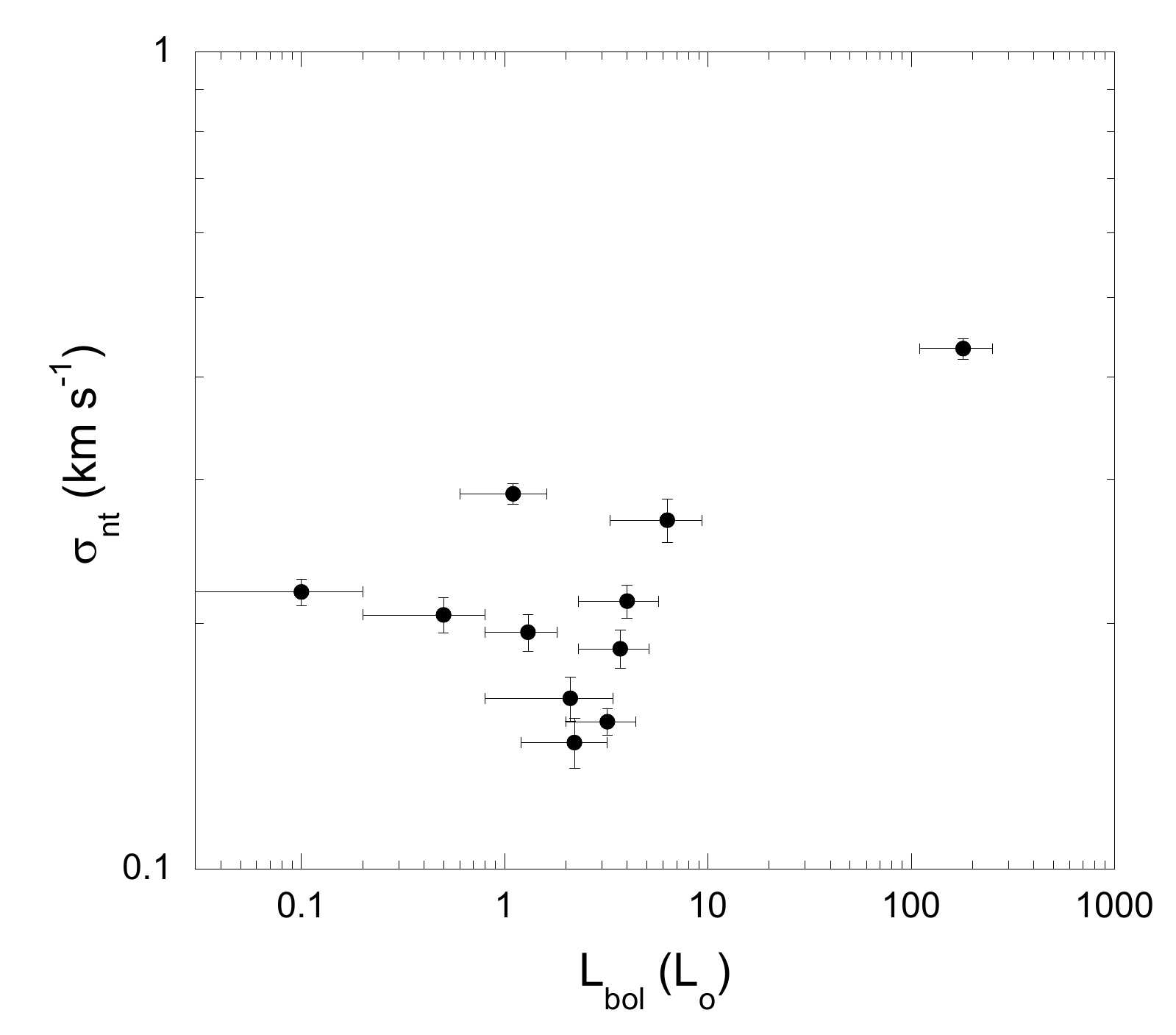}
\caption{Nonthermal velocity dispersion vs.
bolometric luminosity of the YSO \citep{2020ApJS..251...20D}.  \label{fig:sigma-Lbol}}
\end{figure*}

At the bottom of Table \ref{tbl:vir}, we list parameter statistics for all the cores together, and starless and star-forming cores separately.
The starless and star-forming cores have very similar radii, 
suggesting no evolutionary change in the core radius is detected at 0.027-pc linear resolution.  
Furthermore, their nonthermal velocity dispersions, masses, and virial masses are not appreciably different.
Therefore, no evidence for the dissipation of turbulence 
is identified at the employed spatial resolution, as already noted by \cite{2020ApJS..249...33K}.
The absence of differences in the velocity dispersion
between the starless and star-forming core
contradicts the results of earlier studies \citep{1986ApJ...307..337B, 1989ApJS...71...89B, 1989ApJ...346..168Z, 1993ApJ...404..643T}.
One possibility is that largely improved sensitivities allow
the identification of  
low-luminosity young stellar objects (YSOs), which were previously undetectable.
To test this possibility, we plotted the nonthermal velocity dispersion against the bolometric luminosity of the YSO
taken from \cite{2020ApJS..251...20D}
(Figure \ref{fig:sigma-Lbol}). 
Except for the core with the most luminous YSO (G211.47$-$19.27South, $L$ =  180$\pm$70 $L_{\sun}$), 
no clear positive correlation was observed.
Our data were less affected by star formation activities because the tracer {\nth} employed
is insensitive to shocked gas \citep{1977ApJ...211..755T,1993ApJ...406L..29W,1996ARA&A..34..111B,2002ApJ...572..238C}, unlike CS and NH$_3$ employed in previous studies.
Our samples contain warm $T_{bol}$ YSOs, which are not too young to observe the impacts of YSO activities.

\begin{figure}
\epsscale{0.7}
\figurenum{13}
\includegraphics[bb=0 0 600 600, width=14cm]{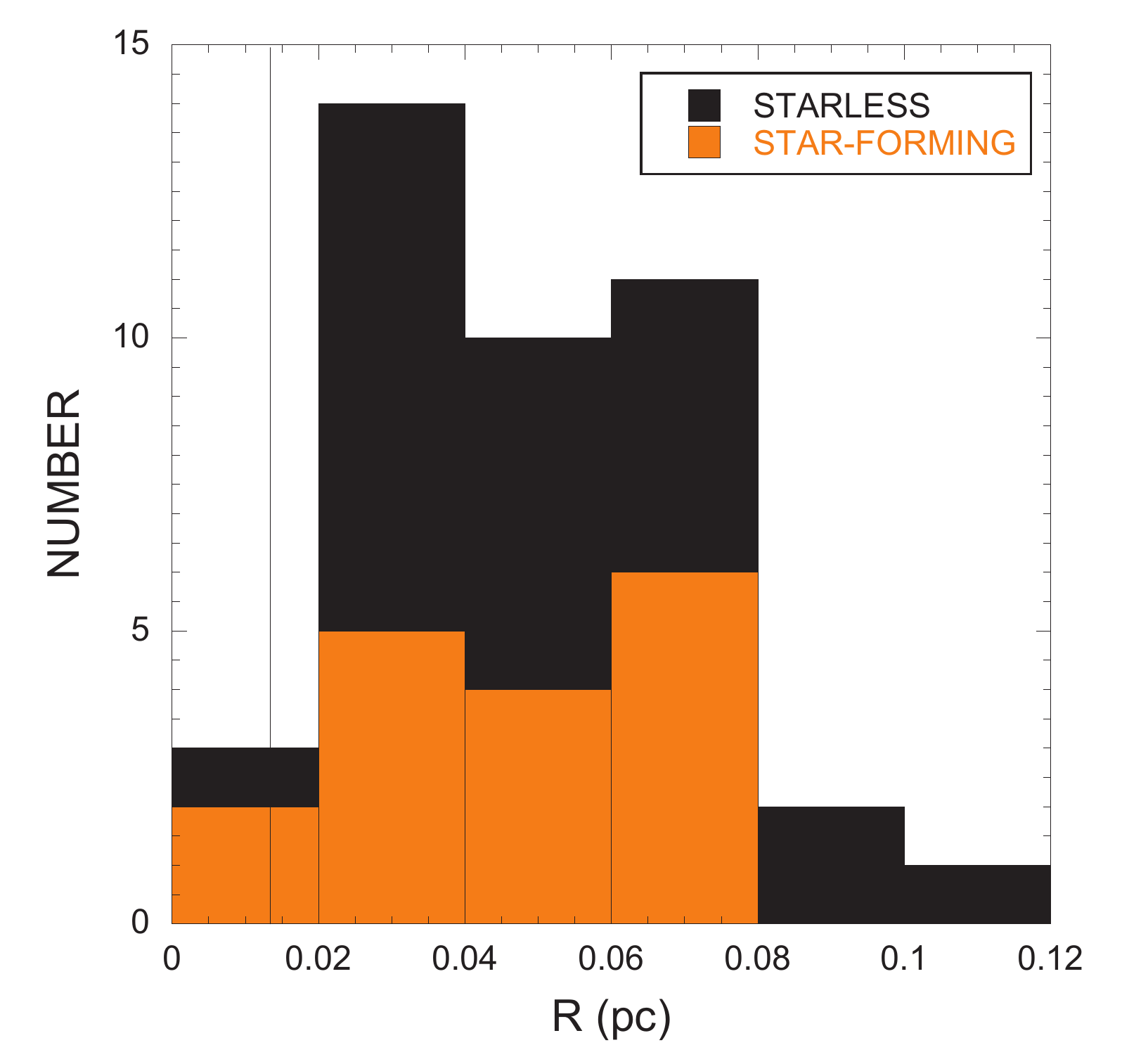}
\caption{Histogram of the  
deconvolved
radii 
of the SCUBA-2 cores in the Orion region. Black and orange colors indicate the starless and star-forming cores, respectively. 
The vertical solid line represents the telescope beam radius. \label{fig:r-hist}}
\end{figure}

We also compared the deconvolved radii \citep{2018ApJS..236...51Y} between the starless and star-forming cores. 
Figure \ref{fig:r-hist} shows the histogram of the SCUBA-2 radius $R$ (dust) of 
the two categories. 
The starless and star-forming cores have very similar radii.
We corrected the semi-major axis $a$ and semi-minor axis $b$
for the telescope beam in quadrature,
and calculated the deconvolved radius $R$:

\begin{equation}
R = [(a^2-({\rm HPBW}/2)^2)(b^2-({\rm HPBW}/2)^2)]^{1/4}.
\end{equation}

The SCUBA-2 beam radius is 7 arcsec or 0.0135 pc; 
thus, the cores are well resolved. 
This similarity is not consistent with the result from H$^{13}$CO$^+$ observations toward Taurus
by \citet{1994Natur.368..719M},
who concluded that starless cores are larger than star-forming cores, 
but consistent with that from N$_2$H$^+$ observations by \citet{2004ApJ...606..333T} for the same cores.
A possible reason for this difference is the depletion of H$^{13}$CO$^+$ molecules in the starless cores,
which could cause flat-topped radial intensity profiles toward
cold starless cores, and correspondingly lead to larger radii \citep{2004ApJ...606..333T}. 
It seems that when we use a molecular-line tracer affected by depletion,
we see a size difference between the starless and star-forming core.
In contrast, when we use the dust continuum or a tracer that is less affected by depletion,
it seems difficult to observe differences in radii with HPBWs of 0.01$-$0.03 pc.
When we adopt Welch's t-test for the null hypothesis that the starless and star-forming cores have equal radii,
we obtain a p-value of 0.55, which is considerably larger than a standard threshold of 0.05 for statistical significance.
Thus, we cannot statistically conclude that these two categories have different radii.

It is suggested that sizes of identified cores are often very close to the telescope beam size, and this can be well explained if cores
have power-law density profiles.
\citet{1991ApJ...382..555L} indicated that the observed emission size tends to be 
10$-$80{\%} larger than the original beam size if we assume the power-law radial intensity distribution. 
\citet{2003ApJS..145..111Y} showed anticorrelation between the power-law index and the deconvolved emission size (see their Fig. 27).  
Then, in such cases, the observed radius represents the radial density slope rather than the physical boundary.
The mean HWHM radius ($23\farcs0~\pm~11\farcs4$ or 0.045$\pm$0.022 pc) and beam radius (7$\arcsec$ or 0.0135 pc) suggest a power-law index $p$ of $<$1.4 for the radial density profile 
$\rho~(r) \propto r^{-p}$.
Therefore, the SCUBA-2 core in Orion studied by \cite{2018ApJS..236...51Y} arguably has a shallow radial density profile.

\section{DISCUSSION \label{sec:dis}}

\subsection{Physical parameters of the starless cores against CEF2.0 \label{sec:cef}}

We investigated how the physical parameters of the starless cores compare to evolutionary indications based on chemistry, 
as encoded by CEF2.0,
which is the chemical evolution tracer based the deuterium fraction.
Table \ref{tbl:df} lists the deuterium fraction and CEF2.0 for all cores which allow column density estimation.
Note that CEF2.0 is defined only for the starless cores in Orion, but this Table contains both the starless and star-forming cores, and also the cores outside Orion.
\cite{2020ApJS..249...33K} pointed out that the deuterium fraction will be seriously underestimated for distant ($>$ 1 kpc) cores due to beam dilution.  The adopted distance is listed in Table \ref{tbl:field}.
Table \ref{tbl:yso} in the Appendix contains the YSO association information.
The CEF2.0 values of the Orion starless cores were calculated from
the column density ratios of {\ntd}/{\nth} and DNC/HN$^{13}$C from our single-pointing observations \citep{2020ApJS..249...33K}.
We question whether there is any evidence that stable cores change into unstable cores by changing physical properties such as turbulence.

\begin{figure*}
\epsscale{0.7}
\figurenum{14}
\includegraphics[bb=0 0 600 600, width=14cm]{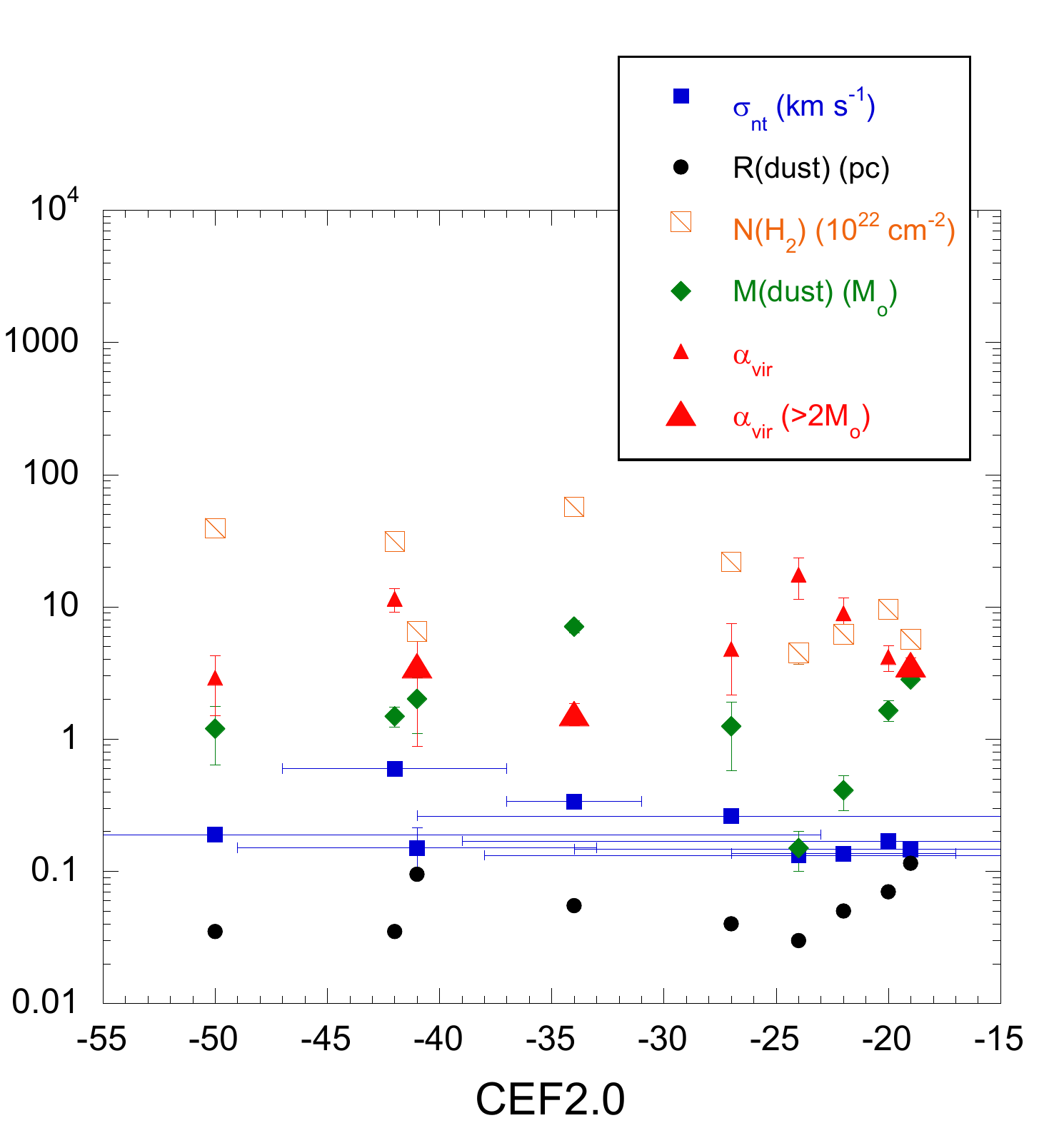}
\caption{Nonthermal velocity dispersions, radii, H$_2$ column densities, masses, and virial parameters of the starless core vs. CEF2.0. 
The core radii, column densities, and masses are taken from \cite{2018ApJS..236...51Y}.
The error bar for CEF2.0 is plotted only for the nonthermal velocity dispersion for clarity. \label{fig:CEF-trend}}
\end{figure*}

Figure \ref{fig:CEF-trend} shows the plot of the nonthermal velocity dispersions, radii, H$_2$ column densities, masses, 
and virial parameters of the Orion starless cores against CEF2.0,
in which no significant trend is observed.
Radius $R$ and mass $M$ increase with increasing CEF2.0 from CEF2.0 = $-$24 to $-$19, but 
this rise is probably not significant 
when we take into account the error in CEF2.0.
The virial parameter $\alpha_{vir}$  
of the starless cores
is as large as 1$-$20.
If we allow 1$\sigma$ uncertainty, four cores (44\%) out of the nine 
starless cores
in this figure may fit the range $0.5 < \alpha_{vir} < 2$.
Large virial parameters for starless cores suggest 
that most of them are either pressure-confined or transient.
\cite{2020arXiv200607325C} showed the evolution of unbound cores
into bound ones in their simulation.

\cite{2005ApJ...619..379C} noted that the
chemical properties (including the deuterium fraction) and physical properties
(density, line width, etc.) of starless cores as a whole
can trace the evolutionary stage, although
they did not find very tight dependencies between these properties.
Their samples included starless cores at younger stages such as 
B68, TMC-1, L492, L1498, and L1512 \citep{2005ApJ...632..982S, 2006ApJ...646..258H}, 
which have values of $-$100 $\lesssim$ CEF2.0 $\lesssim$ $-$60 \citep{2020ApJS..249...33K}, implying lower deuterium fractions. 
They concluded that evolved starless cores with higher $N$(\ntd)/$N$(\nth) tend to have
higher central H$_2$ column densities and more compact density profiles.
They employed multiple tracers including {\nth}, {\ntd}, and the dust continuum emission,
and the above physical parameters were derived at linear resolutions of 0.006$-$0.02 pc, 
1.5$-$6 times better than our linear resolution (0.03$-$0.04 pc).
The difference in the linear resolution and/or the CEF2.0 range may at least partially explain the difference in conclusion.

\cite{2012A&A...537A...4M} observed Bok globules in NH$_3$ and CCS to study their chemical evolution.
They did not find any globules with extremely high CCS abundances
and concluded that all of the observed Bok globules are in a relatively evolved chemical state.
It is possible that SCUBA-2 cores with high column densities selected in our study 
are similarly biased to evolved cores.

We did not observe intense CCS emission even at the starless core stage.
Relatively late chemical evolution stages may be one reason, but
we also suspect that serious depletion occurs in CCS at cold starless cores,
which leads to low values of the CCS column density.
Some previous studies of core chemistry evolution 
used the abundance at the molecular line emission maximum
\citep{2014PASJ...66...16T, 2014PASJ...66..119O},
which can be different from that at density peaks traced by the dust continuum emission.
In such cases, depletion of CCS can be alleviated.
We used the SCUBA-2 position as the core centers where CCS depletion could be severe.

\floattable
\begin{deluxetable}{lccccc}
\tablecaption{Deuterium Fraction and CEF2.0 \label{tbl:df}}
\tablecolumns{6}
\tablenum{6}
\tablewidth{0pt}
\tablehead{
\colhead{SCUBA-2 core} &
\colhead{$V_{LSR}$} &
\colhead{$N$({\ntd})/$N$({\nth})}&
\colhead{$N$(DNC)/$N$(HN$^{13}$C)}&
\colhead{YSO}&
\colhead{CEF2.0}\\
\colhead{} &
\colhead{km s$^{-1}$} &
\colhead{}&
\colhead{}&
\colhead{}&
\colhead{} 
}
\startdata
G192.32$-$11.88North	&	12.1 	&	0.15 	$\pm$	0.08 	&	5.2 	$\pm$	3.7 	&	Y	&				\\
G192.32$-$11.88South	&	12.2 	&	0.17 	$\pm$	0.02 	&	3.8 	$\pm$	2.7 	&	Y	&				\\
G203.21$-$11.20West1	&	10.7 	&	0.30 	$\pm$	0.04 	&		$<$	4.1 	&	Y	&				\\
G203.21$-$11.20West2	&	10.2 	&	0.21 	$\pm$	0.05 	&		$<$	4.0 	&	Y	&				\\
G204.4$-$11.3A2East	&	1.6 	&	0.27 	$\pm$	0.07 	&		$<$	4.2 	&	Y	&				\\
G204.4$-$11.3A2West	&	1.7 	&	0.09 	$\pm$	0.04 	&		. . .		&	Y	&				\\
G206.12$-$15.76	&	8.5 	&		. . .		&		$<$	4.8 	&	Y	&				\\
G206.93$-$16.61West1	&	9.3 	&		$<$	0.15 	&	1.0 	$\pm$	0.7 	&	Y	&				\\
G206.93$-$16.61West3	&	9.3 	&	0.04 	$\pm$	0.02 	&	1.7 	$\pm$	1.2 	&	Y	&				\\
G206.93$-$16.61West4	&	10.1 	&	0.16 	$\pm$	0.04 	&	5.1 	$\pm$	3.7 	&	N	&	-27 	$\pm$	14 	\\
G206.93$-$16.61West5	&	9.0 	&	0.15 	$\pm$	0.02 	&		. . .		&	N	&	-34 	$\pm$	3 	\\
G206.93$-$16.61West6	&	10.4 	&		$<$	0.23 	&	2.7 	$\pm$	1.9 	&	Y	&				\\
G207.36$-$19.82North2	&	11.2 	&	0.22 	$\pm$	0.05 	&		$<$	6.8 	&	Y	&				\\
G207.36$-$19.82South	&	11.3 	&	0.11 	$\pm$	0.04 	&		. . .		&	N	&	-41 	$\pm$	8 	\\
G208.68$-$19.20North1	&	11.1 	&		$<$	0.08 	&	2.6 	$\pm$	1.8 	&	Y	&				\\
G208.68$-$19.20North2	&	11.2 	&	0.11 	$\pm$	0.01 	&	3.1 	$\pm$	2.2 	&	Y	&				\\
G208.68$-$19.20North3	&	11.1 	&	0.06 	$\pm$	0.01 	&	2.7 	$\pm$	1.9 	&	Y	&				\\
G209.05$-$19.73North	&	8.3 	&	0.27 	$\pm$	0.06 	&	5.7 	$\pm$	4.2 	&	N	&	-19 	$\pm$	15 	\\
G209.05$-$19.73South	&	7.9 	&	0.23 	$\pm$	0.13 	&	5.9 	$\pm$	4.4 	&	N	&	-20 	$\pm$	19 	\\
G209.29$-$19.65North1	&	8.6 	&	0.06 	$\pm$	0.03 	&	2.0 	$\pm$	1.4 	&	N	&	-58 	$\pm$	17 	\\
G209.29$-$19.65South1	&	7.5 	&	0.11 	$\pm$	0.02 	&		. . .		&	N	&	-42 	$\pm$	5 	\\
G209.29$-$19.65South2	&	7.8 	&		$<$	0.08 	&	3.4 	$\pm$	2.5 	&	N	&	-38 	$\pm$	36 	\\
G209.29$-$19.65South2	&	9.0 	&	0.14 	$\pm$	0.08 	&	0.7 	$\pm$	0.5 	&	N	&	-67 	$\pm$	18 	\\
G209.77$-$19.40East1	&	8.1 	&	0.05 	$\pm$	0.01 	&	1.8 	$\pm$	1.3 	&	Y	&				\\
G209.77$-$19.40East2	&	8.0 	&		$<$	0.14 	&	2.5 	$\pm$	1.8 	&	N	&	-50 	$\pm$	27 	\\
G209.77$-$19.40East3	&	7.8 	&	0.25 	$\pm$	0.04 	&	4.7 	$\pm$	3.4 	&	N	&	-23 	$\pm$	14 	\\
G209.77$-$19.40East3	&	8.3 	&		$<$	0.63 	&	1.9 	$\pm$	1.4 	&	N	&	-61 	$\pm$	28 	\\
G209.94$-$19.52North	&	8.1 	&	0.33 	$\pm$	0.07 	&	3.6 	$\pm$	2.5 	&	Y	&				\\
G209.94$-$19.52South1	&	7.5 	&	0.24 	$\pm$	0.13 	&	6.3 	$\pm$	4.5 	&	N	&	-18 	$\pm$	15 	\\
G209.94$-$19.52South1	&	8.1 	&	0.24 	$\pm$	0.03 	&	4.1 	$\pm$	2.9 	&	N	&	-26 	$\pm$	14 	\\
G210.82$-$19.47North1	&	5.3 	&	0.32 	$\pm$	0.03 	&	4.0 	$\pm$	2.9 	&	Y	&				\\
G210.82$-$19.47North2	&	5.3 	&	0.20 	$\pm$	0.02 	&	5.3 	$\pm$	3.8 	&	N	&	-24 	$\pm$	14 	\\
G211.16$-$19.33North2	&	3.7 	&	0.10 	$\pm$	0.04 	&		$<$	2.5 	&	Y	&				\\
G211.16$-$19.33North3	&	3.4 	&	0.24 	$\pm$	0.05 	&		$<$	3.5 	&	N	&	-22 	$\pm$	5 	\\
G211.16$-$19.33North5	&	4.3 	&	0.31 	$\pm$	0.06 	&		$<$	3.2 	&	Y	&				\\
G211.47$-$19.27North	&	4.1 	&	0.10 	$\pm$	0.02 	&		$<$	3.1 	&	Y	&				\\
G212.10$-$19.15North1	&	4.3 	&	0.35 	$\pm$	0.07 	&		$<$	6.1 	&	Y	&				\\
G212.10$-$19.15North3	&	4.2 	&	0.06 	$\pm$	0.03 	&	2.8 	$\pm$	2.0 	&	Y	&				\\
G212.10$-$19.15South	&	3.9 	&	0.39 	$\pm$	0.06 	&		$<$	3.5 	&	Y	&				\\
SCOPEG159.18$-$20.09	&	6.3 	&		. . .		&	13.8 	$\pm$	9.7 	&	Y	&				\\
SCOPEG159.22$-$20.11	&	6.7 	&		. . .		&	10.5 	$\pm$	7.5 	&	Y	&				\\
SCOPEG173.17$+$02.36	&	-18.9 	&		. . .		&	1.1 	$\pm$	0.8 	&	Y	&				\\
SCOPEG173.18$+$02.35	&	-19.0 	&		. . .		&	1.3 	$\pm$	0.9 	&	Y	&				\\
SCOPEG173.19$+$02.35	&	-19.3 	&		. . .		&	1.6 	$\pm$	1.2 	&	Y	&				\\
SCOPEG001.37$+$20.95	&	0.8 	&		. . .		&	14.6 	$\pm$	10.3 	&	N	&				\\
SCOPEG017.38$+$02.26	&	10.7 	&		. . .		&	1.7 	$\pm$	1.3 	&	Y	&				\\
SCOPEG017.37$+$02.24	&	10.5 	&		. . .		&	1.1 	$\pm$	0.8 	&	Y	&				\\
SCOPEG017.36$+$02.23	&	10.4 	&		. . .		&	0.6 	$\pm$	0.4 	&	Y	&				\\
SCOPEG014.18$-$00.23	&	40.5 	&		. . .		&	1.1 	$\pm$	0.8 	&	Y	&				\\
SCOPEG016.93$+$00.25	&	24.2 	&		. . .		&	2.3 	$\pm$	1.6 	&	N	&				\\
SCOPEG016.93$+$00.24	&	24.3 	&		. . .		&	0.6 	$\pm$	0.4 	&	N	&				\\
SCOPEG016.93$+$00.24	&	26.3 	&		. . .		&	0.8 	$\pm$	0.6 	&	N	&				\\
SCOPEG016.93$+$00.22	&	23.6 	&		. . .		&	1.4 	$\pm$	1.0 	&	Y	&				\\
SCOPEG033.74$-$00.01	&	105.0 	&		. . .		&	0.4 	$\pm$	0.3 	&	Y	&				\\
SCOPEG035.48$-$00.29	&	45.3 	&		. . .		&	1.8 	$\pm$	1.3 	&	Y	&				\\
SCOPEG035.52$-$00.27	&	45.1 	&		. . .		&	1.6 	$\pm$	1.1 	&	Y	&				\\
SCOPEG035.48$-$00.31	&	44.9 	&		. . .		&	1.6 	$\pm$	1.1 	&	Y	&				\\
SCOPEG034.75$-$01.38	&	45.6 	&		. . .		&	0.9 	$\pm$	0.6 	&	Y	&				\\
SCOPEG069.80$-$01.67	&	12.4 	&		. . .		&	1.2 	$\pm$	0.8 	&	Y	&				\\
SCOPEG069.81$-$01.67	&	12.1 	&		. . .		&	1.2 	$\pm$	0.9 	&	Y	&				\\
SCOPEG105.37$+$09.84	&	-9.7 	&		. . .		&	1.8 	$\pm$	1.3 	&	Y	&				\\
SCOPEG107.16$+$05.45	&	-10.2 	&		. . .		&	5.5 	$\pm$	3.9 	&	Y	&				\\
SCOPEG107.18$+$05.43	&	-10.8 	&		. . .		&	1.8 	$\pm$	1.3 	&	Y	&				\\
\enddata
\end{deluxetable}

\subsection{Core evolution against the bolometric temperature}

\begin{figure*}
\epsscale{0.7}
\figurenum{15}
\includegraphics[bb=0 0 600 600, width=12cm]{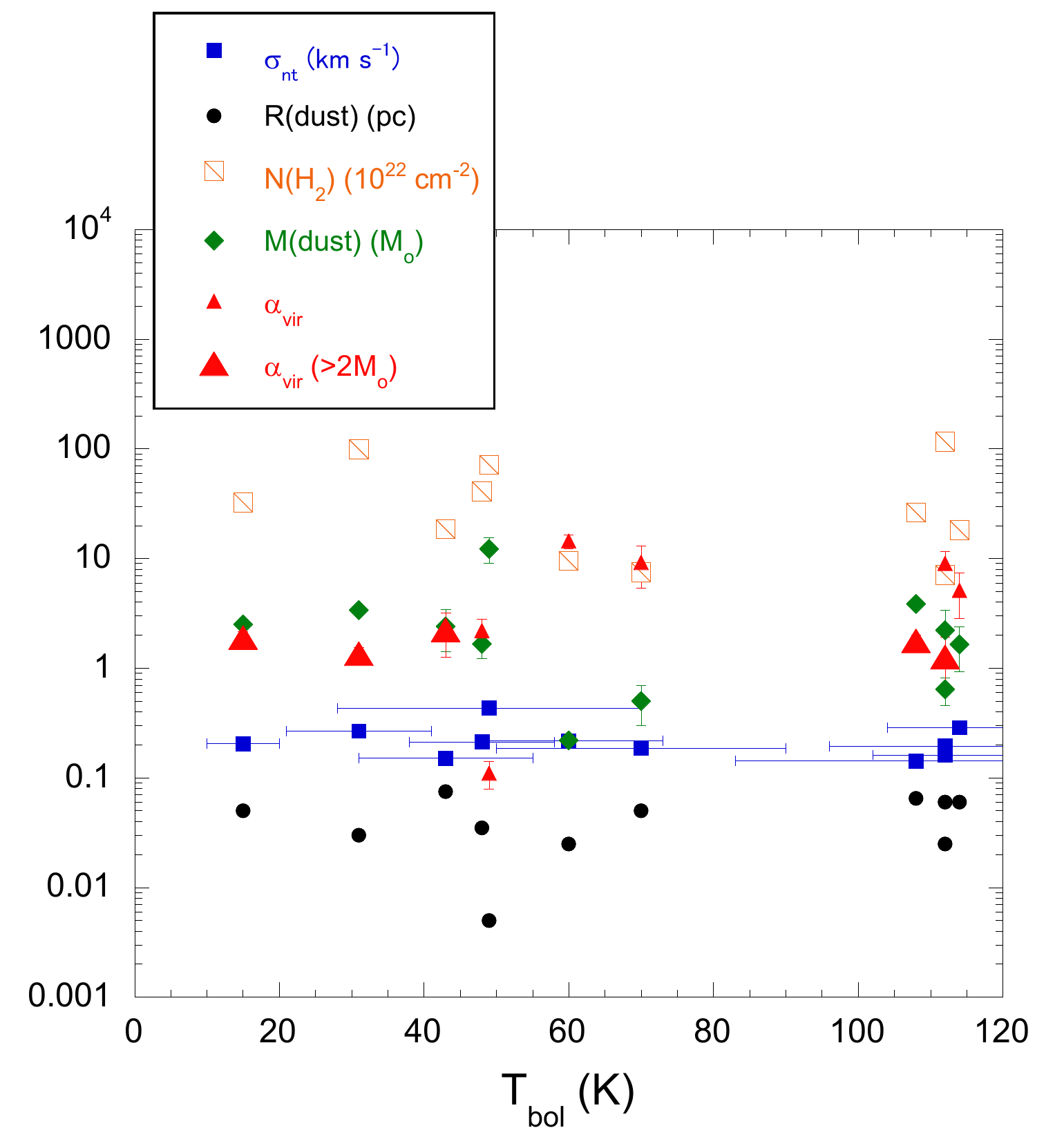}
\caption{Nonthermal velocity dispersions, radii, H$_2$ column densities, masses, and virial parameters 
of the star-forming cores
vs. bolometric temperature $T_{bol}$  of the YSOs \citep{2020ApJS..251...20D}. 
The error bar for $T_{bol}$ is plotted only for the nonthermal velocity dispersion for clarity. \label{fig:Tbol-trend}}
\end{figure*}

We investigated the physical parameters of the star-forming cores against 
the bolometric temperature of associated YSOs \citep{2020ApJS..251...20D}. 
The bolometric temperature is a reliable empirical tracer of a star-forming core's evolution \citep{1995ApJ...445..377C}.
The bolometric temperature $T_{bol}$ can be obtained from
the flux-weighted mean frequencies in the observed
spectral energy distributions (SEDs) \citep{1993ApJ...413L..47M}.
\cite{2020ApJS..251...20D} used their ALMA~1.3 mm fluxes as well as fluxes 
at other wavelengths from
\cite{2003yCat.2246....0C}, \cite{2007MNRAS.379.1599L}, \cite{2010AJ....140.1868W}, \cite{2012AJ....144..192M},
\cite{2013ApJ...767...36S}, \cite{2015ApJ...798..128T}, \cite{2015PASJ...67...50D}, and \cite{2018ApJS..236...51Y}.
Figure \ref{fig:Tbol-trend} shows the nonthermal velocity dispersions, radii, H$_2$ column densities, masses, and virial parameters
vs. $T_{bol}$ for the star-forming cores \citep{2020ApJS..251...20D}.
We did not observe any significant trends in this Figure.
The virial parameter $\alpha_{vir}$ 
of the star-forming core
ranged from 0.1 to 20.
Six (55\%) of the eleven 
star-forming cores
in this Figure may fit the range  $\alpha_{vir}$ = 0.5$-$2 
when we allow a 1$\sigma$ uncertainty.
The core that forms a protostar should be gravitationally bound or pressure confined at least when the protostar forms.
Large virial parameters of the star-forming core suggest either the confining external pressure,
fast core dispersal after star formation, strong tidal forces from the ambient gas,
or systematic bias in estimation of the parameter.
Furthermore, the associated outflow can stir up the core to some extent through magnetic fields,
although {\nth} is less affected by shocks.

\begin{figure*}
\epsscale{1.1}
\figurenum{16}
\includegraphics[bb=0 0 600 600, width=14cm]{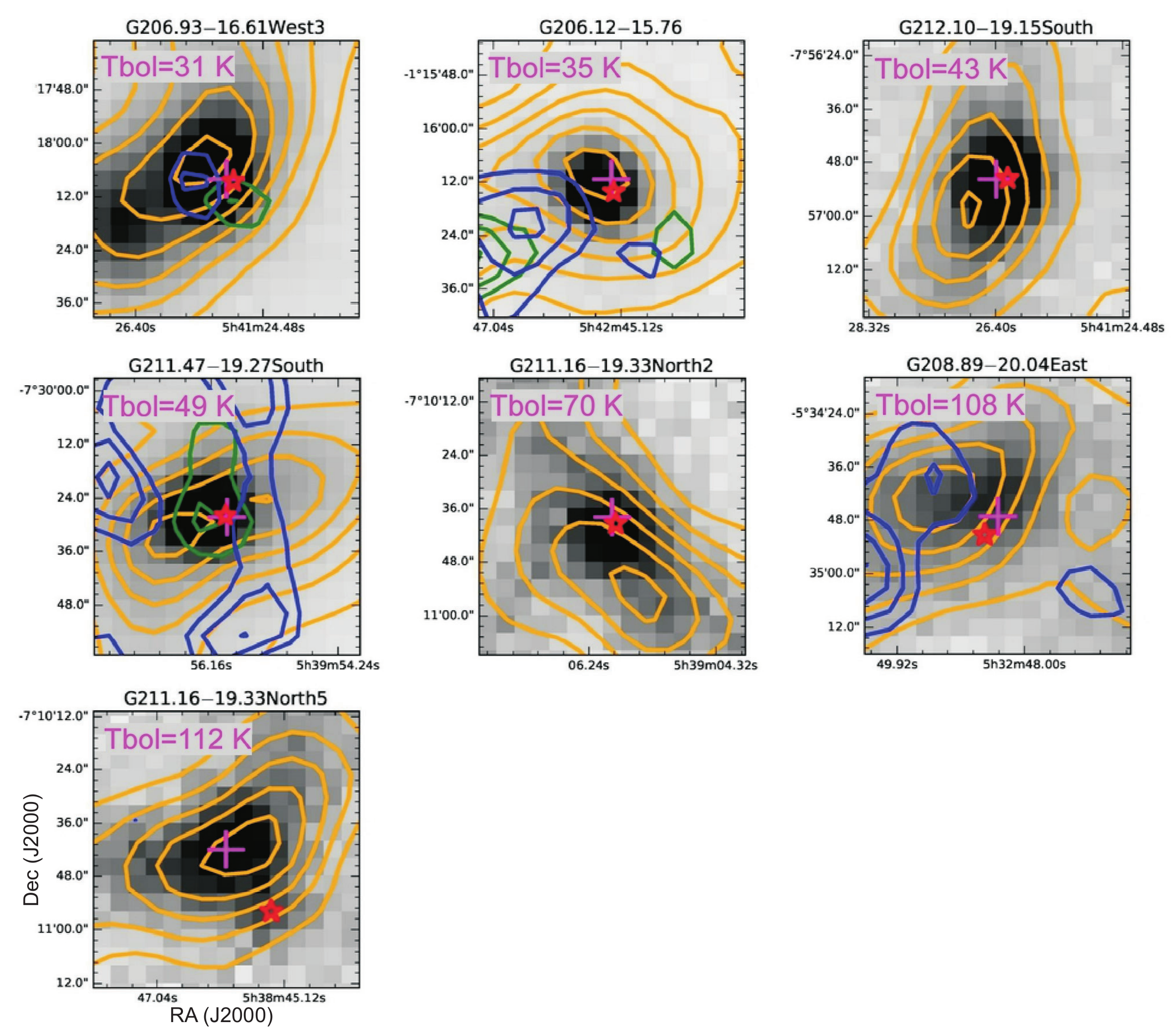}
\caption{Close-up maps sorted in order of evolutionary stages using  $T_{\rm bol}$ for representative star-forming cores. Each map represents the integrated intensity contour map superposed on the 850-{\micron} dust continuum emission map. The orange, green, and blue contours represent the {\nth}, CCS-H, and CCS-L  integrated-intensity contour maps, respectively. 
The innermost to outermost contour levels correspond to 95\%, 80\%, 65\%, 50\%, 35\%, 20\%, 5\%, 
and 3~$\sigma$, respectively. 
The peak value and $\sigma$ value of integrated intensity are listed in Table \ref{tbl:map}. 
The cross and star symbols represent a SCUBA-2 core and a protostar, respectively. 
The error for $T_{\rm bol}$ is listed in Table \ref{tbl:yso} in the Appendix.
\label{fig:spmap}}
\end{figure*}

We also investigated the distribution of CCS (young-gas tracer) and {\nth} (evolved-gas tracer) vs. $T_{bol}$. 
Figure \ref{fig:spmap} shows the close-up maps of representative star-forming cores 
aligned in order of 
the bolometric temperature $T_{bol}$  of the associated YSO as an evolutionary tracer. 
Comparisons including HC$_3$N on the full-sized maps of the all star-forming Orion cores 
with $T_{bol}$ can be seen in Table \ref{tbl:yso}, Figures1 to 7, and Figure Set 17. Again,
no clear trend is seen.

\subsection{Relationship with the association with the Filament}

Some of the observed cores are apparently located in filaments. 
Because filaments likely play an important role in 
core mass accretion, we investigate whether there are any differences between
cores inside or outside filaments.
Association with a filament was judged from the SCUBA-2 and {\nth} distributions,
and is shown in Table \ref{tbl:vir}.
Table \ref{tbl:fila} summarizes the statistics of the Orion core.
Here, we do not distinguish the starless and star-forming cores.
The nonthermal velocity dispersions and radii are similar between the two categories.
The cores associated with filaments are slightly less massive and have smaller virial parameters,
although the standard deviation is large and so the differences may not be significant.

\floattable
\begin{deluxetable}{llllll}
\tablecaption{Core Properties with and without the filaments \label{tbl:fila}}
\tablecolumns{6}
\tablenum{7}
\tablewidth{0pt}
\tablehead{
\colhead{Filament} &
\colhead{Calculation} &
\colhead{$\sigma_{nt}$} &
\colhead{$R$ (dust)} &
\colhead{$M$ (dust)}&
\colhead{$\alpha_{vir}$} \\
\colhead{} &
\colhead{} &
\colhead{km s$^{-1}$} &
\colhead{pc} &
\colhead{$M_{\sun}$}&
\colhead{}
}
\startdata
Y & mean    & 0.24$\pm$0.11  &  0.05$\pm$0.02 & 1.58$\pm$1.41 & 6.29$\pm$4.56 \\
Y & median & 0.20                  &  0.05                &  1.23               & 4.93                 \\
N & mean    & 0.25$\pm$0.09 &  0.04$\pm$0.03 & 3.03$\pm$3.66 & 4.54$\pm$4.62 \\
N & median & 0.22                  &  0.04                &  2.01               &  3.18                \\
\enddata
\end{deluxetable}

\subsection{Kinematics}

Details of the kinematics of the cores discussed here are beyond the scope of this paper, but
we consider here two example fields, each containing representative examples, G206.12 and G211.16.
We selected them from Orion cores, as they have more accurate distances \citep{2017ApJ...834..142K,1997A&A...323L..49P}
and are less affected by beam dilution \citep{2020ApJS..249...33K}.

One field G206.12 has only one SCUBA-2 core inside the mapped area  
(Figure \ref{fig:850_4cont_G206_12}),
whereas the other field G211.16 contains six SCUBA-2 cores (Figure \ref{fig:850_4cont_G211_16}).
The {\nth} emission was used for analysis.
We calculated the intensity-weighted radial velocity (moment 1) using the hyperfine group
$J = 1\rightarrow0$ 
$F_1$ = $2\rightarrow1$ containing the brightest hyperfine component to obtain better S/N ratios, and the velocity dispersion (moment 2) using
the isolated hyperfine component $J = 1\rightarrow0$,
$F_1, F = 0, 1\rightarrow1, 2$, which is optically thinner and less affected by the neighboring satellites.

\begin{figure*}
\epsscale{0.5}
\figurenum{17}
\includegraphics[bb=0 0 600 1200, width=10cm]{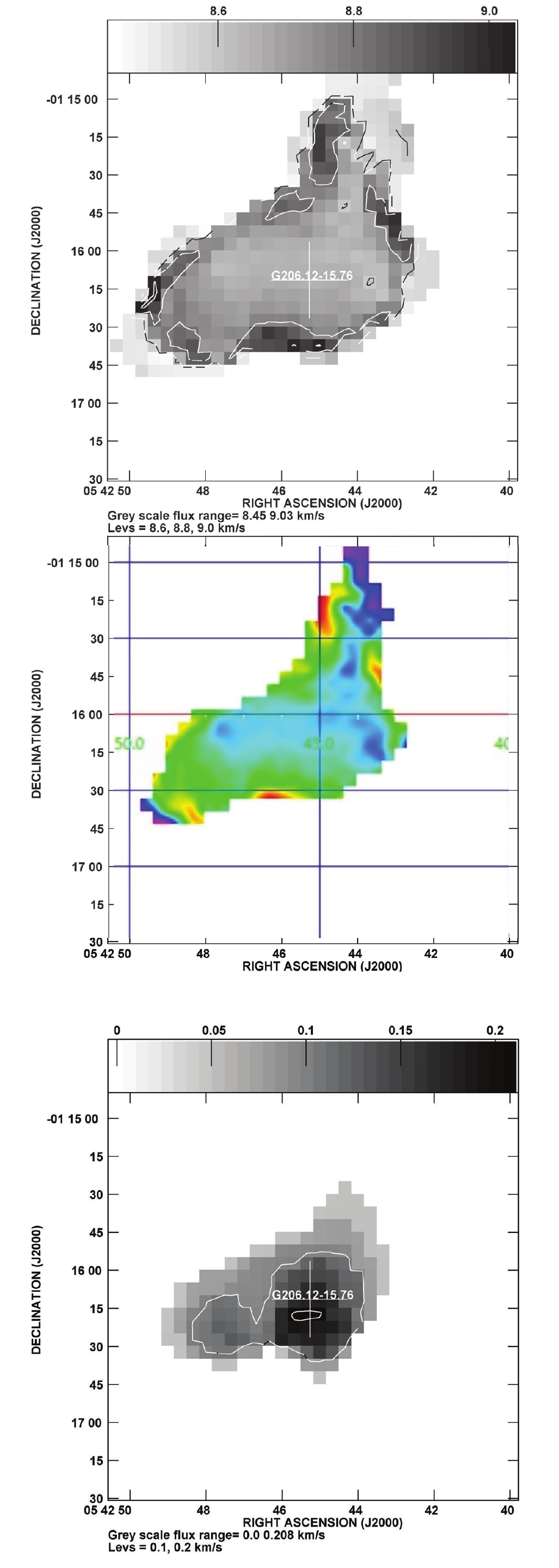}
\caption{The top and middle panels show the
intensity-weighted velocity (moment 1) maps in grey-scale
and in pseudo-color, respectively, toward field G206.12 (Figure \ref{fig:850_4cont_G206_12}).
The bottom panel shows the intensity-weighted velocity dispersion (moment 2) map
in grey-scale toward the same field. 
\label{fig:G206_12MOM}}
\end{figure*}

\begin{figure*}
\epsscale{0.9}
\figurenum{18}
\includegraphics[bb=0 0 600 1200, width=10cm]{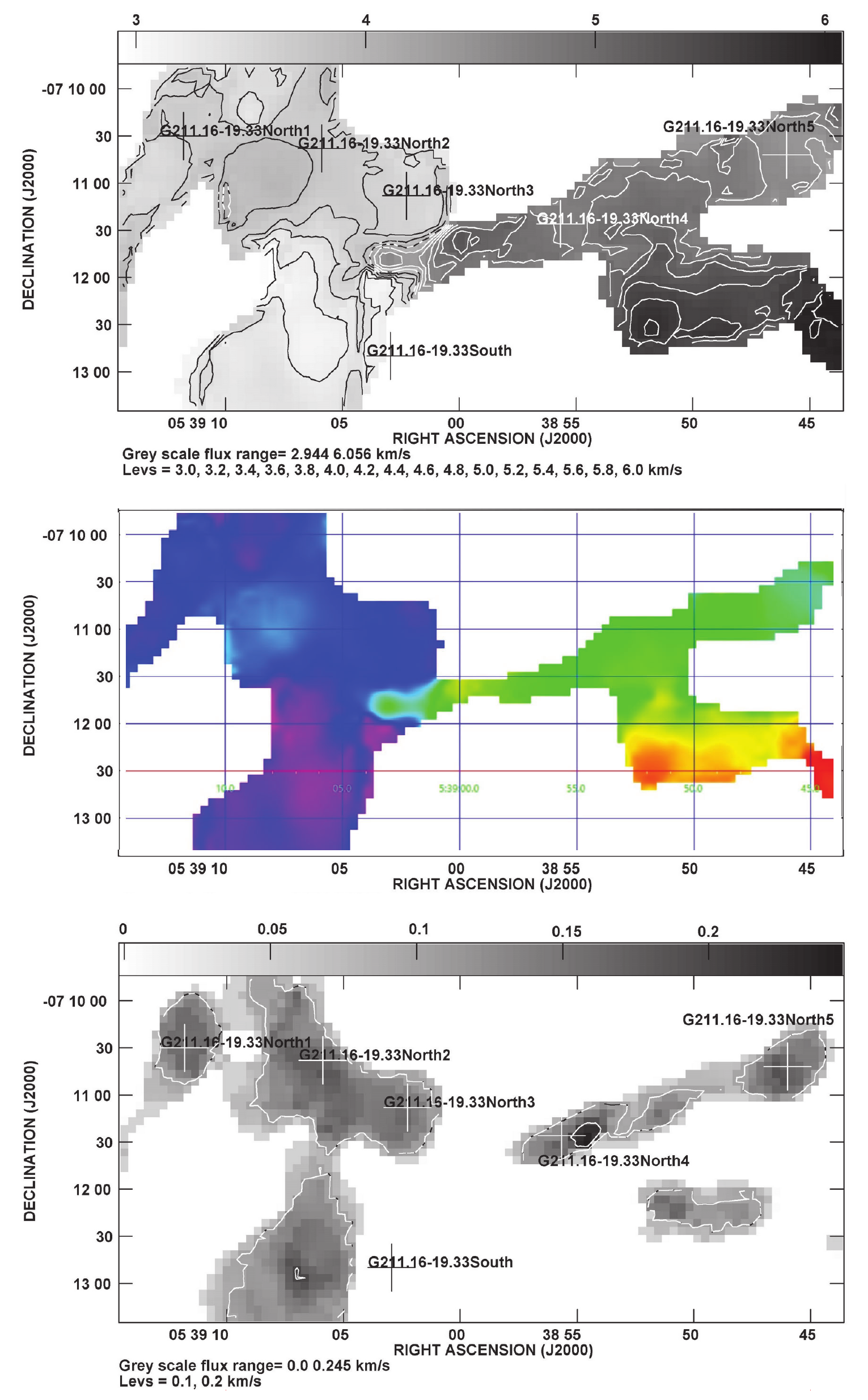}
\caption{The same as Figure \ref{fig:G206_12MOM} but for field G211.16 (Figure \ref{fig:850_4cont_G211_16}).
\label{fig:G211_16MOM}}
\end{figure*}

Field G206.12 (Figure \ref{fig:G206_12MOM}) does not exhibit any significant velocity gradients.
The velocity dispersion seems to increase toward the core center harboring the YSO.
This shift may represent the influence from the YSO.
In field G211.16 (Figure \ref{fig:G211_16MOM}), however, the velocity field shows distinct velocities for filaments.

\section{SUMMARY \label{sec:sum}}
We performed OTF mapping observations of 44 fields containing 107 SCUBA-2 cores 
in the {\nth}, {\hcn}, 94 GHz CCS, and 82 GHz CCS lines using the Nobeyama 45-m telescope. 
The distribution of {\nth} and {\hcn} line emissions is similar to the 
respective 850-{\micron} distribution, 
whereas CCS lines are undetected or distributed in clumpy structures surrounding the peak position of the 850-{\micron} dust continuum emission. 
Occasionally (12\%), we detected the CCS emission, which is considered an early-type gas tracer, toward YSOs.  
The nonthermal velocity dispersions of the SCUBA-2 cores in Orion are very similar between the starless and star-forming core populations, 
suggesting that evolution toward star formation does not immediately affect nonthermal velocity dispersion.
\vspace{5mm}



\acknowledgments
K.T. and T.S. were supported by JSPS KAKENHI Grant Number 20H05645. 
J.H. thanks the National Natural Science Foundation of China under grant Nos. 11873086 and U1631237 and support by the Yunnan Province of China (No.2017HC018), 
and his work is sponsored in part by the Chinese Academy of Sciences (CAS), 
through a grant to the CAS South America Center for Astronomy (CASSACA) in Santiago, Chile.

\vspace{5mm}
\facilities{No:45m}

\software{AIPS \citep{1996ASPC..101...37V}, APLpy \citep{2012ascl.soft08017R}, Astropy \citep{2013A&A...558A..33A, 2018AJ....156..123A}, 
NOSTAR \citep{2008PASJ...60..445S}, SAOImageDS9 \citep{2003ASPC..295..489J}, spectral-cube \citep{2016ascl.soft09017R}}

\appendix

\section{Maps}

Tables \ref{tbl:field} and \ref{tbl:yso} summarize the field name, map-center coordinates, map size, 
distance,
region, SCUBA-2 core name, core coordinates, 
name of associated YSO, 
and YSO coordinates as well as additional information.
We list cores in the order employed in \cite{2020ApJS..249...33K} for consistency.
Namely, we first list the Orion sources in the order of increasing Galactic longitude, 
and then the other sources in the order of increasing Right Ascension. 
Figure Set 19  (37 images), which is  available in the online journal, shows the integrated-intensity contour maps of 
the 37 fields in the {\nth}, {\hcn}, {\cch}, and {\ccl} lines
in the order of Tables \ref{tbl:maplist} and \ref{tbl:field}. 
Note that Figures 1 to 7 are located in the main text.
The velocity integration ranges, peak intensity, and rms noise level for the integrated intensity contour map are summarized in Table \ref{tbl:map}. 
The velocity range for {\nth} is selected for the hyperfine group
$J = 1\rightarrow0$ 
$F_1$ = 2$\rightarrow$1 containing the brightest hyperfine components.

As explained in \S\ref{sec:line}, we
fitted two-dimensional Gaussians to the spatial emission features of the Orion cores in the four lines.
The coordinates of the emission feature centers and the deconvolved radii of the emission are listed in Tables \ref{tbl:core4line1} and \ref{tbl:core4line2}.

\startlongtable
\begin{longrotatetable}
\begin{deluxetable*}{lCCCCclCC} 
\tablecaption{Mapped fields and SCUBA-2 cores \label{tbl:field}}
\tablewidth{0pt}
\tabletypesize{\scriptsize}
\tablenum{A1}
\tablehead{
\colhead{Field} &  \multicolumn{2}{c}{Map Center} &   
\colhead{Map Size} & \colhead{Distance} & \colhead{Region} &
\multicolumn{3}{c}{SCUBA-2 core} \\
\cline{2-3}
\cline{7-9}
\colhead{} & 
\colhead{R.A.(J2000)} & \colhead{Dec.(J2000)} &  
\colhead{} & \colhead{} & \colhead{} &
\colhead{Core Name} & 
\colhead{R.A.(J2000)} & \colhead{Dec.(J2000)} \\ 
\colhead{} & 
\colhead{hh:mm:ss.ss} & \colhead{dd:mm:ss.s} & \colhead{arcmin$\times$arcmin} & \colhead{kpc}& \colhead{}& \colhead{} & \colhead{hh:mm:ss.ss} & \colhead{dd:mm:ss.s} 
}
\decimalcolnumbers
\startdata
G192 & 05:29:54.72 & +12:16:44.4 & 3.0\times3.0 & 0.38 & OL & G192.32$$-$$11.88North & 05:29:54.47 & +12:16:56.0 \\
 &  &  &  &  &  & G192.32$$-$$11.88South & 05:29:54.74 & +12:16:32.0 \\
\hline 
G203 & 05:53:43.39 & +03:22:47.0 & 6.0\times3.0 & 0.39 & OB & G203.21$$-$$11.20East1 & 05:53:51.11 & +03:23:04.9 \\
 &  &  &  &  &  & G203.21$$-$$11.20East2 & 05:53:47.90 & +03:23:08.9 \\
 &  &  &  &  &  & G203.21$$-$$11.20West1 & 05:53:42.83 & +03:22:32.9 \\
 &  &  &  &  &  & G203.21$$-$$11.20West2 & 05:53:39.62 & +03:22:24.9 \\
\hline 
G204 & 05:55:38.88 & +02:11:18.3 & 4.0\times4.0 & 0.39 & OB & G204.4$$-$$11.3A2East & 05:55:38.43 & +02:11:33.3 \\
 &  &  &  &  &  & G204.4$$-$$11.3A2West & 05:55:35.49 & +02:11:01.3 \\
\hline 
G206.12 & 05:42:45.27 & -01:16:11.4 & 3.0\times3.0 & 0.39 & OB & G206.12$$-$$15.76 & 05:42:45.27 & -01:16:11.4 \\
\hline 
G206.21 & 05:41:38.48 & -01:36:27.5 & 4.0\times4.0 & 0.39 & OB & G206.21$$-$$16.17North & 05:41:39.28 & -01:35:52.9 \\
\hline 
G206.93 & 05:41:27.53 & -02:18:18.3 & 4.0\times8.0 & 0.42 & OB & G206.93$$-$$16.61West1 & 05:41:25.57 & -02:16:04.3 \\
 &  &  &  &  &  & G206.93$$-$$16.61West3 & 05:41:25.04 & -02:18:08.1 \\
 &  &  &  &  &  & G206.93$$-$$16.61West4 & 05:41:25.84 & -02:19:28.4 \\
 &  &  &  &  &  & G206.93$$-$$16.61West5 & 05:41:28.77 & -02:20:04.3 \\
 &  &  &  &  &  & G206.93$$-$$16.61West6 & 05:41:29.57 & -02:21:16.1 \\
\hline 
G207 & 05:30:49.36 & -04:11:22.0 & 4.0\times4.0 & 0.39 & OA & G207.36$$-$$19.82North1 & 05:30:50.94 & -04:10:35.6 \\
 &  &  &  &  &  & G207.36$$-$$19.82North2 & 05:30:50.67 & -04:10:15.6 \\
 &  &  &  &  &  & G207.36$$-$$19.82North3 & 05:30:46.40 & -04:10:27.6 \\
 &  &  &  &  &  & G207.36$$-$$19.82North4 & 05:30:44.81 & -04:10:27.6 \\
 &  &  &  &  &  & G207.36$$-$$19.82South & 05:30:46.81 & -04:12:29.4 \\
\hline 
G208.68 & 05:35:20.43 & -05:00:52.7 & 3.0\times3.0 & 0.39 & OA & G208.68$$-$$19.20North1 & 05:35:23.37 & -05:01:28.7 \\
 &  &  &  &  &  & G208.68$$-$$19.20North2 & 05:35:20.45 & -05:00:53.0 \\
 &  &  &  &  &  & G208.68$$-$$19.20North3 & 05:35:18.03 & -05:00:20.6 \\
\hline 
G208.89 & 05:32:36.62 & -05:35:13.7 & 8.0\times4.0 & 0.39 & OA & G208.89$$-$$20.04East & 05:32:48.40 & -05:34:47.1 \\
\hline 
G209.05 & 05:34:03.42 & -05:32:24.3 & 3.0\times6.0 & 0.39 & OA & G209.05$$-$$19.73North & 05:34:03.96 & -05:32:42.5 \\
 &  &  &  &  &  & G209.05$$-$$19.73South & 05:34:03.12 & -05:34:11.0 \\
\hline 
G209.29North & 05:35:00.16 & -05:40:02.4 & 3.0\times3.0 & 0.39 & OA & G209.29$$-$$19.65North1 & 05:35:00.25 & -05:40:02.4 \\
\hline 
G209.29South & 05:34:53.87 & -05:45:52.2 & 3.0\times3.0 & 0.39 & OA & G209.29$$-$$19.65South1 & 05:34:55.99 & -05:46:03.2 \\
 &  &  &  &  &  & G209.29$$-$$19.65South2 & 05:34:53.81 & -05:46:12.8 \\
 &  &  &  &  &  & G209.29$$-$$19.65South3 & 05:34:49.87 & -05:46:11.6 \\
\hline 
G209.77 & 05:36:34.94 & -06:02:18.3 & 4.0\times4.0 & 0.43 & OA & G209.77$$-$$19.40East1 & 05:36:32.45 & -06:01:16.7 \\
 &  &  &  &  &  & G209.77$$-$$19.40East2 & 05:36:32.19 & -06:02:04.7 \\
 &  &  &  &  &  & G209.77$$-$$19.40East3 & 05:36:35.94 & -06:02:44.7 \\
\hline 
G209.94North & 05:36:11.55 & -06:10:44.8 & 3.0\times3.0 & 0.43 & OA & G209.94$$-$$19.52North & 05:36:11.55 & -06:10:44.8 \\
\hline 
G209.94South & 05:36:24.96 & -06:14:04.7 & 3.0\times3.0 & 0.43 & OA & G209.94$$-$$19.52South1 & 05:36:24.96 & -06:14:04.7 \\
\hline 
G210 & 05:38:00.00 & -06:57:29.0 & 3.0\times3.0 & 0.43 & OA & G210.82$$-$$19.47North1 & 05:37:56.56 & -06:56:35.1 \\
 &  &  &  &  &  & G210.82$$-$$19.47North2 & 05:37:59.84 & -06:57:09.9 \\
\hline 
G211.16 & 05:38:59.06 & -07:11:36.4 & 8.0\times4.0 & 0.43 & OA & G211.16$$-$$19.33North1 & 05:39:11.80 & -07:10:29.9 \\
 &  &  &  &  &  & G211.16$$-$$19.33North2 & 05:39:05.89 & -07:10:37.9 \\
 &  &  &  &  &  & G211.16$$-$$19.33North3 & 05:39:02.26 & -07:11:07.9 \\
 &  &  &  &  &  & G211.16$$-$$19.33North4 & 05:38:55.67 & -07:11:25.9 \\
 &  &  &  &  &  & G211.16$$-$$19.33North5 & 05:38:46.00 & -07:10:41.9 \\
 &  &  &  &  &  & G211.16$$-$$19.33South & 05:39:02.94 & -07:12:49.9 \\
\hline 
G211.47 & 05:39:56.63 & -07:30:23.6 & 4.0\times4.0 & 0.43 & OA & G211.47$$-$$19.27North & 05:39:57.27 & -07:29:38.3 \\
 &  &  &  &  &  & G211.47$$-$$19.27South & 05:39:55.92 & -07:30:28.3 \\
\hline 
G211.72 & 05:40:19.93 & -07:34:43.3 & 4.0\times8.0 & 0.43 & OA & G211.72$$-$$19.25North & 05:40:13.72 & -07:32:16.8 \\
 &  &  &  &  &  & G211.72$$-$$19.25South1 & 05:40:19.04 & -07:34:28.8 \\
\hline 
G212 & 05:41:22.52 & -07:54:07.3 & 4.0\times8.0 & 0.43 & OA & G212.10$$-$$19.15North1 & 05:41:21.56 & -07:52:27.7 \\
 &  &  &  &  &  & G212.10$$-$$19.15North2 & 05:41:23.98 & -07:53:48.5 \\
 &  &  &  &  &  & G212.10$$-$$19.15North3 & 05:41:24.82 & -07:55:08.5 \\
 &  &  &  &  &  & G212.10$$-$$19.15South & 05:41:26.39 & -07:56:51.8 \\
\hline 
G159 & 03:33:18.19 & +31:08:12.6 & 4.0\times4.0 & 0.3 & H & SCOPEG159.21$$-$$20.13 & 03:33:16.08 & +31:06:50.4 \\
 &  &  &  &  &  & SCOPEG159.18$$-$$20.09 & 03:33:17.76 & +31:09:32.4 \\
 &  &  &  &  &  & SCOPEG159.22$$-$$20.11 & 03:33:21.36 & +31:07:26.4 \\
\hline 
G171 & 04:28:39.36 & +26:51:32.4 & 3.0\times3.0 & 0.14 & H & SCOPEG171.50$$-$$14.91 & 04:28:39.36 & +26:51:32.4 \\
\hline 
G172 & 05:36:52.97 & +36:10:22.4 & 4.0\times4.0 & 1.3 & H & SCOPEG172.88$$+$$02.26 & 05:36:51.60 & +36:10:40.8 \\
 &  &  &  &  &  & SCOPEG172.88$$+$$02.27 & 05:36:53.76 & +36:10:33.6 \\
 &  &  &  &  &  & SCOPEG172.89$$+$$02.27 & 05:36:54.96 & +36:10:12.0 \\
\hline 
G173 & 05:38:00.71 & +35:58:27.9 & 3.0\times3.0 & 1.3 & H & SCOPEG173.17$$+$$02.36 & 05:38:00.48 & +35:58:58.8 \\
 &  &  &  &  &  & SCOPEG173.18$$+$$02.35 & 05:38:01.68 & +35:58:15.6 \\
 &  &  &  &  &  & SCOPEG173.19$$+$$02.35 & 05:38:01.68 & +35:57:39.6 \\
\hline 
G178 & 05:39:06.87 & +30:05:41.0 & 3.0\times3.0 & 0.96 & G & SCOPEG178.27$$-$$00.60 & 05:39:06.48 & +30:05:24.0 \\
 &  &  &  &  &  & SCOPEG178.28$$-$$00.60 & 05:39:07.44 & +30:04:44.4 \\
\hline 
G006 & 15:54:08.64 & -02:52:40.8 & 4.0\times4.0 & 0.11 & H & SCOPEG006.01$$+$$36.74 & 15:54:08.64 & -02:52:44.4 \\
\hline 
G001 & 16:34:32.75 & -15:47:06.9 & 4.0\times4.0 & 0.12 & H & SCOPEG001.37$$+$$20.95 & 16:34:35.28 & -15:46:55.2 \\
\hline 
G17 & 18:14:19.75 & -12:43:45.6 & 4.0\times7.0 & 2.56 & H & SCOPEG017.38$$+$$02.26 & 18:14:18.96 & -12:43:58.8 \\
 &  &  &  &  &  & SCOPEG017.38$$+$$02.25 & 18:14:21.12 & -12:44:38.4 \\
 &  &  &  &  &  & SCOPEG017.37$$+$$02.24 & 18:14:22.56 & -12:45:25.2 \\
 &  &  &  &  &  & SCOPEG017.36$$+$$02.23 & 18:14:24.00 & -12:45:54.0 \\
\hline 
G14 & 18:16:57.56 & -16:41:35.7 & 7.0\times7.0 & 3.07 & G & SCOPEG014.20$$-$$00.18 & 18:16:55.44 & -16:41:45.6 \\
 &  &  &  &  &  & SCOPEG014.23$$-$$00.17 & 18:16:58.80 & -16:39:50.4 \\
 &  &  &  &  &  & SCOPEG014.18$$-$$00.23 & 18:17:05.28 & -16:43:44.4 \\
\hline 
G16.96 & 18:20:41.53 & -14:05:15.6 & 6.0\times5.0 & 1.87 & G & SCOPEG016.93$$+$$00.28 & 18:20:35.76 & -14:04:15.6 \\
 &  &  &  &  &  & SCOPEG016.93$$+$$00.27 & 18:20:39.84 & -14:04:51.6 \\
 &  &  &  &  &  & SCOPEG016.93$$+$$00.25 & 18:20:43.20 & -14:05:16.8 \\
 &  &  &  &  &  & SCOPEG016.93$$+$$00.24 & 18:20:44.88 & -14:05:31.2 \\
 &  &  &  &  &  & SCOPEG016.92$$+$$00.23 & 18:20:46.56 & -14:06:14.4 \\
 &  &  &  &  &  & SCOPEG016.93$$+$$00.22 & 18:20:50.64 & -14:06:00.0 \\
\hline 
G16.36 & 18:22:38.21 & -14:58:32.3 & 11.0\times6.0 & 3.15 & G & SCOPEG016.30$$-$$00.53 & 18:22:20.16 & -15:00:14.4 \\
 &  &  &  &  &  & SCOPEG016.34$$-$$00.59 & 18:22:37.20 & -15:00:00.0 \\
 &  &  &  &  &  & SCOPEG016.38$$-$$00.61 & 18:22:47.52 & -14:58:37.2 \\
 &  &  &  &  &  & SCOPEG016.42$$-$$00.64 & 18:22:58.08 & -14:57:00.0 \\
\hline 
G24 & 18:34:18.16 & -7:48:34.4 & 6.0\times6.0 & 9.21 & G & SCOPEG024.02$$+$$00.24 & 18:34:13.44 & -07:48:32.4 \\
 &  &  &  &  &  & SCOPEG024.02$$+$$00.21 & 18:34:18.72 & -07:49:44.4 \\
\hline 
G33 & 18:52:55.79 & +00:41:36.9 & 3.0\times6.0 & 6.5 & G & SCOPEG033.74$$-$$00.01 & 18:52:57.12 & +00:43:01.2 \\
\hline 
G35 & 18:57:07.96 & +02:09:27.6 & 3.0\times6.0 & 2.21 & G & SCOPEG035.48$$-$$00.29 & 18:57:06.96 & +02:08:24.0 \\
 &  &  &  &  &  & SCOPEG035.52$$-$$00.27 & 18:57:08.40 & +02:10:48.0 \\
 &  &  &  &  &  & SCOPEG035.48$$-$$00.31 & 18:57:11.28 & +02:07:30.0 \\
\hline 
G34 & 18:59:43.05 & +00:59:50.1 & 7.0\times7.5 & 2.19 & G & SCOPEG034.75$$-$$01.38 & 18:59:41.04 & +00:59:06.0 \\
\hline 
G57 & 19:23:52.75 & +23:07:36.5 & 4.0\times4.0 & 0.8 & H & SCOPEG057.11$$+$$03.66 & 19:23:49.20 & +23:07:58.8 \\
 &  &  &  &  &  & SCOPEG057.10$$+$$03.63 & 19:23:56.88 & +23:06:28.8 \\
\hline 
G69 & 20:13:32.71 & +31:22:10.7 & 3.0\times3.0 & 2.48 & G & SCOPEG069.80$$-$$01.67 & 20:13:32.40 & +31:21:50.4 \\
 &  &  &  &  &  & SCOPEG069.81$$-$$01.67 & 20:13:33.84 & +31:22:01.2 \\
\hline 
G74 & 20:17:59.29 & +35:56:03.2 & 4.0\times4.0 & 4.34 & G & SCOPEG074.10$$+$$00.11 & 20:17:56.40 & +35:55:22.8 \\
 &  &  &  &  &  & SCOPEG074.11$$+$$00.11 & 20:17:58.56 & +35:55:51.6 \\
\hline 
G82 & 20:51:20.80 & +41:24:08.7 & 6.5\times6.5 & 1.54 & G & SCOPEG082.36$$-$$01.83 & 20:51:16.56 & +41:22:58.8 \\
 &  &  &  &  &  & SCOPEG082.40$$-$$01.84 & 20:51:24.96 & +41:24:46.8 \\
 &  &  &  &  &  & SCOPEG082.41$$-$$01.84 & 20:51:27.36 & +41:25:22.8 \\
 &  &  &  &  &  & SCOPEG082.42$$-$$01.84 & 20:51:28.80 & +41:25:48.0 \\
\hline 
G91 & 21:00:39.09 & +52:27:59.8 & 4.0\times4.0 & 0.8 & H & SCOPEG091.85$$+$$04.12 & 21:00:38.40 & +52:27:57.6 \\
\hline 
G92 & 21:04:03.97 & +52:34:06.6 & 4.0\times4.0 & 0.8 & H & SCOPEG092.27$$+$$03.79 & 21:04:04.56 & +52:33:43.2 \\
\hline 
G105 & 21:43:08.48 & +66:05:34.0 & 4.0\times8.0 & 1.15 & H & SCOPEG105.37$$+$$09.84 & 21:43:00.72 & +66:03:21.6 \\
 &  &  &  &  &  & SCOPEG105.41$$+$$09.88 & 21:43:05.28 & +66:06:54.0 \\
\hline 
G93 & 21:44:55.74 & +47:39:59.3 & 3.0\times3.0 & 0.49 & H & SCOPEG093.53$$-$$04.26 & 21:44:52.08 & +47:40:30.0 \\
 &  &  &  &  &  & SCOPEG093.54$$-$$04.28 & 21:44:57.60 & +47:39:57.6 \\
\hline 
G107 & 22:21:26.27 & +63:37:20.5 & 4.0\times4.0 & 0.76 & H & SCOPEG107.16$$+$$05.45 & 22:21:18.00 & +63:37:33.6 \\
 &  &  &  &  &  & SCOPEG107.18$$+$$05.43 & 22:21:33.60 & +63:37:19.2 \\
\enddata
\tablecomments{Column 1: Field name of mapped region, Column 2-3: Map center in the equatorial coordinate system J2000, Column 4: Map size in arcsecond $\times$ arcsecond,  Column 5: Distance in kpc, Column 6: Region: `OL' indicates a core in $\lambda$ Orionis, `OA' indicates a core in Orion A, `OB' indicates a core in Orion B, `G' indicates a core at Galactic plane ($|b|<2\arcdeg$), `H' indicates cores at high latitudes ($|b|\geq2\arcdeg$), Column 7: SCUBA-2 core name, Column 8-9: Coordinates of SCUBA-2 core in J2000.}
\end{deluxetable*}
\end{longrotatetable}

\startlongtable
\begin{deluxetable*}{lccCCCCC} 
\tablecaption{Basic information of YSOs associated with SCUBA-2 cores \label{tbl:yso}}
\tablewidth{0pt}
\tabletypesize{\scriptsize}
\tablenum{A2}
\tablehead{
\colhead{SCUBA-2 core} & \colhead{Class} & \colhead{YSO name} & 
\colhead{R.A.(J2000)} & \colhead{Dec.(J2000)} & \colhead{$T_{\rm bol}$} & \colhead{$L_{\rm bol}$} & \colhead{$\theta_{\rm core-YSO}$}  
\\
\colhead{} & \colhead{} & \colhead{} &
\colhead{hh:mm:ss.ss} & \colhead{dd:mm:ss.s} & \colhead{K} & \colhead{$L_{\sun}$} &\colhead{arcsec}
}
\decimalcolnumbers
\startdata
G192.32$-$11.88North & P0 & [LZK2016] G192N & 05:29:54.16 & +12:16:53.1 & . . .  &  . . .  & 5.4 \\
G192.32$-$11.88South & P & [LZK2016] G192S & 05:29:54.35 & +12:16:29.7 & 60$\pm$13 &  0.1$\pm$0.1  & 6.2 \\
G203.21$-$11.20East1 & S & . . . & . . . & . . . & . . . & . . . & . . . \\
G203.21$-$11.20East2 & S & . . . & . . . & . . . & . . . & . . . & . . . \\
G203.21$-$11.20West1 & P0 & Herschel J055342.5+032236 & 05:53:42.51 & +03:22:35.7 & . . . &  . . .  & 5.6 \\
G203.21$-$11.20West2 & P0 & Herschel J055339.5+032225 & 05:53:39.55 & +03:22:25.4 & 15$\pm$5 &  0.5$\pm$0.3  & 1.2 \\
G204.4$-$11.3A2East & P & Herschel J055538.2+021135 & 05:55:38.23 & +02:11:35.2 & . . . &  . . .  & 3.5 \\
G204.4$-$11.3A2West & P & Herschel J055535.3+021103 & 05:55:35.36 & +02:11:03.2 & . . . &  . . .  & 2.7 \\
G206.12$-$15.76 & P0 & HOPS 400 & 05:42:45.23 & -01:16:14.2 & 35$\pm$9 &  3.0$\pm$1.4  & 2.9 \\
G206.21$-$16.17North & S & . . . & . . . & . . . & . . . & . . . & . . . \\
G206.93$-$16.61West1 & PI & HOPS 300 & 05:41:24.21 & -02:16:06.5 &  . . . &  . . .  & 20.5 \\
G206.93$-$16.61West3 & P0 & HOPS 399 & 05:41:24.94 & -02:18:08.5 & 31$\pm$10 &  6.3$\pm$3.0  & 1.6 \\
G206.93$-$16.61West4 & S & . . . & . . . & . . . & . . . & . . . & . . . \\
G206.93$-$16.61West5 & S & . . . & . . . & . . . & . . . & . . . & . . . \\
G206.93$-$16.61West6 & P0 & HOPS 398 & 05:41:29.40 & -02:21:17.1 & . . . &  . . .  & 2.7 \\
G207.36$-$19.82North1 & PI & VISION J05305129-0410322 & 05:30:51.29 & -04:10:32.2 & . . . &  . . .  & 6.2 \\
G207.36$-$19.82North2 & PI & VISION J05305129-0410322 & 05:30:51.29 & -04:10:32.2 & . . . &  . . .  & 19.0 \\
G207.36$-$19.82North3 & S & . . . & . . . & . . . & . . . & . . . & . . . \\
G207.36$-$19.82North4 & S & . . . & . . . & . . . & . . . & . . . & . . . \\
G207.36$-$19.82South & S & . . . & . . . & . . . & . . . & . . . & . . . \\
G208.68$-$19.20North1 & P0 & HOPS 87 & 05:35:23.47 & -05:01:28.7 & 38$\pm$13 &  36.7$\pm$14.5  & 1.5 \\
G208.68$-$19.20North2 & PF & HOPS 89 & 05:35:19.96 & -05:01:02.6 & 112$\pm$10 &  2.1$\pm$1.3  & 12.1 \\
G208.68$-$19.20North3 & PF & HOPS 92 & 05:35:18.32 & -05:00:33.0 & 158$\pm$20 & 22.0$\pm$8.7 & 13.1 \\
G208.89$-$20.04East & PI & WISE J053248.59-053451.2 & 05:32:48.60 & -05:34:51.3 & 108$\pm$25 &  2.2$\pm$1.0  & 5.1 \\
G209.05$-$19.73North & S & . . . & . . . & . . . & . . . & . . . & . . . \\
G209.05$-$19.73South & S & . . . & . . . & . . . & . . . & . . . & . . . \\
G209.29$-$19.65North1 & S & . . . & . . . & . . . & . . . & . . . & . . . \\
G209.29$-$19.65South1 & S & . . . & . . . & . . . & . . . & . . . & . . . \\
G209.29$-$19.65South2 & S & . . . & . . . & . . . & . . . & . . . \\G209.29$-$19.65South2 & S & . . . & . . . & . . . & . . . & . . . & . . . \\
G209.29$-$19.65South3 & PII & [MGM2012] 1221 & 05:34:49.08 & -05:46:04.8 & . . . &  . . .  & 13.6 \\
G209.77$-$19.40East1 & PF & HOPS 192 & 05:36:32.45 & -06:01:16.2 & . . . &  . . .  & 0.5 \\
G209.77$-$19.40East2 & S & . . . & . . . & . . . & . . . & . . . & . . . \\
G209.77$-$19.40East3 & S & . . . & . . . & . . . & . . . & . . . & . . . \\
G209.94$-$19.52North & PII & [MGM2012] 1025 & 05:36:09.66 & -06:10:30.6 & . . . &  . . .  & 31.6 \\
G209.94$-$19.52South1 & S & . . . & . . . & . . . & . . . & . . . & . . . \\
G210.82$-$19.47North1 & PI & HOPS 157 & 05:37:56.57 & -06:56:39.2 & . . . &  . . .  & 4.1 \\
G210.82$-$19.47North2 & S & . . . & . . . & . . . & . . . & . . . & . . . \\
G211.16$-$19.33North1 & PF & HOPS 129 & 05:39:11.85 & -07:10:35.0 & . . . &  . . .  & 5.2 \\
G211.16$-$19.33North2 & PI & HOPS 133 & 05:39:05.83 & -07:10:39.4 & 70$\pm$20 &  3.7$\pm$1.4 & 1.7 \\
G211.16$-$19.33North3 & S & . . . & . . . & . . . & . . . & . . . & . . . \\
G211.16$-$19.33North4 & P & Herschel J053854.1-071123 & 05:38:54.13 & -07:11:22.8 & . . . &  . . .  & 23.1 \\
G211.16$-$19.33North5 & PI & HOPS 135 & 05:38:45.31 & -07:10:55.9 & 112$\pm$16 &  1.3$\pm$0.5  & 17.3 \\
G211.16$-$19.33South & PI & HOPS 130 & 05:39:02.96 & -07:12:52.3 & . . . &  . . .  & 2.4 \\
G211.47$-$19.27North & P0 & HOPS 290 & 05:39:57.41 & -07:29:33.4 & 48$\pm$10 & 4.0$\pm$1.7  & 5.3 \\
G211.47$-$19.27South & P0 & HOPS 288 & 05:39:55.94 & -07:30:28.0 & 49$\pm$21 &  180.0$\pm$70.0  & 0.5 \\
G211.72$-$19.25North & PII & WISE J054013.78-073216.0 & 05:40:13.79 & -07:32:16.1 & . . . &  . . .  & 1.2 \\
G211.72$-$19.25South1 & S & . . . & . . . & . . . & . . . & . . . & . . . \\
G212.10$-$19.15North1 & P & Herschel J054120.5-075237 & 05:41:20.55 & -07:52:36.6 & . . . &  . . .  & 17.5 \\
G212.10$-$19.15North2 & PI & HOPS 263 & 05:41:23.68 & -07:53:46.8 & 114$\pm$10 &  1.1$\pm$0.5  & 4.7 \\
G212.10$-$19.15North3 & PI & HOPS 254 & 05:41:24.52 & -07:55:07.3 & . . . &  . . .  & 4.7 \\
G212.10$-$19.15South & P0 & HOPS 247 & 05:41:26.22 & -07:56:51.6 & 43$\pm$12 & 3.2$\pm1.2  & 2.5 \\
SCOPEG159.21$-$20.13 & P0 & [EES2009] Per-emb 10 & 03:33:16.45 & +31:06:52.5 & . . . &  . . .  & 5.2 \\
SCOPEG159.18$-$20.09 & P0 & SSTc2d J033317.8+310931 & 03:33:17.85 & +31:09:31.9 & . . . &  . . .  & 1.3 \\
SCOPEG159.22$-$20.11 & P0 & [SDA2014] West41 & 03:33:21.30 & +31:07:27.0 & . . . &  . . .  & 1.0 \\
SCOPEG171.50$-$14.91 & PI & NAME LDN 1521F IRS & 04:28:38.95 & +26:51:35.1 & . . . &  . . .  & 6.1 \\
SCOPEG172.88$+$02.26 & P & WISE J053651.31+361037.1 & 05:36:51.31 & +36:10:37.2 & . . . &  . . .  & 5.1 \\
SCOPEG172.88$+$02.27 & P & WISE J053653.75+361033.5 & 05:36:53.76 & +36:10:33.6 & . . . &  . . .  & 0.0 \\
SCOPEG172.89$+$02.27 & P & WISE J053654.91+361008.2 & 05:36:54.91 & +36:10:08.3 & . . . &  . . .  & 3.7 \\
SCOPEG173.17$+$02.36 & PI & WISE J053800.11+355903.7 & 05:38:00.12 & +35:59:03.7 & . . . &  . . .  & 6.6 \\
SCOPEG173.18$+$02.35 & P & Herschel J053801.5+355817 & 05:38:01.54 & +35:58:17.2 & . . . &  . . .  & 2.3 \\
SCOPEG173.19$+$02.35 & P & Herschel J053801.3+355734 & 05:38:01.32 & +35:57:33.5 & . . . &  . . .  & 7.5 \\
SCOPEG178.27$-$00.60 & P & Herschel J053906.4+300524 & 05:39:06.47 & +30:05:24.4 & . . . &  . . .  & 0.4 \\
SCOPEG178.28$-$00.60 & P & Herschel J053907.3+300452 & 05:39:07.31 & +30:04:51.9 & . . . &  . . .  & 7.7 \\
SCOPEG006.01$+$36.74 & S & . . . & . . . & . . . & . . . & . . . & . . . \\
SCOPEG001.37$+$20.95 & S & . . . & . . . & . . . & . . . & . . . & . . . \\
SCOPEG017.38$+$02.26 & P & WISE J181418.81-124358.7 & 18:14:18.82 & -12:43:58.7 & . . . &  . . .  & 2.1 \\
SCOPEG017.38$+$02.25 & PI & MSX6C G017.3765+02.2512 & 18:14:21.10 & -12:44:33.0 & . . . &  . . .  & 5.4 \\
SCOPEG017.37$+$02.24 & P & WISE J181422.18-124521.2 & 18:14:22.18 & -12:45:21.2 & . . . &  . . .  & 6.8 \\
SCOPEG017.36$+$02.23 & P & WISE J181424.00-124548.6 & 18:14:24.00 & -12:45:48.6 & . . . &  . . .  & 5.4 \\
SCOPEG014.20$-$00.18 & P & WISE J181656.04-164141.5 & 18:16:56.04 & -16:41:41.6 & . . . &  . . .  & 9.6 \\
SCOPEG014.23$-$00.17 & S & . . . & . . . & . . . & . . . & . . . & . . . \\
SCOPEG014.18$-$00.23 & P & Herschel G014.1897-0.2273 & 18:17:05.61 & -16:43:28.3 & . . . &  . . .  & 16.8 \\
SCOPEG016.93$+$00.28 & P & MSX6C G016.9261+00.2854 & 18:20:35.50 & -14:04:16.0 & . . . &  . . .  & 3.8 \\
SCOPEG016.93$+$00.27 & P & Herschel G016.9253+0.2662 & 18:20:39.57 & -14:04:51.2 & . . . &  . . .  & 4.0 \\
SCOPEG016.93$+$00.25 & S & . . . & . . . & . . . & . . . & . . . & . . . \\
SCOPEG016.93$+$00.25 & S & . . . & . . . & . . . & . . . & . . . & . . . \\
SCOPEG016.93$+$00.24 & S & . . . & . . . & . . . & . . . & . . . & . . . \\
SCOPEG016.93$+$00.24 & S & . . . & . . . & . . . & . . . & . . . & . . . \\
SCOPEG016.92$+$00.23 & S & . . . & . . . & . . . & . . . & . . . & . . . \\
SCOPEG016.93$+$00.22 & P & Herschel G016.9288+0.2180 & 18:20:50.48 & -14:06:01.8 & . . . &  . . .  & 2.9 \\
SCOPEG016.30$-$00.53 & S & . . . & . . . & . . . & . . . & . . . & . . . \\
SCOPEG016.34$-$00.59 & PI & WISE J182236.21-145956.5 & 18:22:36.21 & -14:59:56.6 & . . . &  . . .  & 14.7 \\
SCOPEG016.38$-$00.61 & P & Herschel J182247.7-145848 & 18:22:47.71 & -14:58:47.5 & . . . &  . . .  & 10.7 \\
SCOPEG016.42$-$00.64 & S & . . . & . . . & . . . & . . . & . . . & . . . \\
SCOPEG016.42$-$00.64 & S & . . . & . . . & . . . & . . . & . . . & . . . \\
SCOPEG024.02$+$00.24 & S & . . . & . . . & . . . & . . . & . . . & . . . \\
SCOPEG024.02$+$00.21 & S & . . . & . . . & . . . & . . . & . . . & . . . \\
SCOPEG033.74$-$00.01 & P & Herschel J185256.6+004316 & 18:52:56.68 & +00:43:15.5 & . . . &  . . .  & 15.8 \\
SCOPEG035.48$-$00.29 & P & SSTGLMC G035.4858-00.2876 & 18:57:07.40 & +02:08:29.2 & . . . &  . . .  & 8.4 \\
SCOPEG035.52$-$00.27 & P & Herschel G035.5235-0.2728 & 18:57:08.34 & +02:10:54.5 & . . . &  . . .  & 6.5 \\
SCOPEG035.48$-$00.31 & P & MIREX G035.4823-00.3086 & 18:57:11.50 & +02:07:43.6 & . . . &  . . .  & 14.0 \\
SCOPEG034.75$-$01.38 & P & Herschel J185941.2+005908 & 18:59:41.23 & +00:59:08.1 & . . . &  . . .  & 3.5 \\
SCOPEG057.11$+$03.66 & S & . . . & . . . & . . . & . . . & . . . & . . . \\
SCOPEG057.10$+$03.63 & P & WISE J192356.78+230633.0 & 19:23:56.78 & +23:06:33.1 & . . . &  . . .  & 4.5 \\
SCOPEG069.80$-$01.67 & PI & [MJR2015] 3117 & 20:13:32.55 & +31:22:00.4 & . . . & . . .  10.2 \\
SCOPEG069.81$-$01.67 & P & WISE J201333.64+312206.3 & 20:13:33.64 & +31:22:06.3 & . . . &  . . .  & 5.7 \\
SCOPEG074.10$+$00.11 & P & Herschel J201756.3+355525 & 20:17:56.30 & +35:55:25.3 & . . . &  . . .  & 2.8 \\
SCOPEG074.11$+$00.11 & P & WISE J201758.56+355552.3 & 20:17:58.57 & +35:55:52.4 & . . . &  . . .  & 0.8 \\
SCOPEG082.36$-$01.83 & S & . . . & . . . & . . . & . . . & . . . & . . . \\
SCOPEG082.40$-$01.84 & P & [MJR2015] 3335 & 20:51:24.82 & +41:24:49.5 & . . . &  . . .  & 3.2 \\
SCOPEG082.41$-$01.84 & S & . . . & . . . & . . . & . . . & . . . & . . . \\
SCOPEG082.42$-$01.84 & S & . . . & . . . & . . . & . . . & . . . & . . . \\
SCOPEG091.85$+$04.12 & P & WISE J210038.77+522757.5 & 21:00:38.76 & +52:27:57.5 & . . . &  . . .  & 3.3 \\
SCOPEG092.27$+$03.79 & PI & [MJR2015] 3554 & 21:04:04.50 & +52:33:47.2 & . . . & 4.0 \\
SCOPEG105.37$+$09.84 & PI & [SS2009] NGC 7129-S3-U419 & 21:43:01.78 & +66:03:24.4 & . . . &  . . .  & 7.0 \\
SCOPEG105.41$+$09.88 & PIII & 2MASS J21430502+6606533 & 21:43:04.98 & +66:06:53.3 & . . . &  . . .  & 2.0 \\
SCOPEG093.53$-$04.26 & P & HHL 73 IRS 1 & 21:44:52.01 & +47:40:30.8 & . . . &  . . .  & 1.1 \\
SCOPEG093.54$-$04.28 & S & . . . & . . . & . . . & . . . & . . . & . . . \\
SCOPEG107.16$+$05.45 & PII & [MJR2015] 3797 & 22:21:20.15 & +63:37:37.8 & . . . &  . . .  & 15.0 \\
SCOPEG107.18$+$05.43 & P & [HLB98] Onsala 164 & 22:21:33.20 & +63:37:22.0 & . . . &  . . .  & 3.9 \\
\enddata
\tablecomments{Column 1: SCUBA-2 core name, Column 2: Core classification: `S' indicates a starless core, `P' indicates a protostellar core, `P0' indicates a core including class 0 protostar, `PI' indicates a core including a class I protostar, `PF' indicates a core including a flat-spectrum protostar,  `PII' indicates a core including a class II protostar, Column 3: name of young stellar object (YSO), Column 4-5: Coordinate of YSO in Equatorial system (J2000), Column 6: Bolometric temperature, Column 7: Separation between SCUBA-2 core and YSO in arcseconds.}
\end{deluxetable*}

\startlongtable
\begin{longrotatetable}
\begin{deluxetable*}{lCCCCCCCCCCCC} 
\tablecaption{Parameters for integrated intensity maps for {\nth}, {\hcn}, {\cch}, and {\ccl} lines \label{tbl:map}}
\tablewidth{0pt}
\tabletypesize{\scriptsize}
\tablenum{A3}
\tablehead{
\colhead{Field} &  \multicolumn{2}{c}{\nth} &   \multicolumn{2}{c}{Other} &   \multicolumn{2}{c}{\nth} &   \multicolumn{2}{c}{\hcn} &   \multicolumn{2}{c}{\cch} &   \multicolumn{2}{c}{\ccl} \\
\cline{2-5}
\cline{6-13}
\colhead{} &
\colhead{$v_{\rm l}$} & \colhead{$v_{\rm h}$} &
\colhead{$v_{\rm l}$} & \colhead{$v_{\rm h}$} &
\colhead{$\sigma$} &
\colhead{$I_{\rm peak}$} & \colhead{$\sigma$} &
\colhead{$I_{\rm peak}$} & \colhead{$\sigma$} &
\colhead{$I_{\rm peak}$} & \colhead{$\sigma$} &
\colhead{$I_{\rm peak}$} \\
\colhead{} &
\colhead{km s$^{-1}$} &\colhead{km s$^{-1}$} &\colhead{km s$^{-1}$} &\colhead{km s$^{-1}$} &
\colhead{K km s$^{-1}$} &\colhead{K km s$^{-1}$} &\colhead{K km s$^{-1}$} &\colhead{K km s$^{-1}$} &
\colhead{K km s$^{-1}$} &\colhead{K km s$^{-1}$} &\colhead{K km s$^{-1}$} &\colhead{K km s$^{-1}$} 
}
\decimalcolnumbers
\startdata
G192 & 10.7 & 14.0 & 11.4 & 13.1 & 0.05 & 2.26 & 0.04 & 0.46 & 0.05 & 0.15 & 0.04 & 0.13 \\
G203 & 8.9 & 12.1 & 9.3 & 11.1 & 0.07 & 1.59 & 0.07 & 0.37 & 0.06 & 0.24 & 0.06 & 0.31 \\
G204 & 0.3 & 3.2 & 0.8 & 2.3 & 0.06 & 3.15 & 0.05 & 1.00 & 0.05 & 0.25 & 0.05 & 0.28 \\
G206.12 & 7.4 & 10.0 & 7.8 & 9.0 & 0.11 & 1.73 & 0.05 & 0.43 & 0.05 & 0.26 & 0.05 & 0.26 \\
G206.21 & 8.4 & 11.4 & 9.3 & 10.7 & 0.07 & 2.43 & 0.05 & 0.18 & 0.06 & 0.19 & 0.05 & 0.15 \\
G206.93 & 7.8 & 12.3 & 8.4 & 11.3 & 0.07 & 3.67 & 0.07 & 1.11 & 0.08 & 0.32 & 0.07 & 0.28 \\
G207 & 9.3 & 12.8 & 10.3 & 11.5 & 0.10 & 2.40 & 0.06 & 0.18 & 0.07 & 0.21 & 0.06 & 0.19 \\
G208.68 & 9.1 & 13.5 & 9.7 & 12.5 & 0.06 & 9.50 & 0.06 & 1.62 & 0.05 & 0.27 & 0.06 & 0.30 \\
G208.89 & 7.2 & 10.7 & 7.7 & 9.6 & 0.07 & 3.24 & 0.05 & 1.34 & 0.06 & 0.36 & 0.05 & 0.44 \\
G209.05 & 6.8 & 10.1 & 7.4 & 9.0 & 0.06 & 1.15 & 0.04 & 0.36 & 0.05 & 0.17 & 0.04 & 0.20 \\
G209.29North & 6.3 & 10.6 & 7.2 & 9.6 & 0.08 & 2.88 & 0.06 & 0.62 & 0.07 & 0.26 & 0.06 & 0.23 \\
G209.29 & 5.7 & 10.8 & 7.4 & 9.5 & 0.08 & 3.02 & 0.06 & 0.22 & 0.06 & 0.22 & 0.06 & 0.20 \\
G209.77 & 6.8 & 9.9 & 7.4 & 8.8 & 0.06 & 3.27 & 0.04 & 0.34 & 0.05 & 0.19 & 0.04 & 0.19 \\
G209.94North & 6.7 & 10.1 & 7.5 & 9.1 & 0.07 & 2.35 & 0.06 & 0.45 & 0.06 & 0.15 & 0.06 & 0.22 \\
G209.94South & 6.5 & 10.1 & 7.5 & 8.9 & 0.07 & 1.67 & 0.05 & 0.39 & 0.05 & 0.18 & 0.05 & 0.20 \\
G210 & 3.9 & 7.0 & 4.6 & 5.9 & 0.05 & 2.01 & 0.03 & 0.31 & 0.03 & 0.13 & 0.04 & 0.14 \\
G211.16 & 1.8 & 7.5 & 2.3 & 6.6 & 0.06 & 2.34 & 0.07 & 0.76 & 0.07 & 0.30 & 0.07 & 0.32 \\
G211.47 & 2.7 & 7.4 & 3.3 & 6.2 & 0.07 & 4.17 & 0.06 & 1.14 & 0.07 & 0.32 & 0.06 & 0.30 \\
G211.72 & 2.4 & 6.0 & 3.0 & 5.2 & 0.08 & 1.45 & 0.07 & 1.02 & 0.07 & 0.31 & 0.06 & 0.44 \\
G212 & 2.5 & 6.8 & 3.2 & 5.9 & 0.08 & 2.83 & 0.07 & 1.71 & 0.08 & 0.44 & 0.07 & 0.63 \\
G159 & 4.9 & 8.7 & 5.4 & 7.8 & 0.07 & 4.05 & 0.07 & 1.08 & 0.07 & 0.62 & 0.07 & 0.70 \\
G171 & 5.3 & 8.0 & 5.8 & 7.0 & 0.05 & 1.72 & 0.04 & 1.02 & 0.04 & 0.29 & 0.04 & 0.51 \\
G172 & -19.2 & -15.4 & -18.7 & -16.1 & 0.09 & 3.00 & 0.08 & 1.02 & 0.09 & 0.48 & 0.08 & 0.56 \\
G173 & -22.2 & -16.3 & -20.4 & -17.8 & 0.08 & 5.87 & 0.06 & 1.31 & 0.07 & 0.55 & 0.06 & 0.60 \\
G178 & -1.8 & 1.0 & -1.1 & -0.3 & 0.06 & 1.20 & 0.03 & 0.16 & 0.04 & 0.17 & 0.03 & 0.12 \\
G006 & 1.2 & 4.0 & 1.8 & 2.8 & 0.07 & 1.43 & 0.05 & 0.28 & 0.05 & 0.23 & 0.05 & 0.22 \\
G001 & -0.8 & 2.4 & -0.3 & 1.3 & 0.07 & 3.34 & 0.06 & 1.46 & 0.07 & 0.34 & 0.06 & 0.27 \\
G17 & 8.3 & 12.9 & 9.0 & 11.9 & 0.07 & 5.82 & 0.07 & 1.23 & 0.08 & 0.30 & 0.07 & 0.24 \\
G14 & 34.5 & 41.9 & 34.4 & 43.0 & 0.09 & 17.38 & 0.24 & 6.49 & 0.14 & 0.49 & 0.13 & 0.46 \\
G16.96 & 22.3 & 27.3 & 23.1 & 27.4 & 0.08 & 2.89 & 0.09 & 0.45 & 0.11 & 0.35 & 0.09 & 0.30 \\
G16.36 & 35.0 & 44.5 & 37.0 & 44.0 & 0.14 & 6.14 & 0.19 & 1.30 & 0.20 & 0.75 & 0.16 & 0.57 \\
G24 & 102.4 & 109.0 & 103.7 & 107.2 & 0.16 & 7.61 & 0.12 & 0.97 & 0.13 & 0.52 & 0.11 & 0.55 \\
G33 & 100.0 & 109.0 & 102.9 & 108.9 & 0.09 & 9.36 & 0.10 & 3.40 & 0.11 & 0.45 & 0.09 & 0.36 \\
G35 & 41.0 & 49.1 & 43.6 & 47.3 & 0.09 & 4.94 & 0.08 & 0.79 & 0.08 & 0.35 & 0.08 & 0.41 \\
G34 & 43.0 & 49.4 & 43.0 & 48.8 & 0.10 & 11.16 & 0.13 & 5.71 & 0.14 & 0.64 & 0.13 & 0.86 \\
G57 & 10.7 & 14.1 & 10.3 & 12.8 & 0.07 & 1.92 & 0.06 & 0.76 & 0.07 & 0.26 & 0.06 & 0.29 \\
G69 & 10.3 & 15.1 & 10.5 & 14.2 & 0.06 & 3.13 & 0.06 & 1.15 & 0.07 & 0.33 & 0.06 & 0.36 \\
G74 & -4.3 & 1.6 & -3.6 & 0.4 & 0.08 & 3.00 & 0.08 & 0.90 & 0.09 & 0.43 & 0.07 & 0.53 \\
G82 & 2.5 & 7.1 & 3.3 & 6.0 & 0.06 & 1.56 & 0.06 & 0.43 & 0.09 & 0.36 & 0.06 & 0.34 \\
G91 & -4.6 & -0.8 & -4.1 & -2.1 & 0.07 & 1.65 & 0.05 & 1.03 & 0.06 & 0.51 & 0.05 & 0.60 \\
G92 & -4.5 & 0.3 & -3.9 & -0.6 & 0.07 & 2.52 & 0.06 & 1.23 & 0.08 & 0.35 & 0.07 & 0.42 \\
G105 & -11.6 & -6.9 & -11.2 & -7.7 & 0.08 & 6.41 & 0.08 & 0.98 & 0.09 & 0.32 & 0.07 & 0.28 \\
G93 & 2.5 & 6.6 & 2.6 & 5.7 & 0.06 & 3.63 & 0.06 & 1.20 & 0.07 & 0.33 & 0.05 & 0.34 \\
G107 & -12.5 & -8.5 & -12.3 & -9.5 & 0.07 & 3.68 & 0.07 & 0.92 & 0.07 & 0.29 & 0.07 & 0.27 \\
\enddata
\tablecomments{Column 1: Field name of mapped area, Column 2-3: Lower and higher limits of the velocity integration ranges for integrated intensity map of {\nth} line, Column 4-5: Lower and higher limits of velocity ranges for integrated intensity map of {\hcn}, {\cch}, and {\ccl} lines, Column 6-7: r.m.s and peak values of {\nth} integrated intensity, Column 8-9: r.m.s and peak values of {\hcn} integrated intensity, Column 10-11: r.m.s and peak values of {\cch} integrated intensity, and Column 12-13: r.m.s and peak values of {\ccl} integrated intensity.}
\end{deluxetable*}
\end{longrotatetable}

\figsetstart
\figsetnum{19}
\figsettitle{N$_2$H$^+$, HC$_3$N, CCS-H, and CCS-L maps overlaid on the JCMT SCUBA-2 map for the Appendix.}

\figsetgrpstart
\figsetgrpnum{19.1}
\figsetgrptitle{G192
}
\figsetplot{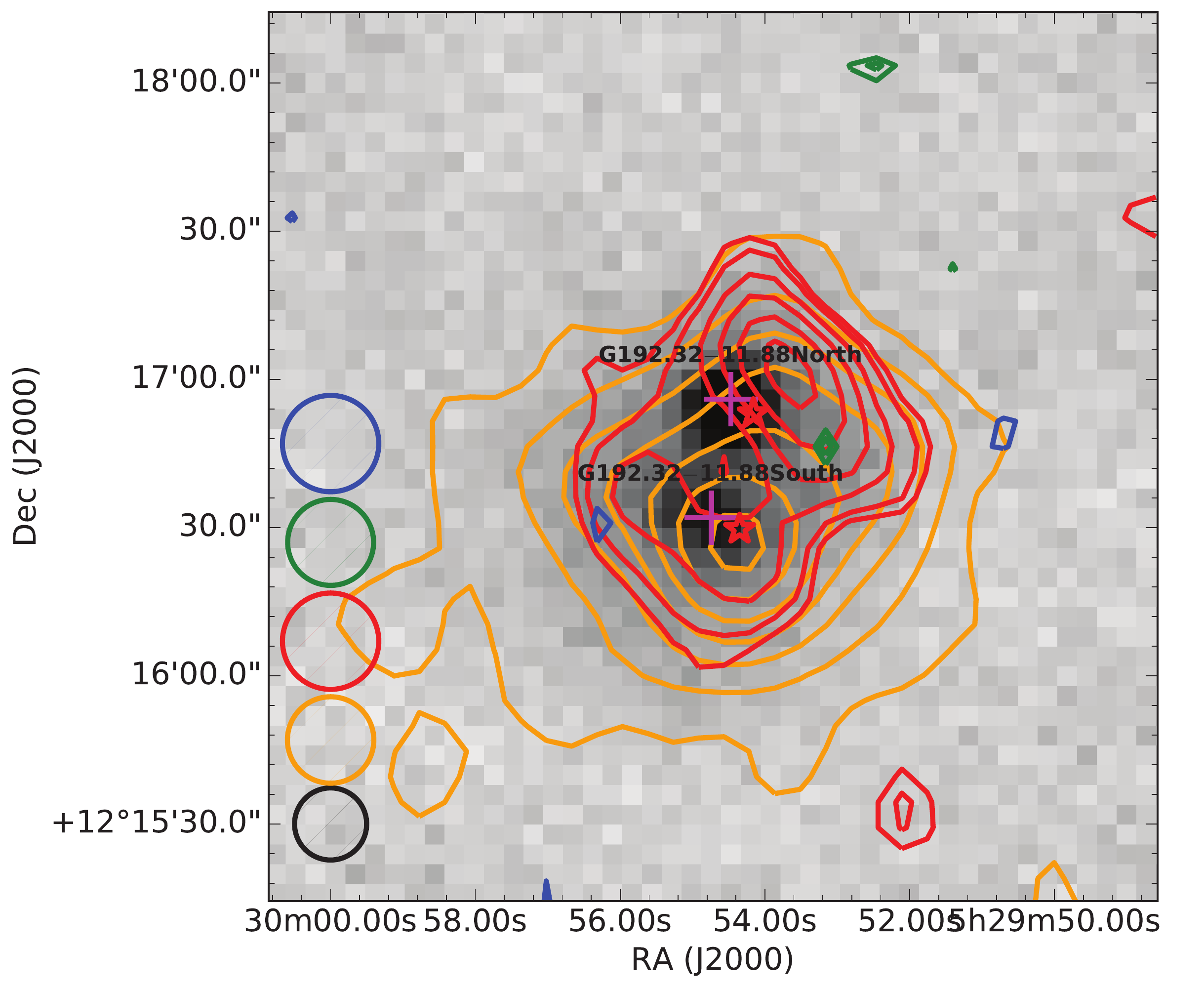}
\figsetgrpnote{Same as Figure 1 but for the specified field}
\figsetgrpend

\figsetend

\begin{figure}
\figurenum{19}
\includegraphics[bb=0 0 600 600, width=10cm]{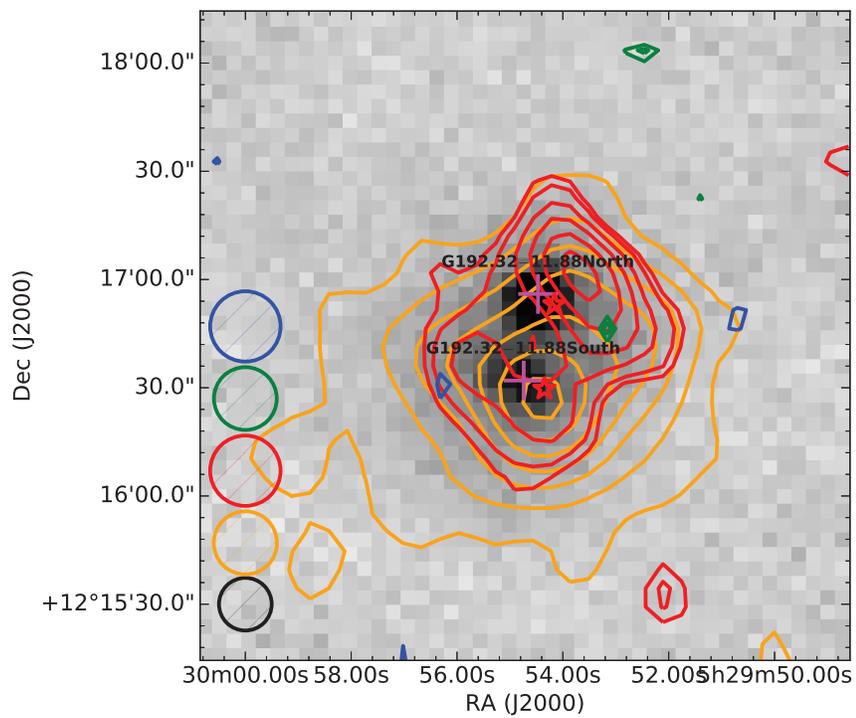}
\caption{Same as Figure 1 but for the specified field}
\end{figure}

\newpage


\begin{longrotatetable}
\begin{deluxetable*}{lCCCCCCCC} 
\tablecaption{Line emission distribution in the Orion region (1) \label{tbl:core4line1}}
\tablewidth{0pt}
\tabletypesize{\scriptsize}
\tablenum{A4}
\tablehead{
\colhead{SCUBA-2 core} &  
\multicolumn{4}{c}{{\nth}} & \multicolumn{4}{c}{HC$_3$N} \\
\cline{2-5}
\cline{6-9}
\colhead{} & 
\colhead{R.A.(J2000)} & \colhead{Dec.(J2000)} &\colhead{$\theta_{\rm offset}$} & \colhead{$R$} & 
\colhead{R.A.(J2000)} & \colhead{Dec.(J2000)} & \colhead{$\theta_{\rm offset}$} & \colhead{$R$}  
\\
\colhead{} & 
\colhead{hh:mm:ss.ss} & \colhead{dd:mm:ss.s} & \colhead{pc} & \colhead{pc} & 
\colhead{hh:mm:ss.ss} & \colhead{dd:mm:ss.s} & \colhead{pc} & \colhead{pc}  
}
\decimalcolnumbers
\startdata
G192.32$-$11.88North	&		. . .	&			. . .	&		. . .	&		. . .	&	05:29:53.90		&		12:17:00.3		&	0.016 		&	0.027 		\\
G192.32$-$11.88South	&	05:29:54.29		&		12:16:33.9		&	0.013 		&	0.045 		&	05:29:54.60		&		12:16:38.9		&	0.004 		&	0.054 		\\
G203.21$-$11.20East1	&	05:53:49.05		&		03:23:08.1		&	0.059 		&	0.080 		&	05:53:51.16		&		03:23:07.3		&	0.001 		&	0.067 		\\
G203.21$-$11.20West1	&		. . .	&			. . .	&		. . .	&		. . .	&		. . .	&			. . .	&		. . .	&		. . .	\\
G203.21$-$11.20West2	&	05:53:41.36		&		03:22:31.2		&	0.049 		&	0.117 		&	05:53:39.50		&		03:22:26.8		&	0.003 		&	0.096 		\\
G204.4$-$11.3A2East	&	05:55:37.98		&		02:11:24.6		&	0.013 		&	0.053 		&	05:55:38.50		&		02:11:19.9		&	0.002 		&	0.060 		\\
G206.12$-$15.76	&	05:42:45.79		&	-01:16:11.4		&	0.015 		&	0.051 		&	05:42:46.43		&	-01:16:15.7		&	0.033 		&	0.053 		\\
G206.21$-$16.17North	&	05:41:39.31		&	-01:35:40.4		&	0.001 		&	0.057 		&		. . .	&			. . .	&		. . .	&		. . .	\\
G206.93$-$16.61West1	&	05:41:25.92		&	-02:16:33.2		&	0.011 		&	0.087 		&	05:41:26.19		&	-02:16:19.2		&	0.019 		&	0.067 		\\
G206.93$-$16.61West3	&	05:41:25.81		&	-02:17:37.5		&	0.024 		&	0.097 		&	05:41:24.88		&	-02:17:10.3		&	0.007 		&	0.095 		\\
G206.93$-$16.61West4	&	05:41:26.15		&	-02:19:20.7		&	0.009 		&	0.059 		&		. . .	&			. . .	&		. . .	&		. . .	\\
G206.93$-$16.61West5	&	05:41:29.97		&	-02:20:10.8		&	0.037 		&	0.080 		&		. . .	&			. . .	&		. . .	&		. . .	\\
G206.93$-$16.61West6	&	05:41:31.03		&	-02:20:57.4		&	0.045 		&	0.080 		&		. . .	&			. . .	&		. . .	&		. . .	\\
G207.36$-$19.82North1	&	05:30:49.62		&	-04:10:28.0		&	0.038 		&	0.063 		&		. . .	&			. . .	&		. . .	&		. . .	\\
G207.36$-$19.82South	&	05:30:47.56		&	-04:12:31.0		&	0.021 		&	0.046 		&		. . .	&			. . .	&		. . .	&		. . .	\\
G208.68$-$19.20North1	&		. . .	&			. . .	&		. . .	&		. . .	&	05:35:23.32		&	-05:01:29.0		&	0.001 		&	0.058 		\\
G208.68$-$19.20North2	&	05:35:19.67		&	-05:00:49.5		&	0.022 		&	0.087 		&		. . .	&			. . .	&		. . .	&		. . .	\\
G208.68$-$19.20North3	&		. . .	&			. . .	&		. . .	&		. . .	&	05:35:17.91		&	-05:00:32.0		&	0.004 		&	0.053 		\\
G208.89$-$20.04East	&	05:32:49.15		&	-05:34:45.3		&	0.021 		&	0.042 		&	05:32:50.13		&	-05:34:52.0		&	0.049 		&	0.053 		\\
G209.05$-$19.73North	&	05:34:04.50		&	-05:32:22.7		&	0.015 		&	0.044 		&	05:34:04.92		&	-05:32:00.4		&	0.028 		&	0.050 		\\
G209.05$-$19.73South	&	05:34:03.97		&	-05:34:10.4		&	0.024 		&	0.049 		&	05:34:04.00		&	-05:34:25.0		&	0.025 		&	0.025 		\\
G209.29$-$19.65North1	&	05:34:58.41		&	-05:40:22.7		&	0.052 		&	0.097 		&		. . .	&			. . .	&		. . .	&		. . .	\\
G209.29$-$19.65South2	&	05:34:54.62		&	-05:46:02.1		&	0.023 		&	0.100 		&		. . .	&			. . .	&		. . .	&		. . .	\\
G209.29$-$19.65South3	&	05:34:50.47		&	-05:45:55.4		&	0.017 		&	0.083 		&		. . .	&			. . .	&		. . .	&		. . .	\\
G209.77$-$19.40East1	&	05:36:32.04		&	-06:01:16.2		&	0.013 		&	0.041 		&	05:36:31.70		&	-06:01:22.3		&	0.024 		&	0.056 		\\
G209.77$-$19.40East2	&	05:36:31.96		&	-06:01:59.4		&	0.007 		&	0.057 		&		. . .	&			. . .	&		. . .	&		. . .	\\
G209.77$-$19.40East3	&	05:36:34.68		&	-06:02:40.5		&	0.040 		&	0.127 		&	05:36:34.88		&	-06:02:55.5		&	0.033 		&	0.137 		\\
G209.94$-$19.52North	&	05:36:11.11		&	-06:10:49.6		&	0.014 		&	0.058 		&	05:36:11.11		&	-06:10:49.8		&	0.014 		&	0.062 		\\
G209.94$-$19.52South1	&	05:36:25.18		&	-06:14:03.3		&	0.007 		&	0.099 		&	05:36:24.92		&	-06:13:44.1		&	0.002 		&	0.087 		\\
G210.82$-$19.47North1	&	05:37:57.06		&	-06:56:33.3		&	0.016 		&	0.051 		&	05:37:57.23		&	-06:56:36.1		&	0.021 		&	0.065 		\\
G211.16$-$19.33North1	&	05:39:11.70		&	-07:10:33.6		&	0.003 		&	0.080 		&	05:39:11.06		&	-07:10:29.1		&	0.023 		&	0.069 		\\
G211.16$-$19.33North2	&	05:39:06.05		&	-07:10:56.2		&	0.005 		&	0.090 		&	05:39:07.73		&	-07:10:21.4		&	0.058 		&	0.114 		\\
G211.16$-$19.33North3	&	05:39:02.95		&	-07:11:15.4		&	0.022 		&	0.065 		&	05:39:03.88		&	-07:11:45.1		&	0.051 		&	0.076 		\\
G211.16$-$19.33North4	&	05:38:53.20		&	-07:11:17.2		&	0.078 		&	0.077 		&	05:38:53.59		&	-07:11:16.0		&	0.066 		&	0.057 		\\
G211.16$-$19.33North5	&	05:38:46.46		&	-07:10:43.6		&	0.014 		&	0.049 		&	05:38:46.66		&	-07:10:42.1		&	0.021 		&	0.056 		\\
G211.16$-$19.33South	&	05:39:07.14		&	-07:12:57.2		&	0.132 		&	0.100 		&	05:39:07.20		&	-07:12:57.5		&	0.134 		&	0.091 		\\
G211.47$-$19.27North	&	05:39:56.39		&	-07:30:15.8		&	0.028 		&	0.136 		&	05:39:55.19		&	-07:29:40.2		&	0.066 		&	0.105 		\\
G211.47$-$19.27South	&	05:39:55.19		&	-07:30:13.1		&	0.023 		&	0.114 		&	05:39:56.07		&	-07:30:17.0		&	0.005 		&	0.082 		\\
G211.72$-$19.25North	&	05:40:13.65		&	-07:32:20.3		&	0.002 		&	0.029 		&	05:40:13.34		&	-07:32:15.3		&	0.012 		&	0.041 		\\
G211.72$-$19.25South1	&		. . .	&			. . .	&		. . .	&		. . .	&	05:40:13.98		&	-07:33:56.5		&	0.160 		&	0.025 		\\
G212.10$-$19.15North1	&	05:41:22.32		&	-07:52:47.2		&	0.024 		&	0.066 		&	05:41:22.88		&	-07:52:42.9		&	0.042 		&	0.030 		\\
G212.10$-$19.15North2	&	05:41:24.52		&	-07:53:53.1		&	0.017 		&	0.051 		&	05:41:24.94		&	-07:53:56.5		&	0.030 		&	0.054 		\\
G212.10$-$19.15North3	&	05:41:21.90		&	-07:55:16.6		&	0.092 		&	0.099 		&	05:41:24.43		&	-07:55:08.7		&	0.012 		&	0.050 		\\
G212.10$-$19.15South	&	05:41:26.54		&	-07:56:53.3		&	0.005 		&	0.041 		&		. . .	&			. . .	&		. . .	&		. . .	\\
\enddata
\tablecomments{Column 1: SCUBA-2 core name, 
Column 2-3: Coordinates of peak intensity for an {\nth} core, 
Column 4: Separation between SCUBA-2 core and {\nth} core in pc, 
Column 5: deconvolved radius of {\nth} core in pc, 
Column 6-7: Coordinates of peak intensity for an HC$_3$N core, 
Column 8: Separation between SCUBA-2 core and HC$_3$N core in pc, and
Column 9: deconvolved radius of HC$_3$N core in pc. 
}
\end{deluxetable*}
\end{longrotatetable}

\begin{longrotatetable}
\begin{deluxetable*}{lCCCCCCCC} 
\tablecaption{Line emission distribution in the Orion region (2) \label{tbl:core4line2}}
\tablewidth{0pt}
\tabletypesize{\scriptsize}
\tablenum{A5}
\tablehead{
\colhead{SCUBA-2 core} &  
\multicolumn{4}{c}{CCS-H} & \multicolumn{4}{c}{CCS-L} \\
\cline{2-5}
\cline{6-9}
\colhead{} & 
\colhead{R.A.(J2000)} & \colhead{Dec.(J2000)} &\colhead{$\theta_{\rm offset}$} & \colhead{$R$} & 
\colhead{R.A.(J2000)} & \colhead{Dec.(J2000)} & \colhead{$\theta_{\rm offset}$} & \colhead{$R$}  
\\
\colhead{} & 
\colhead{hh:mm:ss.ss} & \colhead{dd:mm:ss.s} & \colhead{pc} & \colhead{pc} & 
\colhead{hh:mm:ss.ss} & \colhead{dd:mm:ss.s} & \colhead{pc} & \colhead{pc}  
}
\decimalcolnumbers
\startdata
G203.21$-$11.20West2	&		. . .	&			. . .	&		. . .	&		. . .	&		. . .	&			. . .	&		. . .	&		. . .	\\
G204.4$-$11.3A2East	&	05:55:38.67		&		02:11:18.2		&	0.007 		&	0.064 		&	05:55:38.75		&		02:11:20.2		&	0.009 		&	0.062 		\\
G206.12$-$15.76	&	05:42:47.26		&	-	01:16:19.4		&	0.056 		&	0.087 		&	05:42:47.04		&	-	01:16:15.6		&	0.050 		&	0.074 		\\
G206.21$-$16.17North	&		. . .	&			. . .	&		. . .	&		. . .	&		. . .	&			. . .	&		. . .	&		. . .	\\
G206.93$-$16.61West1	&		. . .	&			. . .	&		. . .	&		. . .	&		. . .	&			. . .	&		. . .	&		. . .	\\
G206.93$-$16.61West3	&		. . .	&			. . .	&		. . .	&		. . .	&		. . .	&			. . .	&		. . .	&		. . .	\\
G206.93$-$16.61West4	&		. . .	&			. . .	&		. . .	&		. . .	&		. . .	&			. . .	&		. . .	&		. . .	\\
G206.93$-$16.61West5	&		. . .	&			. . .	&		. . .	&		. . .	&		. . .	&			. . .	&		. . .	&		. . .	\\
G206.93$-$16.61West6	&		. . .	&			. . .	&		. . .	&		. . .	&		. . .	&			. . .	&		. . .	&		. . .	\\
G207.36$-$19.82North1	&		. . .	&			. . .	&		. . .	&		. . .	&		. . .	&			. . .	&		. . .	&		. . .	\\
G207.36$-$19.82South	&		. . .	&			. . .	&		. . .	&		. . .	&		. . .	&			. . .	&		. . .	&		. . .	\\
G208.68$-$19.20North1	&		. . .	&			. . .	&		. . .	&		. . .	&		. . .	&			. . .	&		. . .	&		. . .	\\
G208.68$-$19.20North2	&		. . .	&			. . .	&		. . .	&		. . .	&	05:35:18.78		&	-	05:01:07.0		&	0.048 		&	0.125 		\\
G208.68$-$19.20North3	&	05:35:17.32		&	-	05:00:33.1		&	0.020 		&	0.068 		&		. . .	&			. . .	&		. . .	&		. . .	\\
G208.89$-$20.04East	&		. . .	&			. . .	&		. . .	&		. . .	&	05:32:50.77		&	-	05:34:57.4		&	0.068 		&	0.046 		\\
G209.05$-$19.73North	&		. . .	&			. . .	&		. . .	&		. . .	&		. . .	&			. . .	&		. . .	&		. . .	\\
G209.05$-$19.73South	&		. . .	&			. . .	&		. . .	&		. . .	&	05:34:03.82		&	-	05:34:57.8		&	0.020 		&	0.021 		\\
G209.29$-$19.65North1	&		. . .	&			. . .	&		. . .	&		. . .	&	05:34:59.50		&	-	05:40:25.1		&	0.021 		&		. . .	\\
G209.29$-$19.65South2	&		. . .	&			. . .	&		. . .	&		. . .	&		. . .	&			. . .	&		. . .	&		. . .	\\
G209.29$-$19.65South3	&		. . .	&			. . .	&		. . .	&		. . .	&		. . .	&			. . .	&		. . .	&		. . .	\\
G209.77$-$19.40East1	&		. . .	&			. . .	&		. . .	&		. . .	&		. . .	&			. . .	&		. . .	&		. . .	\\
G209.77$-$19.40East2	&		. . .	&			. . .	&		. . .	&		. . .	&		. . .	&			. . .	&		. . .	&		. . .	\\
G209.77$-$19.40East3	&		. . .	&			. . .	&		. . .	&		. . .	&	05:36:34.23		&	-	06:03:19.2		&	0.054 		&	0.141 		\\
G209.94$-$19.52North	&		. . .	&			. . .	&		. . .	&		. . .	&		. . .	&			. . .	&		. . .	&		. . .	\\
G209.94$-$19.52South1	&		. . .	&			. . .	&		. . .	&		. . .	&	05:36:27.14		&	-	06:14:15.0		&	0.069 		&		. . .	\\
G210.82$-$19.47North1	&		. . .	&			. . .	&		. . .	&		. . .	&	05:37:57.16		&	-	06:56:24.4		&	0.019 		&	0.052 		\\
G211.16$-$19.33North1	&		. . .	&			. . .	&		. . .	&		. . .	&		. . .	&			. . .	&		. . .	&		. . .	\\
G211.16$-$19.33North2	&		. . .	&			. . .	&		. . .	&		. . .	&	05:39:09.85		&	-	07:10:40.2		&	0.125 		&	0.123 		\\
G211.16$-$19.33North3	&		. . .	&			. . .	&		. . .	&		. . .	&	05:39:02.10		&	-	07:11:36.1		&	0.006 		&	0.055 		\\
G211.16$-$19.33North4	&	05:38:53.88		&	-	07:11:18.2		&	0.056 		&	0.049 		&	05:38:54.98		&	-	07:11:28.7		&	0.022 		&	0.056 		\\
G211.16$-$19.33North5	&		. . .	&			. . .	&		. . .	&		. . .	&		. . .	&			. . .	&		. . .	&		. . .	\\
G211.16$-$19.33South	&		. . .	&			. . .	&		. . .	&		. . .	&	05:39:05.35		&	-	07:12:17.2		&	0.076 		&		. . .	\\
G211.47$-$19.27North	&		. . .	&			. . .	&		. . .	&		. . .	&	05:39:55.40		&	-	07:29:51.4		&	0.059 		&	0.159 		\\
G211.47$-$19.27South	&	05:39:56.44		&	-	07:30:31.3		&	0.016 		&	0.085 		&	05:39:56.60		&	-	07:30:29.2		&	0.021 		&	0.088 		\\
G211.72$-$19.25North	&		. . .	&			. . .	&		. . .	&		. . .	&		. . .	&			. . .	&		. . .	&		. . .	\\
G211.72$-$19.25South1	&		. . .	&			. . .	&		. . .	&		. . .	&	05:40:14.03		&	-	07:34:00.1		&	0.158 		&	0.041 		\\
G212.10$-$19.15North1	&	05:41:21.73		&	-	07:52:37.9		&	0.005 		&	0.104 		&	05:41:21.90		&	-	07:52:39.4		&	0.011 		&	0.091 		\\
G212.10$-$19.15North2	&		. . .	&			. . .	&		. . .	&		. . .	&		. . .	&			. . .	&		. . .	&		. . .	\\
G212.10$-$19.15North3	&	05:41:24.82		&	-	07:55:07.1		&	0.0001 		&	0.038 		&	05:41:26.68		&	-	07:55:06.5		&	0.059 		&	0.045 		\\
G212.10$-$19.15South	&		. . .	&			. . .	&		. . .	&		. . .	&		. . .	&			. . .	&		. . .	&		. . .	\\
\enddata
\tablecomments{Column 1: SCUBA-2 core name, 
Column 2-3: Coordinates of peak intensity for a CCS-H core, 
Column 4: Separation between SCUBA-2 core and CCS-H core in pc, 
Column 5: deconvolved radius of CCS-H core in pc, 
Column 6-7: Coordinates of peak intensity for a CCS-L core, 
Column 8: Separation between SCUBA-2 core and CCS-L core in pc, and
Column 9: deconvolved radius of CCS-L core in pc. 
}
\end{deluxetable*}
\end{longrotatetable}



\end{document}